\documentclass[fleqn,10pt]{wlscirep}

\usepackage{amsmath}
\usepackage{amssymb}
\usepackage{bm}
\usepackage{graphicx}
\usepackage{braket}
\usepackage{physics}
\usepackage{float}
\usepackage{hyperref}
\usepackage{pdflscape}
\usepackage{tikz}
\usetikzlibrary{arrows.meta, calc}
\usepackage{physics}   
\usepackage{xcolor}
\usepackage{float}

\begin{document}

\title{Machine-Learned Compact Subspace Generation for Quantum Selected Configuration Interaction within Density Matrix Embedding Framework}

\author[1]{\small Ashish Kumar Patra}
\author[1]{\small Anurag K. S. V.}
\author[1, 2]{\small Ruchika Bhat}
\author[1]{\small Sai Shankar P.}
\author[3]{\small Rahul Maitra$^\dagger$,}
\author[, 1]{\small Jaiganesh G.\thanks{Corresponding Author email: jaiganesh@qclairvoyance.in, drjaiganesh15@gmail.com\newline\makebox[1.8em][r]{$^\dagger$}email: rmaitra@chem.iitb.ac.in}}
\affil[1]{\small Qclairvoyance Quantum Labs, Secunderabad, TG 500094, India.}
\affil[2]{\small The University of Arizona, Tucson, AZ 85721, USA.}
\affil[3]{\small Indian Institute of Technology Bombay, Mumbai, MH 400076, India.}


\begin{abstract}
Sample-based Quantum Diagonalization (SQD), an extension of Quantum Selected Configuration Interaction (QSCI), has emerged as a promising hybrid quantum-classical paradigm for computing molecular ground state energies. By leveraging quantum sampling instead of variational optimization, QSCI avoids barren plateaus and enables direct reconstruction of correlated electronic wavefunctions. However, existing configuration recovery techniques primarily enforce symmetry constraints without guaranteeing optimal selection of the most physically relevant configurations, often leading to unnecessarily large subspaces and increased classical diagonalization costs. In this work, we introduce a machine-learned compact subspace generation protocol based on Restricted Boltzmann Machines (RBMs), termed QSCI-RBM, and integrate it within the Density Matrix Embedding Theory (DMET) framework. All quantum sampling underlying this work was performed extensively on IBM
Heron-generation hardware, spanning three distinct processors: IBM Heron R1 (\texttt{ibm\_torino}) and
IBM Heron R2 (\texttt{ibm\_fez}) for two small-molecule validation systems, and IBM Heron R3
(\texttt{ibm\_boston}) for a third, larger small-molecule validation system as well as the full
protein-ligand embedding calculations. The RBM is trained directly on these hardware-sampled configurations to learn the underlying probability distribution of dominant determinants, enabling the targeted generation of high-probability configurations without recourse to perturbative seeding. Validated first on three small-molecule benchmarks of increasing active-space size, we then apply
this framework to the simulation of a protein-ligand complex involving the inhibitor Carmofur bound to the
SARS-CoV-2 main protease (M\textsuperscript{pro}), fragmented into ten DMET impurities, with both DMET-QSCI-RBM and a DMET-SQD baseline reaching full chemical-potential self-consistency within the same four $\mu$-iterations. At the converged chemical potential, DMET-QSCI-RBM achieves an energy within $0.05\,\mathrm{mHa}$ of the DMET-FCI reference while accessing only $0.22\%$ of the symmetry-preserving configuration space (or symmetry space) on average, an order-of-magnitude tighter accuracy than DMET-SQD ($0.55\,\mathrm{mHa}$) achieved while accessing $33.0\%$ of the subspace on average, leading to a $\sim$153-fold reduction in subspace size at converged chemical potentials. These findings establish that RBM-assisted subspace expansion produces significantly more compact, hardware-derived determinant spaces while preserving or exceeding the physical accuracy of standard configuration-recovery-based QSCI, thereby reducing classical computational overhead and enabling the scalable quantum embedding simulation of complex biological systems entirely on real quantum hardware.
\end{abstract}

\maketitle


\section{Introduction}

Accurate simulation of correlated electronic structure remains one of the central challenges in quantum chemistry. Exact diagonalization methods such as Full Configuration Interaction (FCI) yield formally exact solutions to the electronic Schr\"{o}dinger equation within a given basis, but the size of the determinant space grows combinatorially with the number of orbitals and electrons \cite{jensen2007introduction_fci_size}, making even distributed implementations on modern supercomputers intractable beyond small or moderately sized molecular systems. Recent algorithmic breakthroughs have pushed this frontier considerably: distributed FCI implementations have reached one trillion determinants for the propane molecule \cite{propane_fci_Gao2024}, massively parallel CASCI calculations have accessed active spaces as large as 24 electrons in 24 orbitals \cite{Vogiatzis2017}, and lossless categorical compression strategies have most recently extended exact diagonalization to the quadrillion-determinant scale for heavy-element systems \cite{Shayit2025}. Yet despite these heroic classical efforts, the non-polynomial scaling persists, and the Hilbert space of a moderate-sized drug molecule or catalytic center remains orders of magnitude beyond even exascale reach. Classical approximation methods such as Coupled Cluster theory \cite{bartlett_cc_theory_2007} and Selected Configuration Interaction \cite{sci_evangelista_1} partially alleviate this challenge by truncating the many-body expansion to the most energetically relevant configurations, but they still encounter steep scaling walls in strongly correlated regimes where the number of important determinants itself proliferates.

Hybrid quantum-classical algorithms offer a promising pathway to overcome these limitations by offloading the exponentially scaling part of the problem onto quantum hardware while leaving classical post-processing to conventional computers. Fault-tolerant approaches such as Quantum Phase Estimation (QPE) \cite{kitaev1995abelian_qpe} can in principle extract molecular eigenvalues with asymptotically optimal precision scaling, but their reliance on deep, fully error-corrected circuits places them firmly beyond current Noisy Intermediate-Scale Quantum (NISQ)\cite{Preskill_2018} hardware. Variational Quantum Eigensolvers (VQE) \cite{Peruzzo2014_vqe, KSV2025, patra2025survey} were consequently among the first NISQ-compatible algorithms proposed for quantum chemistry and have since been applied to problems ranging from small-molecule ground-state energies \cite{Belaloui2025_current_vqe} to protein–ligand binding energy estimation \cite{wiley_protein_ligand_interaction} and embedded active-space simulations on trapped-ion hardware \cite{Kawashima2021_dmet_vqe_ion}. Despite this progress, VQE and related variational approaches suffer from well-documented optimization pathologies, including barren plateaus in the cost-function landscape that can arise from both circuit depth and hardware noise \cite{mcclean2018barren, Wang2021, KSV2025}, optimizer instability, and heightened sensitivity to measurement overhead, all of which become more severe as circuit depth and qubit count grow.

Sample-based Quantum Diagonalization (SQD), also referred to as Quantum Selected Configuration Interaction (QSCI), represents a fundamentally different paradigm for combining quantum and classical resources \cite{sqd_first_paper, kanno2023qsci}. Rather than variationally optimizing a parameterized circuit, QSCI prepares an approximate quantum state and directly samples bitstrings in the computational basis; classical post-processing then uses these samples to construct a compact many-body Hamiltonian in the sampled subspace, which is diagonalized exactly using classical eigensolvers such as the Davidson method \cite{Davidson1975_iterative_eigensolver}. Because the quantum device is used only for sampling rather than for iterative optimization, this approach sidesteps the barren-plateau and optimizer-instability issues that plague VQE, and it has been demonstrated to reach chemical accuracy for systems well beyond the reach of exact diagonalization, including transition-metal clusters and dissociation curves on real quantum-centric supercomputing architectures \cite{sqd_first_paper}, with recent closed-loop demonstrations coupling quantum processors to full-scale classical supercomputers \cite{Shirakawa2025}. The QSCI/SQD paradigm has since been rapidly extended along several fronts: adaptive input-state construction to reduce ansatz-optimization overhead \cite{Nakagawa2024}, Hamiltonian-simulation-based sampling that accesses chemically relevant subspaces without variational loops \cite{Sugisaki2025, Mikkelsen2025}, excited-state calculations \cite{Barison2025_QCExcitedStates_SQD}, open-shell systems \cite{Liepuoniute2025}, implicit solvation \cite{Kaliakin2025a}, non-covalent interactions \cite{Kaliakin2025b}, materials simulations \cite{Alexeev2024_QCSMaterials_sqd, elgammal2026additivebindingenergiesasphalt}, and Krylov-subspace variants offering polynomial-time convergence guarantees \cite{Yu2025, Yoshioka2025}, including randomized-compilation \cite{piccinelli2025_sqdrift} and unitary-decomposition \cite{Asthana2025} formulations that further reduce algorithmic error. To recover dynamic correlation missing from the selected subspace, QSCI has also been hybridized with phaseless auxiliary-field quantum Monte Carlo \cite{Huggins2022, Yoshida2025, Danilov2025}, and GPU-accelerated classical diagonalization kernels have been developed to support the growing subspace sizes demanded by these workflows \cite{Walkup2026, Doi2026}.

The success of QSCI, however, critically depends on the quality of the sampled configuration subspace~\cite{kanno2023qsci, vaquerosabater2026noiseconfigurationrecoveryimpact, kirby2026observationimprovedaccuracyclassical}. Raw hardware samples are noisy and generally violate physical symmetry constraints such as fixed electron number and total spin; existing approaches therefore use configuration recovery techniques to project the sampled bitstrings back onto the physically valid symmetry sector \cite{sqd_first_paper}. While effective at restoring physical validity, these recovery procedures do not guarantee an optimal selection of the dominant configurations that contribute to the ground-state wavefunction, and recent analyses have highlighted critical limitations of QSCI-type subspace construction, including unfavorable scaling of the required subspace size and sensitivity to sampling noise, with configuration interaction expansions from SQD often found to be less compact than classical selection heuristics such as Heat-Bath Configuration Interaction \cite{critical_lims_sqd_2025, Raisuddin2025}. In practice, this often leads to unnecessarily large subspaces that erode the very quantum advantage the method seeks to demonstrate, motivating a broad search for more compact and physically informed subspace-generation strategies. Efforts along this line include compact wavefunction-based selection using stochastic time evolution \cite{Weaving2025}, the Handover Iterative VQE (HI-VQE), which dynamically exchanges configuration data between quantum samplers and classical solvers to produce wavefunctions substantially more compact than classical HCI for strongly correlated systems \cite{hi_vqe_2024, Yoo2026, Ghasemi2026}, and half-qubit architectures that double the addressable active space by mapping qubits to spatial rather than spin orbitals \cite{Smith2025, McFarthing2026, Yoshida2026}.

In parallel, machine learning has emerged as a powerful tool for learning quantum state distributions and representing many-body wavefunctions. Generative models such as Restricted Boltzmann Machines (RBMs) have demonstrated success in variationally representing strongly correlated quantum states and in efficiently sampling physically relevant configurations \cite{carleo_2017, Nomura_RBM_Learning_2017}. Related generative-learning approaches have also been applied directly to configuration-space wavefunction solvers, including generative-machine-learning-driven configuration interaction (CIGen) \cite{Herzog_CIGen_2023} and machine-learning-guided configuration interaction (MLCI) \cite{Coe_MLCI_2018}, both of which iteratively learn to propose determinants that most efficiently compress the many-body wavefunction. Closer to the QSCI setting itself, Physics-Informed Generative machine learning (PIGen-SQD)~\cite{patra2025physicsinformedgenerativemachine} anchors configuration recovery using a hierarchy of M\o{}ller-Plesset perturbative estimates~\cite{Moller1934} seeded from MP2 amplitudes and extended via symbolic tensor contractions to approximate singles, triples, and quadruples with further proliferation by RBM. This perturbative anchoring accelerates convergence and improves accuracy relative to unseeded configuration recovery, particularly under constrained hardware-sampling budgets. These successes naturally motivate the integration of RBM-based generative sampling directly within the QSCI subspace-construction step, replacing or augmenting configuration-recovery heuristics with a learned model that can directly target compact, high-quality determinant subspaces without relying on perturbative seeding.

Furthermore, addressing molecular systems of pharmaceutically relevant size requires a
fragmentation strategy that keeps the quantum resource requirement of each subproblem
hardware-tractable. Density Matrix Embedding Theory (DMET)
\cite{Knizia2012_DMET}, closely related in spirit to dynamical mean-field theory
\cite{Georges1996_DMFT}, provides a rigorous embedding framework that partitions a
large molecular system into smaller impurity fragments, each coupled self-consistently
to a mean-field bath that preserves entanglement with the rest of the system
\cite{wouters2016dmets_guide}. Recent theoretical analyses further refines this
understanding of DMET's convergence behavior around the non-interacting limit
\cite{Cances2025, Negre2025}. Because each impurity problem involves a classically tractable number of active orbitals , DMET is a natural fragmentation
scheme for coupling quantum subroutines to large-scale chemistry. It has previously
been combined with VQE for protein-ligand interaction quantification
\cite{wiley_protein_ligand_interaction} and with dynamically constructed, operator-commutativity-screened
ansätze~\cite{dibyendu2026dmetcompass}. 

More recently, it has been used with SQD to enable
quantum-centric simulation of extended molecular systems \cite{dmet_sqd}. This
DMET-SQD combination has since been extended to ligand-like molecules
\cite{Patra2026, ksv2026iqm}, used as a parallel fragment solver within the localized active space
self-consistent field method to treat iron-sulfur clusters intractable for standard
LASSCF \cite{Wang2025}, coupled with quantum bootstrap embedding to recover
correlation energy in large molecular systems \cite{Bierman2026}, and pushed toward
macroscopic biological targets through heterogeneous quantum-classical supercomputing
workflows spanning over 12,000 atoms \cite{Shajan2026}. A complementary embedded
wavefunction (EWF) framework \cite{ewf_embedding_2022} has similarly been coupled with SQD and demonstrated on IBM quantum hardware to
compute molecular electronic structure for a 300-atom Trp-cage miniprotein
\cite{iijima2023accuratequantumchemicalcalculations}, achieving accuracy competitive
with CCSD fragment benchmarks \cite{Shajan2026}. Integrating SQD with DMET in this
way enables scalable quantum simulations of large molecular systems while keeping the
qubit count per fragment constant, at the cost of introducing an additional
self-consistency loop over the chemical potential that couples the fragments to their
environment.

Beyond these DMET-centred approaches, several modern fragmentation schemes aim to
preserve higher-order correlation effects while maintaining computational efficiency.
Coupled-cluster downfolding methods construct effective low-dimensional Hamiltonians
by integrating out inactive degrees of freedom using classical CC theory, yielding
size-consistent and systematically improvable reduced models
\cite{kowalski_cc_downfolding, bauman2025coupledclusterdownfoldingtheory,
kowalski2024resourceadaptivequantumflowalgorithms}.
Variational Localized Active Space methods, such as LASSCF
\cite{Hermes2020, Otten2022}, explicitly partition the active space into weakly
entangled local subspaces that can be optimized independently, providing a natural
interface for hybrid classical-quantum workflows~\cite{Wang2025}. Related embedding
paradigms include the Fragment Molecular Orbital (FMO) method, which decomposes a
large system into overlapping fragment pairs and has been interfaced with VQE
\cite{FMO_VQE_Lim_2024}
to enable quantum treatment of individual fragments within a classically computed
electrostatic environment. Many-body expansion (MBE) methods
\cite{jinlong_2023_mbe, Xu2024}
and Dynamical Mean-Field Theory (DMFT)
\cite{Georges1996_DMFT, blumenthal2025_dmft_intuition}
provide further alternatives for capturing non-local correlation and dynamical effects.
Bootstrap Embedding (BE) \cite{ye2020bootstrap, Liu2023, Hardikar2024, cho2025quemb}
builds upon DMET by enforcing consistency of reduced density matrices across fragment
boundaries, and has recently been extended to quantum hardware settings. In parallel,
several approaches have focused on adapting classical fragmentation techniques directly
for quantum simulation algorithms: notable examples include DMET-VQE implementations
on superconducting and trapped-ion platforms
\cite{wiley_protein_ligand_interaction, Kawashima2021_dmet_vqe_ion,
iijima2023accuratequantumchemicalcalculations}, divide-and-conquer VQE (DC-VQE)
\cite{fujii_2022_deep_vqe}
which recursively decomposes the system into subsystems solved by shallow local
circuits, and EWF-SQD \cite{iijima2023accuratequantumchemicalcalculations, Shajan2026}
demonstrated on IBM quantum hardware for protein-scale systems. Collectively, these
works define a growing ecosystem of hybrid methods that aim to balance accuracy,
scalability, and near-term hardware constraints in quantum-enabled electronic
structure calculations.

In this work, we combine these advances into a unified framework termed DMET-QSCI-RBM, in which an RBM-based generative sampler is trained directly on hardware-sampled bitstrings to construct compact, symmetry-respecting configuration subspaces within each DMET impurity fragment, without recourse to MP2-based perturbative anchoring. We apply this framework to the Carmofur/SARS-CoV-2 M$^{\text{pro}}$ protein-ligand complex~\cite{Jin2020_carmofur}, a pharmaceutical-scale system that we fragment into eleven molecular fragments, and we demonstrate that DMET-QSCI-RBM achieves chemical accuracy while accessing a substantially smaller fraction of the symmetry-adapted configuration space than converged DMET-SQD, establishing subspace compactness as a key practical advantage of learned, hardware-driven configuration selection over both configuration-recovery-based and perturbatively seeded alternatives.



\section{Methodology}

\subsection{RBM-Guided Iterative Subspace Expansion}\label{sec: methodology_qsci_rbm}

\begin{figure}
    \centering
    \includegraphics[
        angle=90,
        width=0.55\textheight
    ]{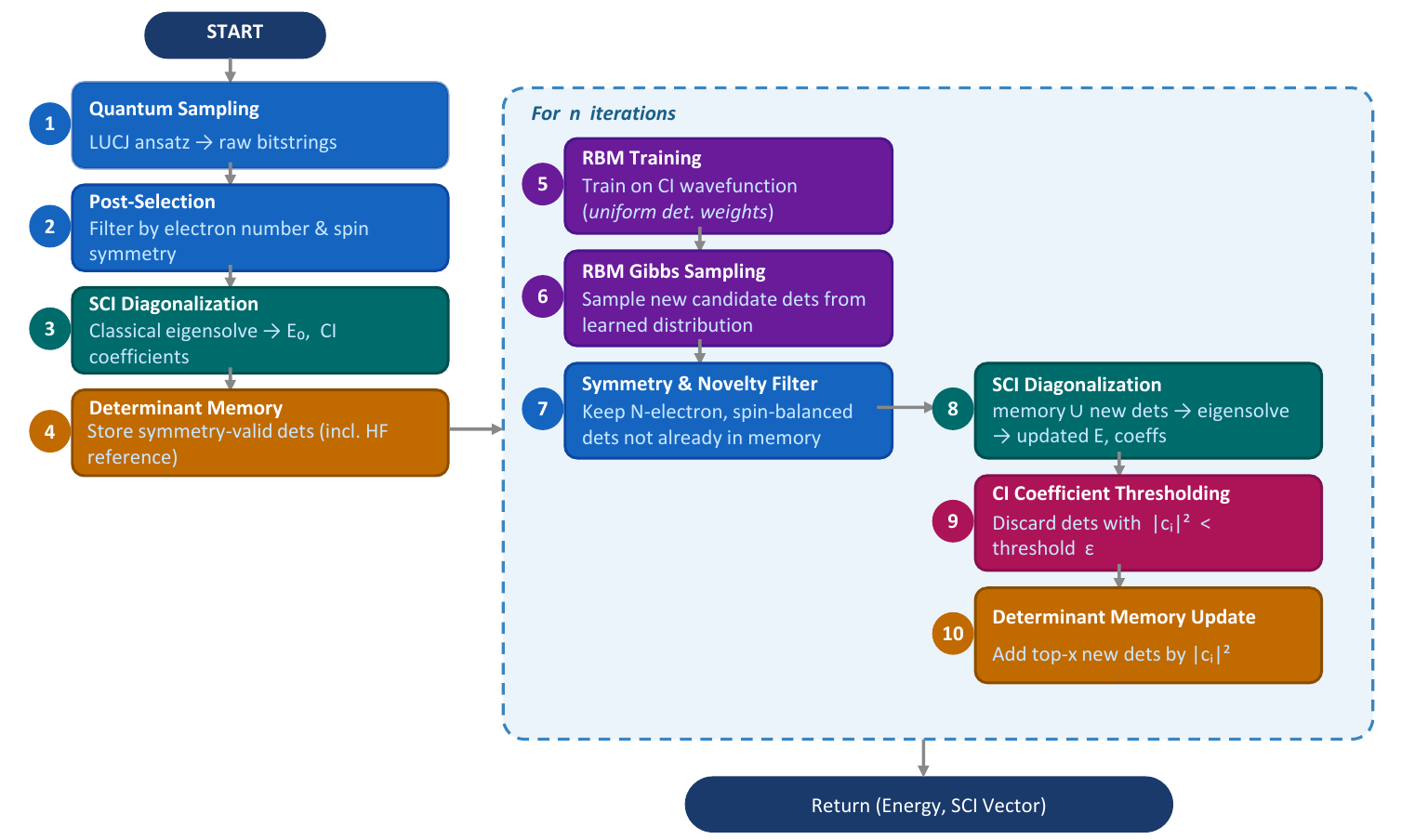}
    \caption{Schematic of the QSCI-RBM iterative workflow: hardware-sampled bitstrings from the
    LUCJ ansatz are post-selected and diagonalized to seed a persistent determinant memory.
    Each iteration trains an RBM on the current wavefunction, Gibbs-samples novel symmetry-valid
    determinants, re-diagonalizes the combined set, and updates the memory with the top
    determinants by $|c_i|^2$ until convergence after $n$ iterations.}
    \label{fig:qsci_rbm_workflow}
\end{figure}

The QSCI-RBM workflow integrates quantum hardware sampling via the Local Unitary Cluster Jastrow
(LUCJ) ansatz with a Restricted Boltzmann Machine (RBM)-guided iterative subspace expansion,
driving classical Selected Configuration Interaction (SCI) diagonalization over a progressively
refined, symmetry-constrained determinant space. The overall procedure is summarized in
Figure~\ref{fig:qsci_rbm_workflow} (Steps 1-10); its defining algorithmic features are highlighted below.

\begin{itemize}

\item \textbf{Post-selection precedes everything else (Steps 1-4).} Raw hardware bitstrings
$\mathbb{S}_{\mathrm{samp}} = \{|\phi_1\rangle, \ldots, |\phi_d\rangle\}$ from LUCJ sampling
(Step~1) are first filtered for particle-number and spin-$z$ conservation (Step~2),
\begin{equation}
    N_\alpha + N_\beta = N_{\mathrm{elec}}, \qquad N_\alpha - N_\beta = 0,
    \label{eq:sym_constraints}
\end{equation}
before any further processing. The surviving symmetry-valid configurations are diagonalized once
(Step~3) to obtain an initial energy and CI coefficients, and are then stored, together with the
mandatory HF reference $|\phi_{\mathrm{HF}}\rangle$, as the initial determinant memory (Step~4),
\begin{equation}
    \mathcal{M}^{(0)} = \left\{|\phi\rangle \in \mathbb{S}_{\mathrm{samp}}
    \;\Big|\; N_\alpha = N_\beta = \tfrac{N_{\mathrm{elec}}}{2} \right\}
    \cup \left\{|\phi_{\mathrm{HF}}\rangle\right\}.
\end{equation}

\item \textbf{The determinant memory is updated at two distinct points within the iterative loop
(Steps 5-10).} First, at Step~7, the symmetry-filtered candidates freshly generated by RBM Gibbs
sampling (Step~6), $\mathcal{G}_{\mathrm{symm}}$, are screened by a combined \emph{symmetry and
novelty filter} which retains only $N$-electron, spin-symmetry obtaining determinants not already present in
memory. These are unioned with the current memory to form the working diagonalization space for that
iteration,
\begin{equation}
    \mathcal{U}^{(t)} = \mathcal{M}^{(t)} \cup \mathcal{G}_{\mathrm{symm}}.
\end{equation}
Second, after spin-string proliferation and SCI diagonalization over $\mathcal{U}^{(t)}$ (Step~8;
Eq.~\eqref{eq:proliferation} below) and CI coefficient thresholding (Step~9), the top-$K$ new
configurations exceeding the threshold $\tau_{\mathrm{sv}}$ are permanently appended to the memory
(Step~10),
\begin{equation}
    \mathcal{M}^{(t+1)} = \mathcal{M}^{(t)} \cup \left\{
        |\phi\rangle \notin \mathcal{M}^{(t)} \;\Big|\;
        \left|\langle\phi|\Psi^{(t)}\rangle\right|^2 > \tau_{\mathrm{sv}}
    \right\}_{\mathrm{top\text{-}}K}.
    \label{eq:memory_update}
\end{equation}
Only this second addition persists across iterations; the first (Step~7) is a temporary expansion
used solely to construct that iteration's diagonalization input, with novelty filtering ensuring
that determinants already resident in memory are never re-admitted or double-counted.

\item \textbf{RBM training uses equal weighting over the top predominant determinants (Step 5).}
At each iteration, the RBM is trained on samples drawn uniformly from the current determinant
memory $\mathcal{M}^{(t)}$ (excluding the HF reference), with every retained determinant assigned
equal sampling weight $w_\phi = 1$ for all $|\phi\rangle \in \mathcal{M}^{(t)} \setminus
\{|\phi_{\mathrm{HF}}\rangle\}$, regardless of its associated CI coefficient magnitude. This is a
deliberate design choice: QSCI-RBM does not attempt to learn the exact ground-state probability
distribution $|c_\phi|^2$, but functions purely as a \emph{sample generator} that has already been
pre-filtered, via the top-$k\%$ post-selection step and the iterative coefficient thresholding of
Eq.~\eqref{eq:memory_update}, to contain only predominant configurations. Equal weighting within
this pre-filtered set then ensures unbiased exploration across the dominant sector at each
iteration, rather than over-concentrating Gibbs sampling around any single high-weight
configuration.

\item \textbf{Spin-string proliferation is applied before diagonalization (Step 8), forming a
second route into the determinant memory.} Rather than diagonalizing directly over the explicit
determinants in $\mathcal{U}^{(t)}$, the unique $\alpha$-configurations $\mathcal{A}_{\mathrm{uniq}}$
and $\beta$-configurations $\mathcal{B}_{\mathrm{uniq}}$ are first extracted from $\mathcal{U}^{(t)}$,
pooled together, and their full Cartesian product is formed as the projection subspace,
\begin{equation}
    \mathcal{P}_{\mathrm{uniq}} = \mathcal{A}_{\mathrm{uniq}} \cup \mathcal{B}_{\mathrm{uniq}},
    \qquad
    \mathbb{S}_{\mathrm{proj}}^{(t)} =
        \mathcal{P}_{\mathrm{uniq}} \otimes \mathcal{P}_{\mathrm{uniq}},
    \label{eq:proliferation}
\end{equation}
prior to SCI diagonalization of the projected Hamiltonian
$\hat{H}_{\mathbb{S}} = \hat{P}_{\mathbb{S}} \hat{H} \hat{P}_{\mathbb{S}}$. This proliferation
step surfaces cross-paired configurations that were never explicitly sampled by hardware or
generated by the RBM, and it is precisely these proliferation-discovered configurations which are
validated by large CI coefficients $|c_\mu|^2$ after diagonalization (Steps 8-9), that
constitute the second, implicit entry pathway into the persistent determinant memory via
Eq.~\eqref{eq:memory_update} (Step 10). Notably, no physics-based perturbative augmentation (e.g.,
MP2- or higher-rank-seeded estimates) is introduced at any stage to supplement this proliferated
space.

\end{itemize}

QSCI-RBM differs from the prior RBM-based configuration recovery scheme, PIGen-SQD~\cite{patra2025physicsinformedgenerativemachine}, at various places including the omission of perturbative augmentation and the equal weighting of training determinants noted above, it seeds the Gibbs chains with unconstrained random bit vectors and imposes the symmetry constraints only afterwards (Steps~6 and~7), replaces the blacklist of negligible-coefficient determinants by the novelty filter and top-$K$ admission of Eq.~\eqref{eq:memory_update}, and runs a fixed iteration budget in place of an energy-based stopping criterion, as seen in the workflow of Figure~\ref{fig:qsci_rbm_workflow}.

\subsection{Integration of QSCI-RBM within DMET}\label{sec:dmet_qsci_rbm_integration}

\begin{figure}
    \centering
    \includegraphics[
        angle=90,
        width=0.65\textheight
    ]{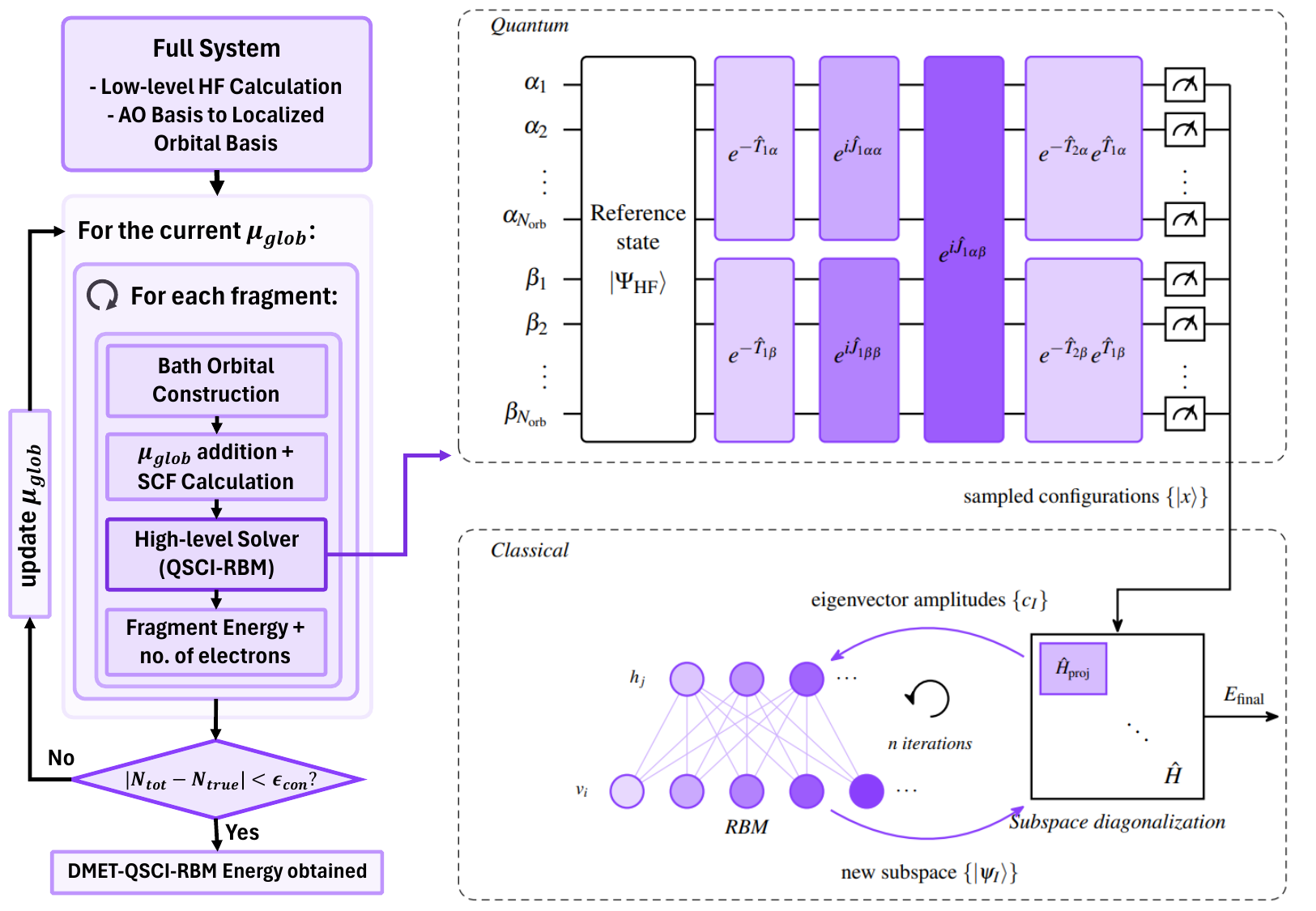}
    \caption{The DMET-QSCI-RBM workflow: the outer DMET self-consistency loop (top) with QSCI-RBM as the
    per-fragment high-level solver (bottom).}
    \label{fig:dmet_qsci_rbm_workflow}
\end{figure}

The theoretical background of DMET is provided in Section~\ref{app: dmet_theory} in the Supplementary Material for reference.The QSCI-RBM solver described above is invoked as the high-level impurity solver within the DMET
self-consistency loop, in place of a conventional FCI or CCSD solver. Figure~\ref{fig:dmet_qsci_rbm_workflow}
summarizes the resulting combined workflow. Starting from a low-level Hartree-Fock calculation of the
full system in the localized orbital basis, each fragment is embedded via bath orbital construction, and
the resulting fragment Hamiltonian is passed to the QSCI-RBM solver, which returns the fragment energy and
electron count. This embedding step is repeated for every fragment at the current global chemical potential
$\mu_{\mathrm{glob}}$, and once all fragments have been solved, $\mu_{\mathrm{glob}}$ is updated via the
Newton-Secant procedure~\cite{scipy_newton} (additional information in Section~\ref{app: dmet_theory} in the Supplementary Information), and the loop is repeated until the
total embedded electron count matches the true electron count to within $\epsilon_{con}$, at which point the
DMET-QSCI-RBM energy is obtained.

Within each fragment solve, the QSCI-RBM pipeline itself proceeds as detailed in
Section~\ref{sec: methodology_qsci_rbm}: a single LUCJ layer of localized orbital rotations $e^{\pm\hat{T}}$
and spin-resolved Jastrow correlators $e^{i\hat{J}}$ acts on the Hartree-Fock reference state, and
measurement of the $\alpha$ and $\beta$ registers yields sampled configurations $\{\ket{x}\}$. These define a
projected Hamiltonian $\hat{H}_{\mathrm{proj}}$, a compact sub-block of the full fragment Hamiltonian
$\hat{H}$, whose diagonalization yields the fragment energy $E_{\mathrm{final}}$ together with eigenvector
amplitudes $\{c_I\}$ used to train an RBM; the RBM's generative sampling then proposes an enriched
configuration subspace for the next diagonalization, and this diagonalization-RBM cycle repeats for $n$
iterations before the converged fragment energy is returned to the outer DMET loop.

\subsection{Experimental Details}
\label{sec:experimental_details}

This section documents the experimental scope, solver configurations, and hardware conditions underlying
the Carmofur/M\textsuperscript{pro} and small-molecule benchmarks: Section~\ref{sec:carmofur_scope} defines the
two solver configurations compared, Section~\ref{sec:spb_rationale} justifies the $\varepsilon_{spb}$ threshold choice.

\subsubsection{Software and Computational Environment}
All classical and quantum-circuit components of the DMET-SQD and DMET-QSCI-RBM
pipelines were implemented using the following software packages. DMET
embedding, bath construction, and the chemical-potential self-consistency loop
for both DMET-SQD and DMET-FCI were carried out using
\verb|Tangelo v0.4.3|\cite{senicourt2022tangeloopensourcepythonpackage}. Local
Unitary Cluster Jastrow (LUCJ) circuits were generated using
\verb|ffsim v0.10.0|\cite{ffsim}, and the resulting quantum circuits were
constructed, transpiled, and executed using \verb|Qiskit v1.4.2|
\cite{javadiabhari2024quantumcomputingqiskit} together with
\verb|Qiskit IBM Runtime v0.36.1|\cite{qiskit_ibm_runtime_2025} for submission
to IBM Quantum hardware. All classical quantum-chemistry computations,
including Hartree-Fock, CCSD, and Selected Configuration Interaction (SCI)
diagonalization, were performed using
\verb|PySCF v2.10.0|\cite{sun2017pyscf}. The Restricted Boltzmann Machine
component of QSCI-RBM was implemented in
\verb|PyTorch v2.10.0|\cite{paszke2019pytorch}. The chemical-potential
root-finding step of the DMET self-consistency loop used the Newton-Secant
method from \verb|scipy v1.15.3|\cite{2020SciPy-NMeth}.

\subsubsection{Experimental Configuration: Solver Scope and Threshold Rationale}
\label{sec:carmofur_scope}
Before presenting the obtained results, we clarify the scope of the Carmofur/M\textsuperscript{pro} experimentation. Two
distinct impurity-solver configurations were benchmarked against a common DMET self-consistency loop: (i)
DMET-SQD with the $\varepsilon_{spb}$ threshold set to $\varepsilon_{spb} = \sqrt{|\mathbb{S}|}$; and (ii)
DMET-QSCI-RBM. Unless otherwise specified, all quantum sampling across both configurations used $100{,}000$
shots per fragment per iteration; hardware error mitigation settings for this and all other experiments in
this work are detailed in Section~\ref{sec:hardware_error_mitigation}.

\paragraph{Rationale for $\varepsilon_{spb}$ Threshold Selection.}
\label{sec:spb_rationale}
The $\varepsilon_{spb} = \sqrt{|\mathbb{S}|}$ setting follows established convention from our group's prior
hardware-based DMET-SQD studies~\cite{ksv2026iqm}, and was adopted here as the standard, literature-consistent
operating point for the DMET-SQD baseline rather than a threshold newly derived for this work. We note that a
more conservative, worst-case a priori bound on the size of the proliferated subspace $S_{\mathrm{proj}}$
relative to the full symmetry space dimension $|\mathbb{S}|$ can be derived by considering the scenario in
which the sampled $\alpha$- and $\beta$-configurations are entirely disjoint following the spin-string
proliferation step of Eq.~\ref{eq:proliferation}; this derivation, provided in Section~\ref{app:spb_derivation}
of the Supplementary Material, yields the tighter bound $\varepsilon_{spb} = \sqrt{|\mathbb{S}|}/2$. The
factor-of-two relaxation to $\varepsilon_{spb} = \sqrt{|\mathbb{S}|}$ used in the main results reflects the
reduced worst-case disjointness expected for closed-shell systems such as Carmofur-M\textsuperscript{pro}, and
should be understood as a practical, convention-consistent choice rather than a claim of optimality; the
derivation in Section~\ref{app:spb_derivation} of the Supplementary Information remains available as a conservative fallback bound for
future work on systems where this closed-shell relaxation may not hold.

The S-CoRe configuration-recovery loop for DMET-SQD was run to convergence for each individual impurity,
i.e., until the accessed subspace and resulting energy stabilized across successive recovery iterations.
DMET-QSCI-RBM was then benchmarked against this DMET-SQD baseline under identical hardware and basis-set
conditions, as detailed in Section~\ref{sec:carmofur_results}.


\section{Validation on Small Molecules}\label{sec:small_molecules_result}

As a benchmark prior to the full protein-ligand simulation, we first validate the
QSCI-RBM framework on two small molecules of increasing size and correlation
complexity: ethylene (C\textsubscript{2}H\textsubscript{4}) and methanolamine
(CH\textsubscript{5}NO), both in the STO-3G basis and run on the IBM Heron Fez
quantum processor.

\subsection{Hardware Error Mitigation for Small-Molecule Validation}
\label{sec:hardware_error_mitigation_smallmol}

For the two small-molecule validation systems of this section, C\textsubscript{2}H\textsubscript{4} and
CH\textsubscript{5}NO, both dynamical decoupling~\cite{dynamical_decoupling_Rahman} and Pauli twirling~\cite{pauli_twirling_foundation_wallman_2016} were enabled to maximize measurement
fidelity, since each system required only a single fragment (no DMET embedding) and could therefore absorb
the additional QPU overhead of twirling. For C\textsubscript{2}H\textsubscript{4}, twirling was configured
with 200 randomizations at 100 shots per randomization (active-circuit strategy). For CH\textsubscript{5}NO,
the twirling budget was adjusted to 700 randomizations at 64 shots per randomization, again using the
active-circuit strategy; dynamical decoupling (XpXm sequence) was applied identically in both cases. This
higher-fidelity regime was computationally affordable at the single-fragment scale of these validation runs.

The exact details of the execution dates and the hardware calibration data pertaining to the IBM Quantum
processors used is provided in Section~\ref{app:execution_dates_calibration} in the Supplementary Material.

\subsection{Ethylene (C\textsubscript{2}H\textsubscript{4})}

\begin{figure}[htbp]
    \centering
    \includegraphics[width=0.94\linewidth]{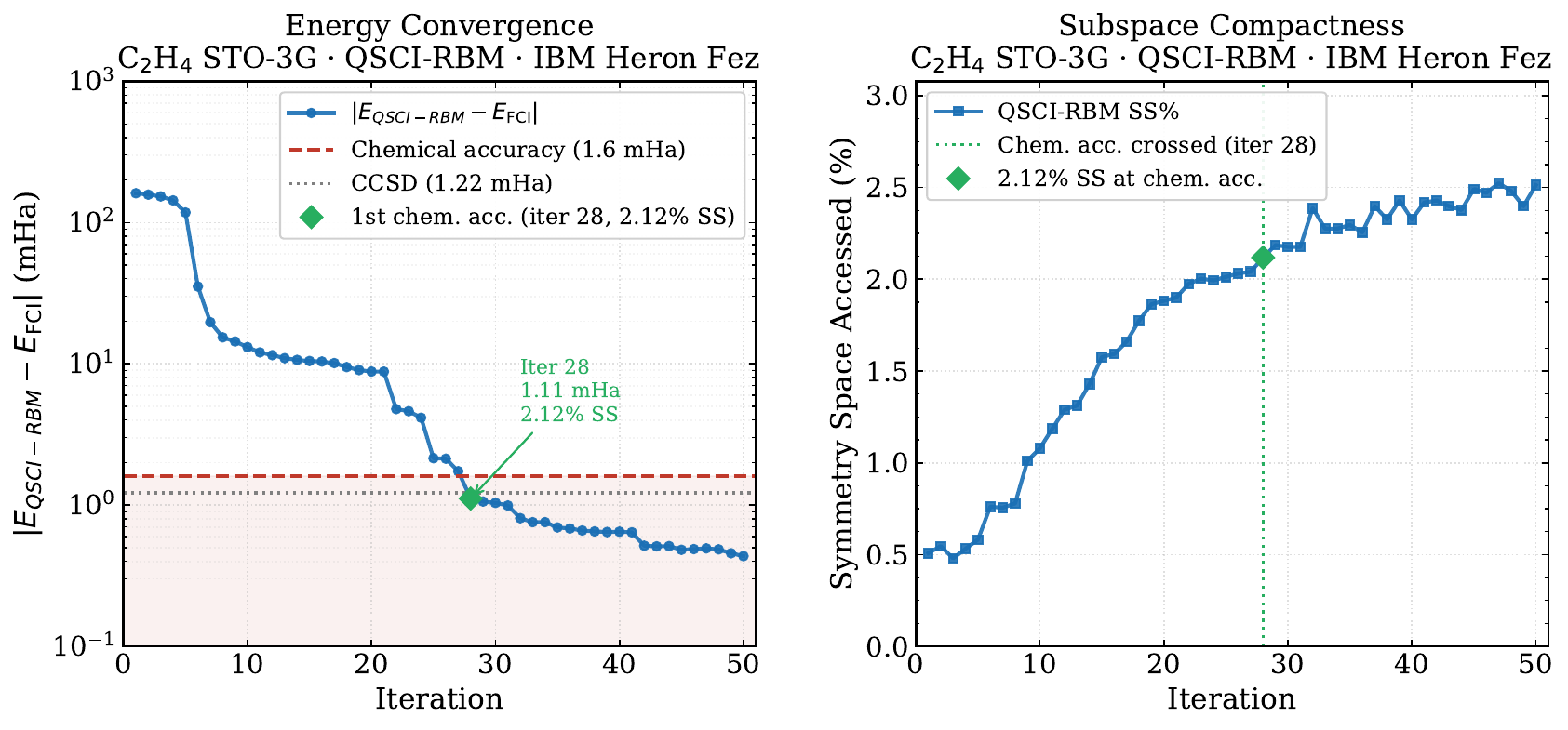}
    \caption{
        \textbf{QSCI-RBM convergence for C\textsubscript{2}H\textsubscript{4} 
        (STO-3G, CAS = 9,018,009, 100K shots, IBM Heron Fez).}
        \textit{Left}: Absolute energy error with respect to FCI as a function of 
        self-consistency iteration (log scale). The dashed red line marks chemical accuracy 
        (1.6\,mHa) and the dotted grey line marks the CCSD reference error (1.22\,mHa); 
        the green diamond indicates the first iteration at which chemical accuracy is 
        achieved (iteration 28, 1.11\,mHa).
        \textit{Right}: Fraction of the symmetry space accessed at each iteration. 
        The subspace grows monotonically until saturating near iteration 28-30, 
        after which mild oscillations are observed around a plateau of $\sim$2.4\%. 
        Chemical accuracy is reached while accessing only 2.12\% of the full configuration 
        space.
    }
    \label{fig:c2h4_convergence}
\end{figure}

The complete active space for C\textsubscript{2}H\textsubscript{4} in the STO-3G
basis comprises 9,018,009 determinants, providing a non-trivial test of subspace
compactness while remaining amenable to exact FCI reference calculations. The
energy convergence and fraction of symmetry space accessed as a function of
self-consistency iteration are shown in Figure~\ref{fig:c2h4_convergence}, using
100K shots.

The absolute energy error with respect to FCI decreases monotonically over the
first $\sim$25 iterations, spanning nearly four orders of magnitude from
$\sim$200\,mHa at iteration 1 to sub-chemical-accuracy at iteration 28, where
the error reaches 1.11\,mHa, which is well below the chemical accuracy threshold of
1.6\,mHa and also below the CCSD reference error of 1.22\,mHa. This is a
notable result: the RBM-guided hardware sampling recovers more correlation energy
than a classical coupled-cluster singles and doubles treatment. Convergence thereafter remains stable, with the
error settling to $\sim$0.4\,mHa in the final iterations, demonstrating that the
RBM-guided subspace expansion consistently recovers correlation energy beyond what
is accessible to CCSD.

Concurrently, the fraction of symmetry space accessed grows monotonically from
$\sim$0.5\% at the first iteration, reflecting the progressive incorporation of
new high-probability determinants generated by Gibbs sampling from the trained
RBM. This growth saturates around iteration 28-30, beyond which the subspace
size exhibits mild oscillations around a plateau of $\sim$2.4\%, indicating that
the RBM has effectively exhausted the dominant sector of the Hilbert space and is
sampling near the boundary of the physically relevant subspace. Crucially,
chemical accuracy is first achieved while accessing only 2.12\% of the full
symmetry space, underscoring the compactness of the RBM-guided subspace relative
to the total configuration space of over nine million determinants.

\subsection{Methanolamine (CH\textsubscript{5}NO)}

\begin{figure}[htbp]
    \centering
    \includegraphics[width=0.94\linewidth]{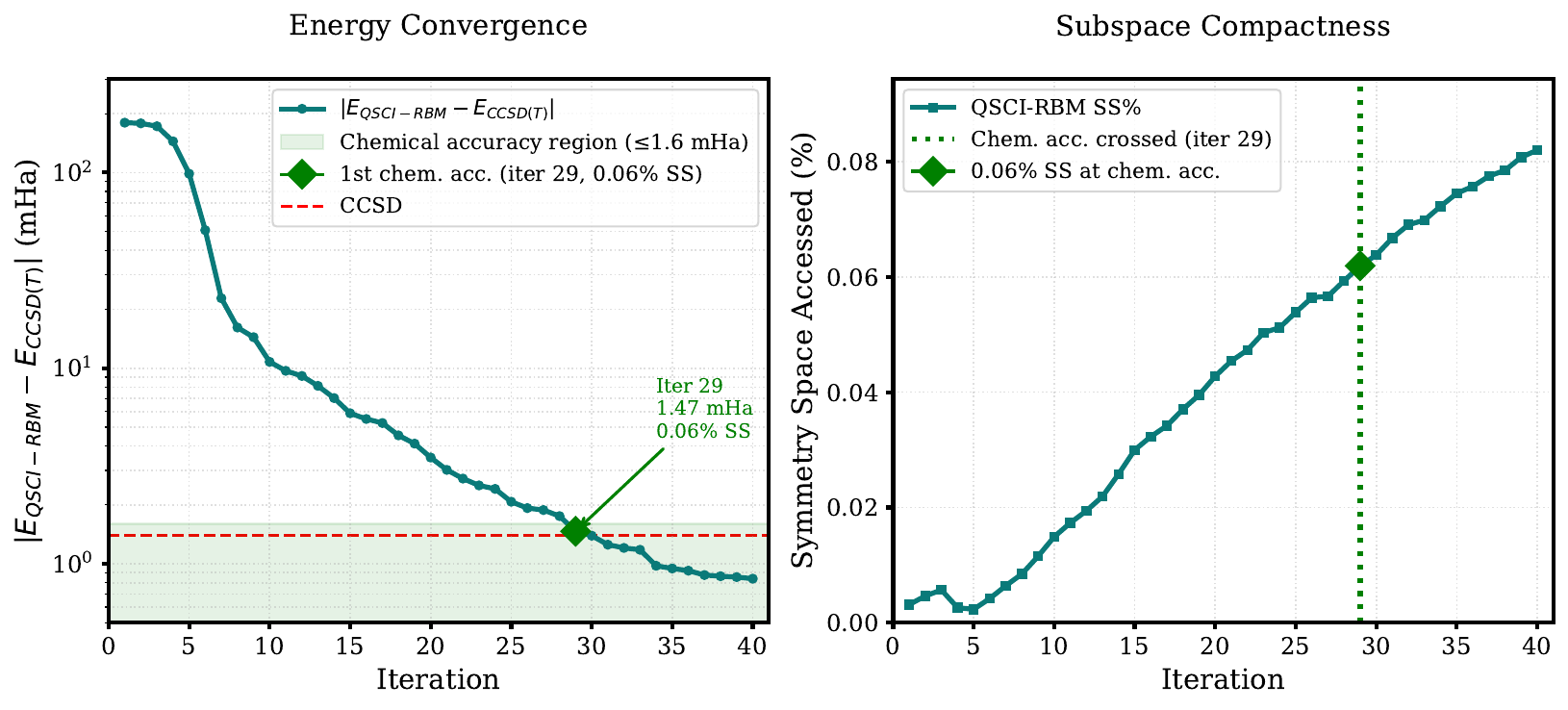}
    \caption{
    \textbf{QSCI-RBM convergence for CH\textsubscript{5}NO 
    (STO-3G,  $|\mathbb{S}|$ $\approx 6.01\times10^{9}$, IBM Heron Fez).}
    \textit{Left}: Absolute energy error with respect to CCSD(T) as a function of 
    self-consistency iteration (log scale). The shaded green band marks chemical accuracy 
    (1.59\,mHa) and the dashed red line marks the CCSD reference error, which itself falls 
    just inside the chemical accuracy region; the green diamond indicates the first 
    iteration at which chemical accuracy is achieved (iteration 29, 1.47\,mHa).
    \textit{Right}: Fraction of the symmetry space accessed at each iteration. 
    The subspace grows monotonically after an initial dip over the first few iterations, 
    with no sharp saturation point, continuing to expand gradually through iteration 40. 
    Chemical accuracy is first reached while accessing only 0.06\% of the full symmetry 
    space.
    }
    \label{fig:ch5no_qsci}
\end{figure}

To assess scalability to a larger active space, we next apply QSCI-RBM to
CH\textsubscript{5}NO in STO-3G, with a CAS of (26e, 20o), corresponding to a
full symmetry-preserving Hilbert space of dimension $\approx 6.01 \times 10^9$, roughly 670$\times$ larger than the C\textsubscript{2}H\textsubscript{4}
case. The convergence profile is shown in Figure~\ref{fig:ch5no_qsci}.

The energy error again exhibits a clear monotonic decrease on a log scale across
the 40 iterations, dropping from $\sim$200\,mHa at iteration 1 to 1.47\,mHa at
iteration 29, where chemical accuracy is first crossed. The CCSD reference error
for this system falls just inside the chemical accuracy region, making
the QSCI-RBM result directly competitive with the classical coupled-cluster
benchmark. Unlike the C\textsubscript{2}H\textsubscript{4} case, the energy
error does not plateau sharply after crossing chemical accuracy but continues to
decrease gradually, suggesting that the RBM has not yet fully saturated the
dominant subspace by iteration 40.

The subspace growth profile for CH\textsubscript{5}NO differs qualitatively from
C\textsubscript{2}H\textsubscript{4}: after an initial dip over the first few
iterations, likely reflecting transient RBM training instability before the
model has seen sufficient symmetry-valid configurations, the symmetry space
accessed grows monotonically and continuously throughout the full 40 iterations,
reaching $\sim$0.085\% by iteration 40 with no clear saturation. Strikingly,
chemical accuracy is first achieved at iteration 29 while accessing only
\textbf{0.06\%} of the full symmetry space of $6.01 \times 10^9$ determinants.
This represents a substantially more compact subspace than the C\textsubscript{2}H\textsubscript{4}
result (0.06\% vs.\ 2.12\%), reflecting the increasingly favourable
compactness-accuracy trade-off of the RBM-guided selection as the Hilbert space
grows: the physically dominant determinants constitute an ever-smaller fraction
of the total space, and the RBM is able to identify and concentrate on this
compact, chemically relevant sector.

\subsection{Water Molecule}

\begin{figure}[htbp]
    \centering
    \includegraphics[width=\textwidth]{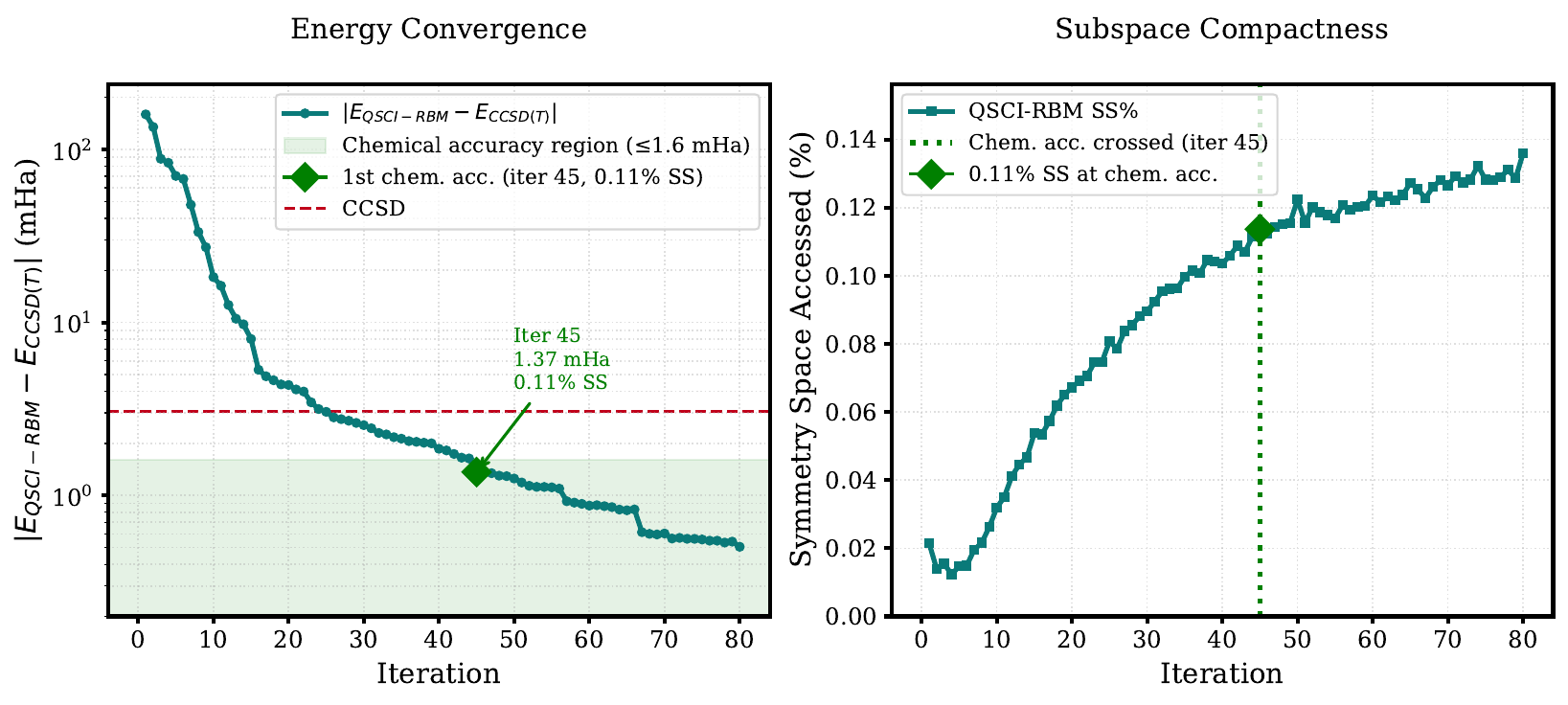}
    \caption{
    \textbf{QSCI-RBM convergence for H\textsubscript{2}O 
    (cc-pVDZ, CAS(10,24), $|\mathbb{S}| = \binom{24}{5}^2 = 1{,}806{,}590{,}016$, 48 qubits, \textbf{IBM Heron R3 Boston}
    ).}
    \textit{Left}: Absolute energy error with respect to CCSD(T) as a function of 
    self-consistency iteration (log scale), shown over 80 iterations. No exact FCI 
    reference exists at this active-space size, and the green diamond marks the first 
    iteration at which chemical accuracy is achieved.
    \textit{Right}: Fraction of the symmetry space accessed at each iteration, showing a
    monotonic but progressively slowing subspace growth.
    }
    \label{fig:h2o_ccpvdz_convergence}
\end{figure}

The H\textsubscript{2}O simulation was performed using the cc-pVDZ basis set\cite{Dunning1989ccpvdz} as shown in Figure~\ref{fig:h2o_ccpvdz_convergence}, with
a full active space of 10 electrons in 24 spatial orbitals (CAS(10,24), 48 qubits under the
Jordan-Wigner mapping), corresponding to a symmetry-preserving Hilbert space of dimension
$\binom{24}{5}^2 = 1{,}806{,}590{,}016$ which is roughly $200\times$ larger than the
C\textsubscript{2}H\textsubscript{4} case, though about $3.3\times$ smaller than the CH\textsubscript{5}NO
active space (CAS(26e,20o), $\binom{20}{13}^2 = 6{,}009{,}350{,}400$) considered above. In
anticipation of slower convergence with larger number of qubits, the QSCI-RBM self-consistency loop was run for 80
iterations, compared to 40--50 for the smaller C\textsubscript{2}H\textsubscript{4} and
CH\textsubscript{5}NO systems, to allow the RBM-guided subspace sufficient iterations to expand
compactly toward the physically dominant sector rather than being cut off prematurely.

Chemical accuracy ($\leq$1.6~mHa) is first reached at iteration 45
(1.37~mHa, 0.11\% of the $\mathbb{S}$ accessed), with a final error of 0.51~mHa observed by iteration 80.
Convergence of the subspace itself is evident in the final 30 or so iterations, where the rate of new
subspace growth per iteration slows markedly, with the accessed fraction plateauing near 0.11 to 0.14\% of
the full symmetry space rather than continuing to expand at the pace observed earlier in the trajectory.
This demonstrates that chemical accuracy is achieved without explosive subspace growth. No exact (FCI)
reference exists at this active-space size similar to the previous case, and errors are reported relative to CCSD/CCSD(T), similar to the case of $CH_5 NO$.

\subsection{Comparison and Hyperparameter Considerations}
Taken together, the results on C\textsubscript{2}H\textsubscript{4}, CH\textsubscript{5}NO, and
H\textsubscript{2}O establish two consistent findings. First, QSCI-RBM achieves
chemical accuracy in all three systems while accessing a remarkably small fraction of
the full symmetry space, 2.12\%, 0.06\%, and 0.11\% respectively, confirming the
compactness of the RBM-guided subspace expansion across qualitatively different
system sizes and correlation landscapes. Second, the energy convergence profiles
are smooth and monotonic throughout, with no signs of instability or collapse,
suggesting that the self-consistency loop between RBM training and QSCI
diagonalization is robust under current hyperparameter settings. An analysis of the
memory accessed and the time utilized is given in Section~\ref{app: water_classical_memory} in the Supplementary Material for the
water molecule, where the experiment was re-run with identical settings.

We note that the hyper-parameters used across these studies, including the RBM
learning rate, number of hidden units, Gibbs sampling chain length, and singular
value threshold, were selected empirically without any
systematic hyperparameter optimisation. The values used in each experiment are
listed in Section~\ref{app : hyperparams} in the Supplementary Information. A dedicated grid search or Bayesian
optimisation over this parameter space would be expected to yield a more compact
subspace, faster convergence, and potentially sub-mHa accuracy at an earlier
iteration, leaving meaningful room for further improvement beyond the results
presented here.

A related, currently fixed hyperparameter is the total number of QSCI-RBM iterations
allotted per calculation. With the exception of the 48-qubit H\textsubscript{2}O case,
which was run for 80 iterations in anticipation of its larger active space, and the
CH\textsubscript{5}NO case, which was run for 40 iterations, every other system considered
in this work, including all fragments of the Carmofur/M\textsuperscript{pro}
protein-ligand complex discussed below, was run for a fixed 50 iterations. This
value is presently a user-supplied threshold rather than a derived quantity, since
the dimension of the subspace ultimately required to reach chemical accuracy is
not known in advance of the calculation. Related methods such as HI-VQE~\cite{hi_vqe_2024}
instead pre-assign a maximum subspace size $k$ before the calculation begins, and
restrict the diagonalization space to at most this size throughout all iterations.
An analogous, computationally-informed thresholding scheme, chosen according to
the classical diagonalization resources available to a given user, could be
introduced into QSCI-RBM in place of the current fixed-iteration-count approach,
and represents a viable direction for future work.

With these benchmarks establishing the validity and compactness of
the QSCI-RBM approach, we proceed to the full DMET-QSCI-RBM treatment of the
Carmofur/SARS-CoV-2 M\textsuperscript{pro} protein-ligand complex.

\section{QSCI-RBM within DMET Framework - Simulation of Carmofur-M\textsuperscript{pro} Complex}
\label{sec:carmofur_results}

Here, we discuss the main result of the paper, where Section~\ref{sec: mol_frag_fractionality} discusses the fragmentation of the molecule and the fractionality spectrum, Section~\ref{sec:f3_f10_hyperparameter_optimization} details the work done in performing the manual hyper-parameter optimization for the extremal fragments based on their fractionality and finally Section~\ref{sec:hardware_error_mitigation} details the error mitigation settings used.

\subsection{Molecular fragmentation and fractionality spectrum}\label{sec: mol_frag_fractionality}

\begin{figure}[htbp]
    \centering
    \includegraphics[width=\linewidth]{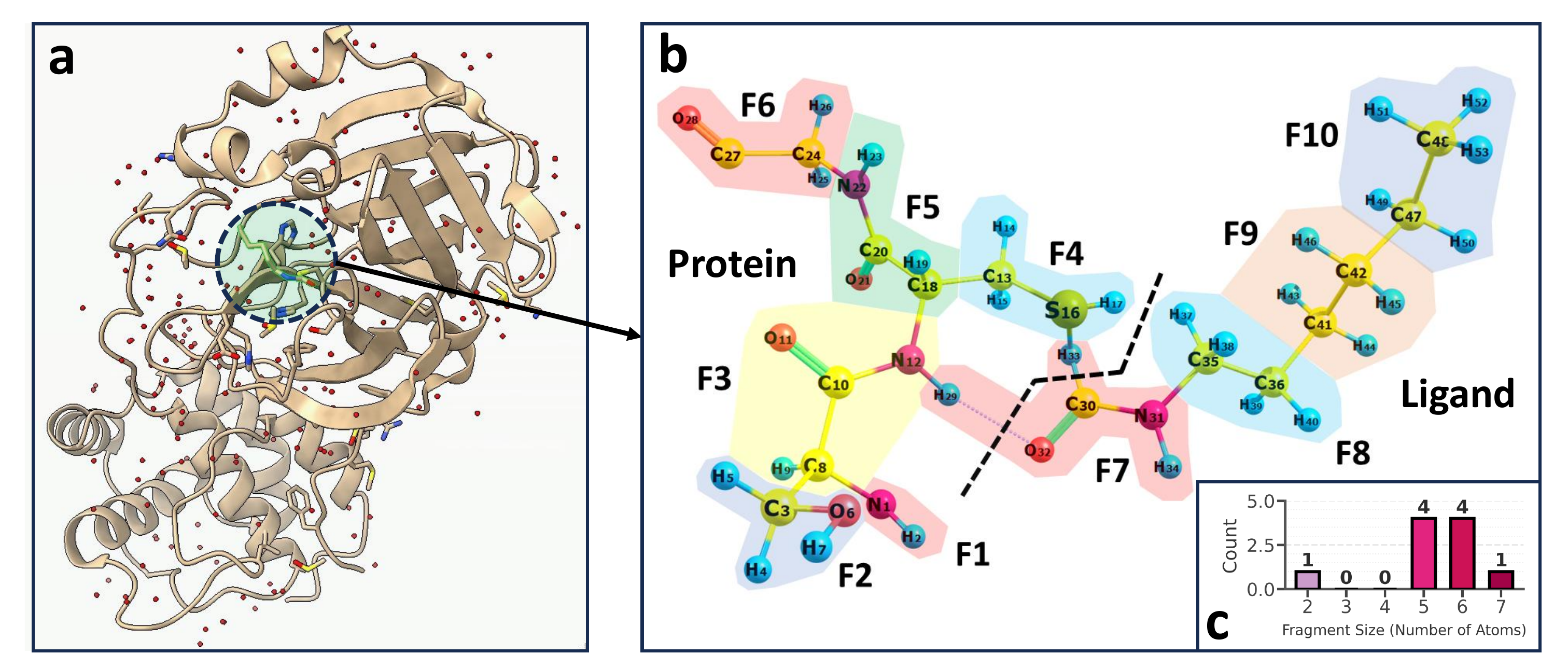}
    \caption{
    \textbf{System setup for the Carmofur-M\textsuperscript{pro} protein-ligand
    complex.}
    (a)~Ribbon representation of the SARS-CoV-2 main protease with Carmofur bound
    at the active site (dashed circle); red spheres indicate explicit water
    molecules.
    (b)~The QM region constructed via the conjugate-cap method ($3\,\text{\AA}$
    isosurface cutoff), partitioned into 10 DMET fragments F1-F10; the dashed
    boundary separates the protein (left) and ligand (right) moieties.
    (c)~Distribution of fragment sizes by atom count.
    }
    \label{fig:carmofur_pl}
\end{figure}

As the target biological application, we consider the SARS-CoV-2 main protease
(M\textsuperscript{pro}) in complex with the covalent inhibitor Carmofur, as
shown in Figure~\ref{fig:carmofur_pl}(a). It should be noted that Carmofur does
not bind M\textsuperscript{pro} as an intact molecule. Nucleophilic attack by
the Cys145 thiolate on the carbamoyl carbon of Carmofur cleaves the compound,
releasing the 5-fluorouracil moiety and leaving only the hexylcarbamoyl
fatty-acid tail covalently bound to Cys145 through a 1.8~\AA{} C-S bond
\cite{Jin2020_carmofur}. The fragment-resolved active space considered in this
work therefore corresponds to the protein in complex with this covalently
bound tail fragment, rather than the full pre-reaction Carmofur molecule. 

To define a tractable QM region, the
molecular fractionation with conjugate caps (MFCC) method~\cite{MFCC_Zhang_2003}
was employed, retaining all protein residues within a $3\,\text{\AA}$ isosurface
distance of the ligand. The resulting QM region encompasses the chemically active
binding pocket together with the full ligand and is illustrated in
Figure~\ref{fig:carmofur_pl}(b). This region was subsequently partitioned into 10
DMET fragments (F1-F10), with fragment boundaries placed along single bonds to
minimise inter-fragment entanglement as much as possible. While more refined
partitioning schemes respecting the full bond topology and molecular geometry
exist~\cite{dmet_sqd, Bowling2025_ProteinLigand_FragQC}, the present manual
fragmentation suffices to rigorously benchmark the DMET-QSCI-RBM framework on a
realistic biological system.

The capping procedure ensures that dangling valences at each boundary are
saturated, preserving the local electronic structure of every
fragment~\cite{MFCC_Zhang_2003}. Fragment sizes range from 2 to 7 atoms, with
the majority containing 5-6 atoms, as summarised in
Figure~\ref{fig:carmofur_pl}(c). Each fragment is treated as an independent DMET impurity problem and solved using the QSCI-RBM solver described above, with DMET
self-consistency~\cite{Knizia2012_DMET, wouters2016dmets_guide} enforced through iterative adjustment of the chemical
potential $\mu_{chem}$ via a Newton-Secant update until the fragment one-particle reduced density matrices match their
mean-field counterparts to within a tolerance of $\varepsilon_{tol} = 5\times10^{-6}$.

\begin{figure}[H]
    \centering
    \includegraphics[width=0.9\linewidth]{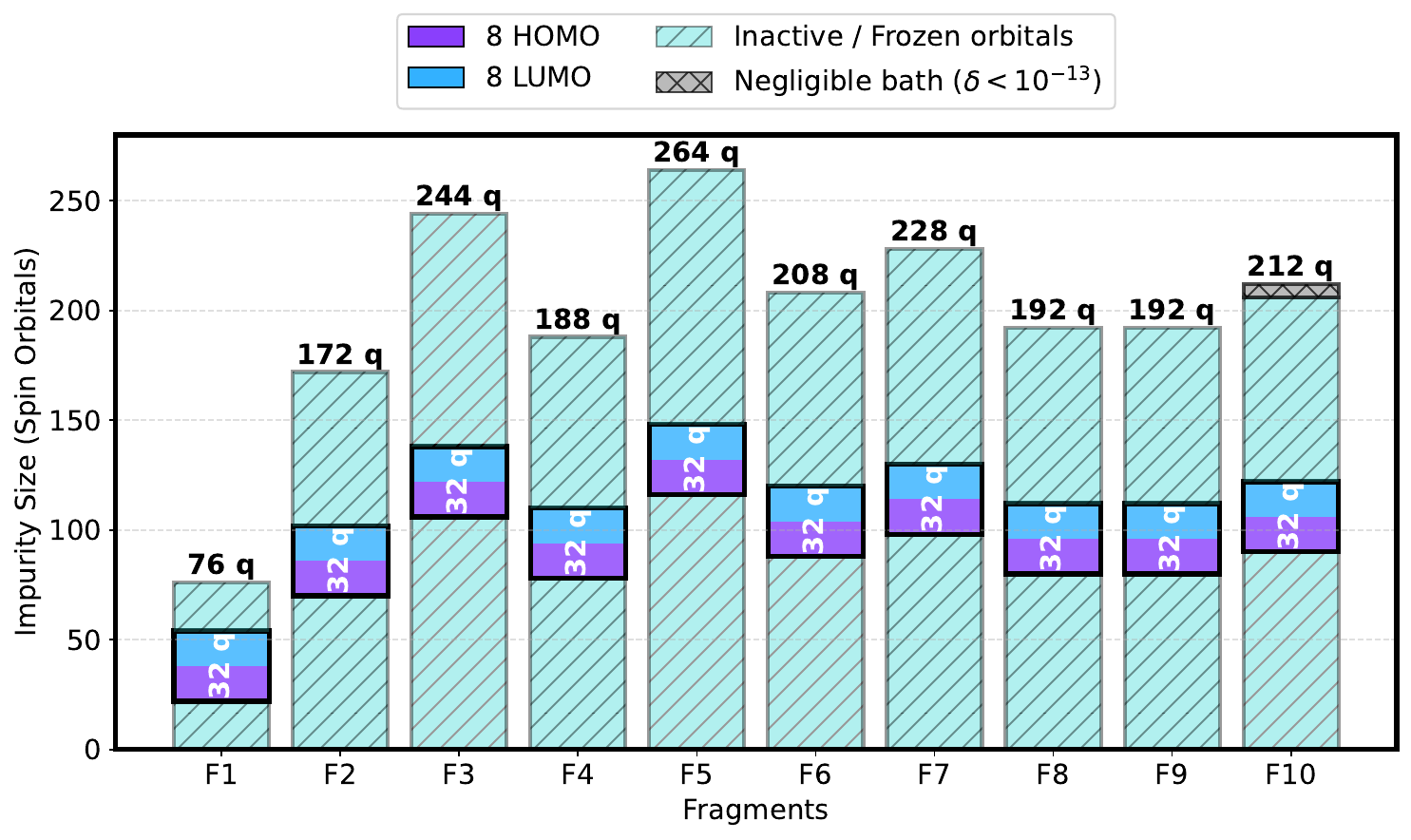}
\caption{
        \textbf{Active space selection for the DMET fragments of the
        Carmofur-M\textsuperscript{pro} complex (cc-pVDZ basis~\cite{Dunning1989ccpvdz}).}
        Each bar shows the total impurity size in spin orbitals for fragments
        F1-F10, with the active space (8 HOMO + 8 LUMO, purple and blue
        respectively) highlighted within each impurity. The remaining orbitals
        are frozen at the Hartree-Fock level (hatched teal). For fragment F10,
        a small subset of nominal bath orbitals with occupation fractionality
        $\delta < 10^{-13}$ are effectively unentangled and are shown separately
        (hatched grey) at the top of the bar, above the frozen virtual orbitals.
        The active space selection yields a uniform 32-qubit quantum circuit per
        fragment, corresponding to a symmetry-preserving Hilbert space of
        dimension $\binom{16}{8}^2 = 165{,}636{,}900$, irrespective of the total
        impurity size.
    }
    \label{fig:carmofur_pl_active_space}
\end{figure}

The QM region was treated using the cc-pVDZ basis set~\cite{Dunning1989ccpvdz}, yielding
impurity sizes ranging from 76 to 264 spin orbitals across the 10 fragments, as
shown in Figure~\ref{fig:carmofur_pl_active_space}. To render each fragment
amenable to quantum simulation within current hardware constraints, an active
space comprising the 8 highest occupied molecular orbitals (HOMO) and 8 lowest
unoccupied molecular orbitals (LUMO) was selected for every impurity Hamiltonian,
yielding a uniform active space of exactly 16 spatial orbitals and $8\alpha +
8\beta$ electrons - equivalently, 32 qubits per fragment under the
Jordan-Wigner mapping~\cite{Jordan1928}. This (16e, 16o) active space defines a
symmetry-preserving Hilbert space of dimension
$\binom{16}{8}^2 = 165{,}636{,}900$ for each fragment. The remaining occupied
and virtual orbitals outside this active window are treated at the
Hartree-Fock level, effectively frozen, and their contributions folded into the
one-body core Hamiltonian as a constant energy offset~\cite{szabo1996_modernqc}.
This uniform active space construction ensures that each fragment circuit has an
identical qubit footprint, greatly simplifying resource estimation and circuit
compilation across all 10 fragments~\cite{dmet_sqd}. The resulting fragment ($A_y$), bath ($B_y$), and frozen core/virtual orbital counts for all 10 fragments are summarized in Section~\ref{app: orb_decomp_pl} in the Supplementary Material.

\begin{figure}[H]
    \centering
    \includegraphics[width=0.95\linewidth]{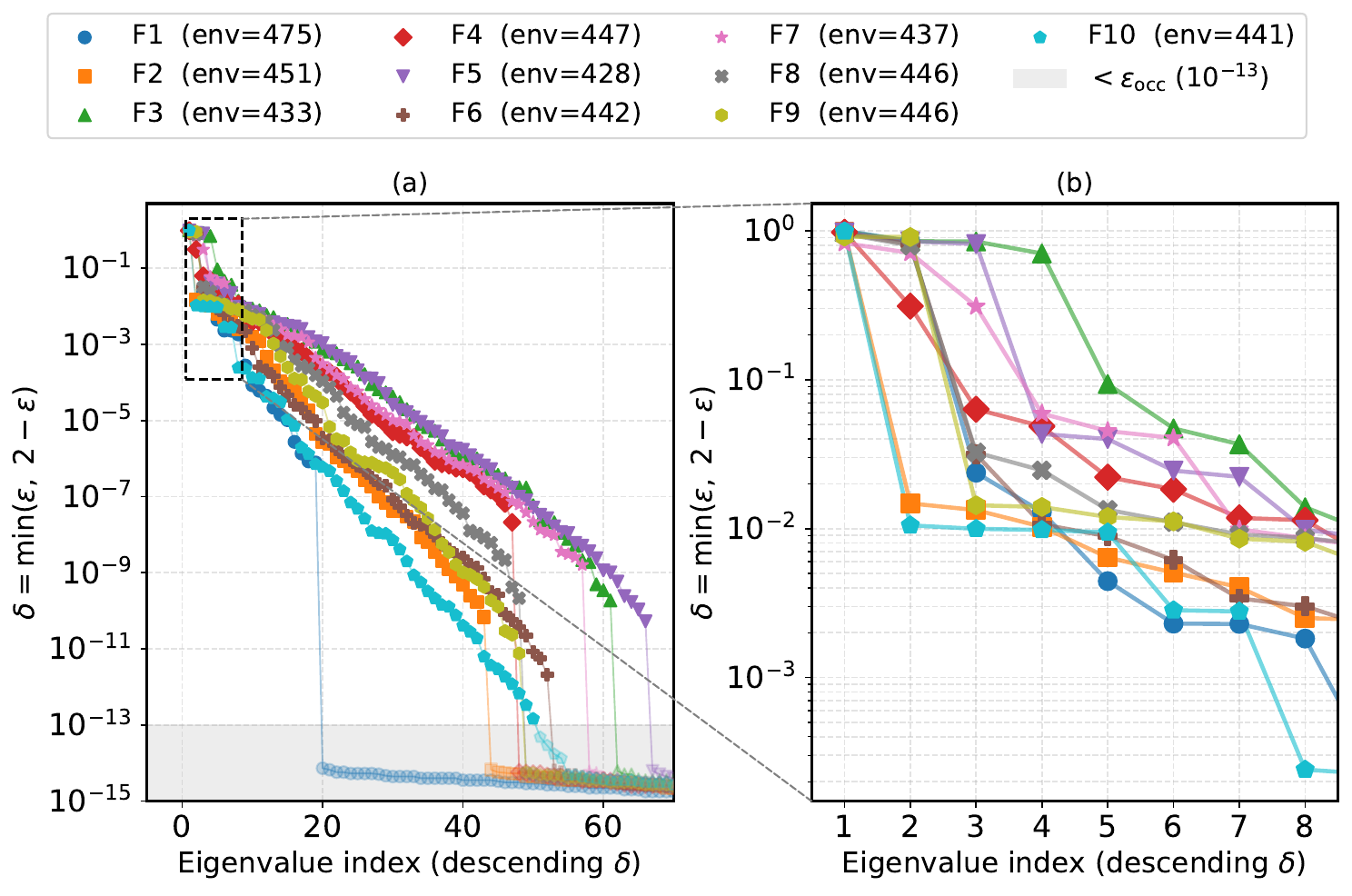}
    \caption{Environment 1-RDM eigenvalue fractionality spectrum $\delta = \min(\varepsilon, 2-\varepsilon)$
    for the 10 DMET fragments of the Carmofur-M\textsuperscript{pro} complex (cc-pVDZ basis~\cite{Dunning1989ccpvdz}). (a) Spectrum per fragment, sorted in descending $\delta$, shown across the full range including values below the occupation threshold; shaded region marks $\delta$ below the occupation threshold
    $\varepsilon_{\mathrm{occ}} = 10^{-13}$, plotted at reduced opacity. (b) Zoomed view of the leading 8 eigenvalues per fragment,
    corresponding to the 8 bath orbitals retained in the (16e, 16o) active space used throughout this work and the dashed box in (a) marks this region.}
    \label{fig:carmofur_10f_eigenval_spectrum}
\end{figure}

Figure~\ref{fig:carmofur_10f_eigenval_spectrum} shows the environment 1-RDM eigenvalue fractionality
spectrum for each of the 10 Carmofur fragments, where $\delta = \min(\varepsilon, 2-\varepsilon)$
measures the deviation of each environment natural-orbital occupation from the nearest integer, with
$\delta \to 2$ indicating strong fragment-bath entanglement and $\delta \to 0$ indicating a nearly
unentangled orbital. The total environment size, or the number of spatial orbitals outside each fragment
from which bath orbitals are drawn, ranges from 428 (F5) to 475 (F1) spatial orbitals across the 10
fragments. Panel (a) of Figure~\ref{fig:carmofur_10f_eigenval_spectrum} shows the full spectrum, with all fragments exhibiting a small number of strongly
fractional eigenvalues followed by a steep roll-off toward the numerical noise floor; the shaded region
marks eigenvalues below the occupation threshold $\varepsilon_{\mathrm{occ}} = 10^{-13}$, below which
orbitals are treated as unentangled and excluded from the impurity problem. Panel (b) of Figure~\ref{fig:carmofur_10f_eigenval_spectrum} zooms in on the
leading 8 eigenvalues per fragment, corresponding to the top 8 bath orbitals ranked by fractionality.
This window was chosen to match the fixed (16e, 16o) active space (8 HOMO + 8 LUMO, 32 qubits) adopted for every
impurity, a size selected so that each fragment circuit remains within current hardware constraints while
still admitting an exact FCI reference for benchmarking. Given total environment sizes of 428-475 spatial
orbitals per fragment, retaining only the top 8 bath orbitals by fractionality means that the remaining
$\sim$420-467 environment orbitals per fragment are approximated as frozen at the mean-field level. 

This
truncation is consistent with the low fractionality ($\delta \lesssim 10^{-2}$) already reached by the 8th
eigenvalue across all fragments in panel (b). More broadly, $\varepsilon_{\mathrm{occ}}$ functions as a
physically grounded, entanglement-aware control parameter governing the trade-off between impurity size
and captured correlation, an effect shown to be especially pronounced in low-symmetry molecular systems
where fragment-bath entanglement is strongly geometry-dependent~\cite{Patra2026}.
Having established the compactness and accuracy of QSCI-RBM on isolated small molecules, we now apply the
full DMET-QSCI-RBM framework to the active region of the Carmofur/SARS-CoV-2 M\textsuperscript{pro} protein-ligand complex,
fragmented into the 10 conjugate-capped impurities F1-F10. Each
fragment carries a uniform (16e, 16o) active space, corresponding to a symmetry-preserving subspace of
dimension $|\mathbb{S}| = \binom{16}{8}^2 = 165{,}636{,}900$, and all quantum sampling was performed on the IBM
Heron R3 QPU (\texttt{ibm\_boston}). We first characterize the baseline DMET-SQD solver under two $\varepsilon_{spb}$ regimes to
establish the accuracy-subspace-size trade-off inherent to configuration-recovery-based QSCI, before turning
to the DMET-QSCI-RBM results and a direct cross-method comparison. 

Across the 10 fragments, F3 and F10 emerge as
the extremal fractionality cases, with F3 retaining the highest bath-orbital fractionality out to the 8th
eigenvalue and F10 decaying the fastest, to the point that a subset of its nominal bath orbitals fall below
$\varepsilon_{\mathrm{occ}}$ and are effectively unentangled. Given this, we restrict the manual hyperparameter
optimization at $\mu = 0$ in the following section to these two bracketing fragments alone.

\subsection{Manual Hyperparameter Optimization on Extremal Fragments}
\label{sec:f3_f10_hyperparameter_optimization}

Building on our previous entanglement-aware embedding analysis, in which we
systematically examined how the bath-orbital fractionality spectrum governs
the trade-off between quantum resource requirements and classical
post-processing cost via the occupation threshold $\varepsilon_{\mathrm{occ}}$~\cite{Patra2026},
we extend this approach here by performing individual fragment-level
hyperparameter analysis for the two extremal-fractionality fragments
identified in Section~\ref{sec: mol_frag_fractionality} which are F3, the ``hardest"
fragment, which retains the highest bath-orbital fractionality out to the
8th eigenvalue, and F10, the ``easiest" fragment, whose fractionality decays
fastest to the point that a subset of its nominal bath orbitals fall below
$\varepsilon_{\mathrm{occ}}$ and are effectively unentangled. These two
fragments bracket the range of entanglement behaviour observed across the
full fragment set, and we consider the hyperparameters obtained from this
targeted analysis to provide a sufficiently informative starting point for
determining viable hyperparameters for the DMET-QSCI-RBM run across all ten
fragments. The hyperparameters obtained in this way, which were then used
for all ten fragments, are listed in
Section~\ref{app : hyperparams} in the Supplementary Information.

\begin{figure}[htbp]
    \centering
    \includegraphics[width=\linewidth]{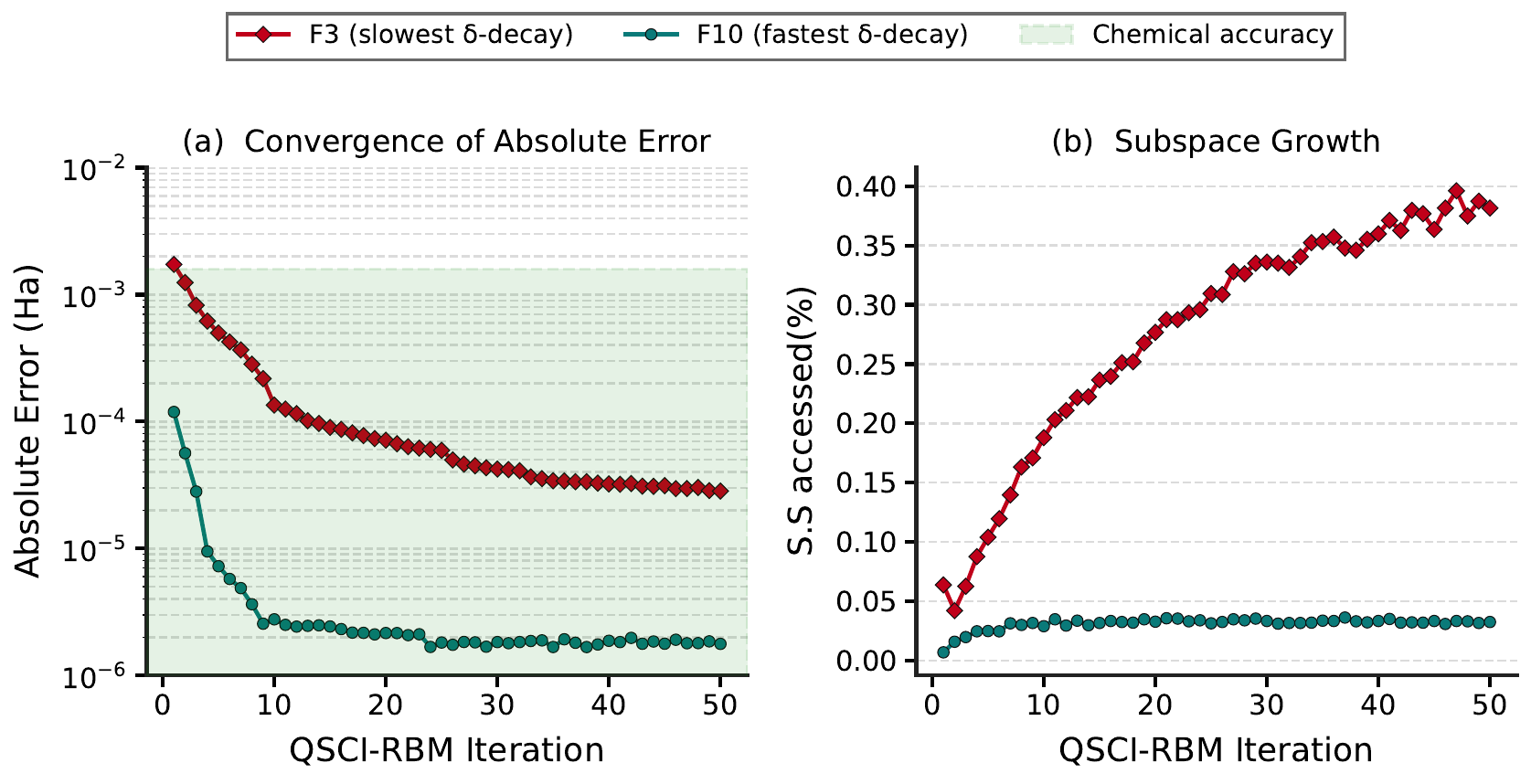}
    \caption{
        \textbf{QSCI-RBM convergence comparison for the two extremal-fractionality
        Carmofur fragments at $\mu = 0$.} F3 (hardest, slowest bath-orbital
        fractionality decay) and F10 (easiest, fastest decay) are shown over 50
        QSCI-RBM iterations. \textbf{(a)} Absolute error relative to the CASCI
        reference, log scale; the shaded green band marks chemical accuracy
        (1 kcal/mol $\approx 1.594\times10^{-3}$ Ha). \textbf{(b)} Symmetry-preserving
        subspace accessed (\%) as a fraction of $\binom{16}{8}^2 = 165{,}636{,}900$,
        per iteration.
    }
    \label{fig:f3_f10_error_subspace_comparison}
\end{figure}

At $\mu = 0$, F10 accesses at most $\sim$0.03\% of the symmetry-preserving
subspace across the 50 QSCI-RBM iterations, while F3 accesses up to
$\sim$0.4\%, an order of magnitude larger. Note that these figures correspond
specifically to $\mu = 0$; as the chemical potential approaches the
self-consistent value $\mu_{\mathrm{chem}}$, the impurity Hamiltonian is
progressively refined to capture additional correlation, and the subspace
fraction required for a given fragment may vary accordingly~\cite{wouters2016dmets_guide}.
In both cases, the final energy after 50 iterations lies comfortably within
chemical accuracy. For F10, this is achieved with a markedly smaller subspace:
its bath-orbital fractionality spectrum falls below $\delta \sim 10^{-3}$ by
the 8th eigenvalue (Figure~\ref{fig:carmofur_10f_eigenval_spectrum} (b)), and the
resulting minimal subspace is nonetheless sufficient to reach an absolute
error below $10^{-5}$ Ha. F3, by contrast, retains $\delta$ just above
$10^{-2}$ at the 8th eigenvalue, and correspondingly accesses roughly an
order of magnitude more subspace ($\sim$0.35\%) while converging to an
absolute error just below $10^{-4}$ Ha. This correlation between the leading
bath-fractionality spectrum and the subspace ultimately required is
consistent with $\varepsilon_{\mathrm{occ}}$ acting as an entanglement-aware
control parameter, as discussed above. Having established this initial manual
hyperparameter optimization on the two bracketing fragments, the resulting
settings are adopted as the starting point for the DMET-QSCI-RBM run across
all ten fragments of the Carmofur-M\textsuperscript{pro} complex.

While the present manual hyperparameter optimization was restricted to the two
extremal fragments (F3 and F10) at $\mu = 0$ for tractability, per-fragment
hyperparameter optimization, tuned individually to the correlation and
subspace requirements of each fragment under classical resource constraints would likely yield further improvements in subspace compactness. However,
performing such optimization independently for all ten fragments at every
chemical-potential iteration of the DMET self-consistency loop would be
computationally cumbersome, and is left for future work.

\subsection{Hardware Sampling Configuration for DMET-Embedded Runs}
\label{sec:hardware_error_mitigation}
All quantum sampling on the Carmofur-M\textsuperscript{pro} complex in this work was performed on the IBM
Boston Heron R3 processor. For both DMET-embedded experiments, DMET-SQD with
$\varepsilon_{spb} = \sqrt{|\mathbb{S}|}$ and DMET-QSCI-RBM, only dynamical decoupling (XpXm) was retained,
and Pauli twirling was deliberately disabled. This reflects a deliberate trade-off between per-shot fidelity
and total QPU execution time, in contrast to the higher-fidelity regime affordable in the single-fragment
validation runs of Section~\ref{sec:hardware_error_mitigation_smallmol}. We note this was a resource-driven
decision rather than a reflection of mitigation efficacy: twirling was observed to substantially increase QPU
execution time per circuit, an overhead that becomes difficult to justify at the scale of the DMET
experiments, where each chemical-potential iteration requires sampling all 10 fragments at $100{,}000$ shots
each, and every solver configuration requires multiple such iterations to reach (or approach)
self-consistency. This total sampling cost scales as
$10~\text{fragments} \times 100{,}000~\text{shots} \times N_{\mu_{chem}-iterations}$ per configuration,
repeated across the two independent solver configurations compared in this work. Given the limited QPU
allocation available for this work, dynamical decoupling alone was used to provide adequate error suppression
for the DMET-scale experiments, while reserving the additional twirling overhead for the single-fragment
validation systems where it was affordable.

For DMET-SQD, the samples-per-batch parameter was set to $\varepsilon_{spb} = \sqrt{|\mathbb{S}|}$. As derived
in Appendix~\ref{app:spb_derivation}, the worst-case bound that guarantees the proliferated diagonalization
subspace never exceeds the full symmetry space, $\varepsilon_{spb} = \sqrt{|\mathbb{S}|}/2$, is a
combinatorial engineering bound obtained under the fully disjoint worst case in which the sampled $\alpha$-
and $\beta$-spin string sets share no overlap.For closed-shell systems, however, this worst case is not realized in practice: as noted in other works~\cite{hi_vqe_2024}, it is often the case that if a configuration of the form
$\ket{\alpha}\otimes\ket{\beta}$ contributes strongly to the ground state, the corresponding
$\ket{\beta}\otimes\ket{\alpha}$ configuration does as well. Consequently, the recovered $\alpha$- and
$\beta$-spin string sets exhibit substantial overlap rather than the disjointness assumed in the worst-case
derivation, so that the effective pool size, and hence the realized diagonalization subspace, remains
controlled even when sampling up to $\varepsilon_{spb} = \sqrt{|\mathbb{S}|}$~\cite{ksv2026iqm}. This
closed-shell relaxation of the worst-case bound provides the practical justification for adopting
$\varepsilon_{spb} = \sqrt{|\mathbb{S}|}$, rather than the more conservative $\sqrt{|\mathbb{S}|}/2$, in this
work.

The exact details of the execution dates and the hardware calibration data pertaining to the IBM Quantum
processors used is provided in Section~\ref{app:execution_dates_calibration} in the Supplementary Material.
 
\subsection{DMET-QSCI-RBM vs.\ DMET-SQD: Convergence and Subspace Compactness}
\label{sec:cross_method_comparison}

Having established the DMET-SQD baseline configuration ($\varepsilon_{spb} = \sqrt{|\mathbb{S}|}$,
Section~\ref{sec:hardware_error_mitigation}), we directly compare its self-consistency trajectory against
DMET-QSCI-RBM for the Carmofur/SARS-CoV-2 M\textsuperscript{pro} complex (10 fragments, cc-pVDZ). Both solvers were evaluated at closely corresponding chemical potentials at each
$\mu$-iteration (Table~\ref{tab:corresponding_mu_energies}), enabling a direct, iteration-by-iteration
comparison of solver accuracy independent of any difference in the outer $\mu$-self-consistency trajectory
itself.

\begin{figure}[htbp]
    \centering
    \includegraphics[width=\linewidth]{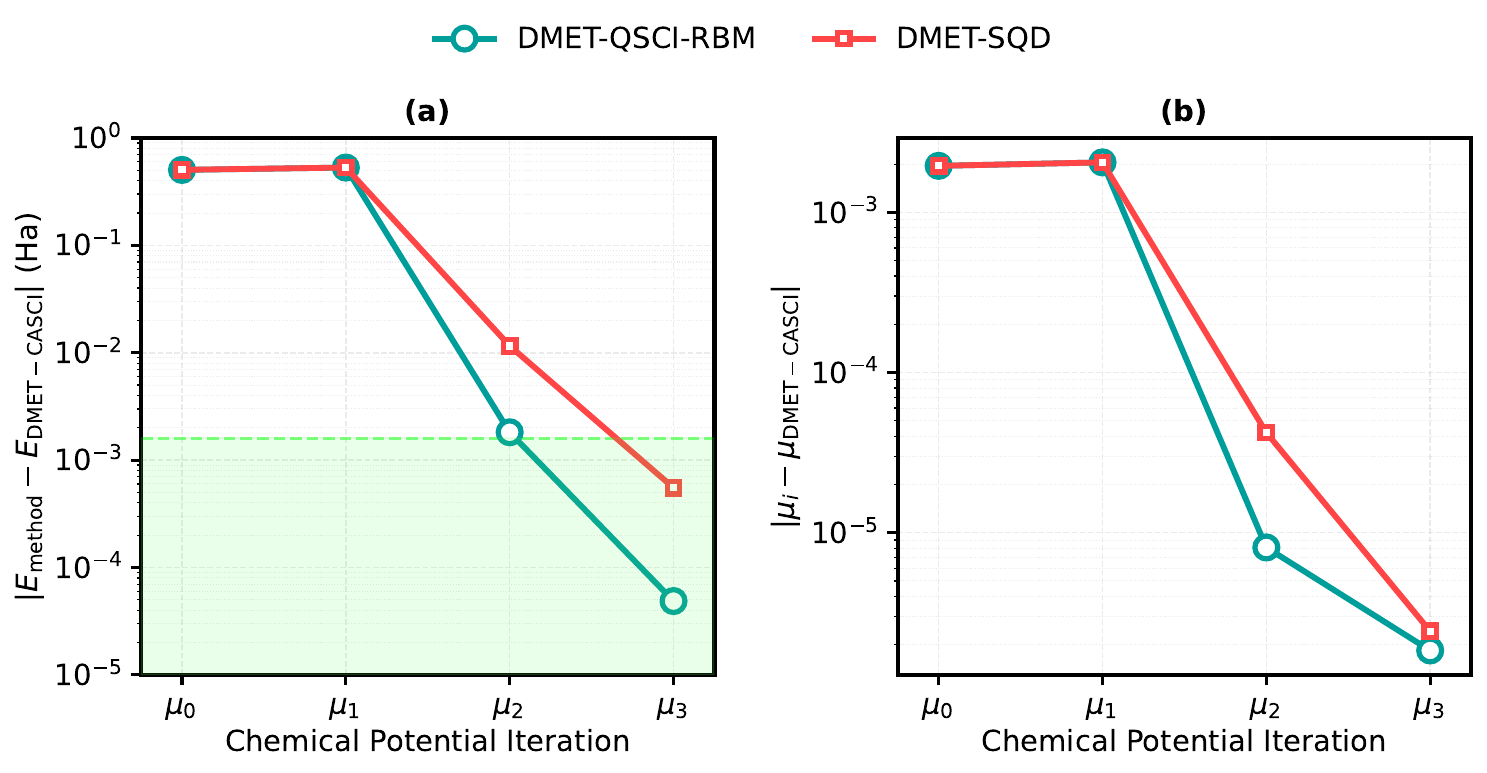}
    \caption{
        \textbf{Convergence of DMET-QSCI-RBM and DMET-SQD at the respective chemical potentials for the
        Carmofur-M\textsuperscript{pro} complex.}
        \textbf{(a)} Absolute energy error relative to DMET-CASCI vs.\ $\mu$-iteration, log scale; shaded band
        marks chemical accuracy (1~kcal/mol $\approx 1.594\times10^{-3}$~Ha).
        \textbf{(b)} Absolute chemical-potential deviation from $\mu_{DMET-FCI}$ vs.\ $\mu$-iteration, log scale.
    }
    \label{fig:DMET_SQD_QSCI_RBM_Comparison}
\end{figure}

Figure~\ref{fig:DMET_SQD_QSCI_RBM_Comparison} shows that both solvers converge over the same four
$\mu$-iterations (self-consistency tolerance $5\times10^{-6}$ on the secant-proposed $\mu$ update), with
near-identical $\mu$-trajectories at every step shown in panel (b). At the far-from-self-consistent early iterations
($\mu_1$, $\mu_2$), both solvers show large energy errors of comparable magnitude
(-505 to -530~mHa), reflecting the poor overlap of these initial guesses with the true self-consistent
chemical potential rather than any solver-specific deficiency. As $\mu$ approaches self-consistency
($\mu_3$, $\mu_4$), DMET-QSCI-RBM consistently achieves tighter energy accuracy than DMET-SQD at similar $\mu$: at $\mu_3$, DMET-QSCI-RBM reaches 1.82~mHa versus DMET-SQD's 11.58~mHa, and at the
converged $\mu_4$, DMET-QSCI-RBM reaches 0.049~mHa versus DMET-SQD's 0.553~mHa which is an $\sim$11-fold
tighter energy error (1 order of magnitude more accurate) when compared to DMET-SQD.

\begin{table}[H]
\centering
\caption{Energy at each $\mu$-iteration for DMET-SQD and DMET-QSCI-RBM, evaluated at closely
corresponding (though not identical) chemical potentials at each iteration, isolating solver
accuracy from outer-loop trajectory differences. Reference:
$E_{\mathrm{DMET-FCI}} = -1573.2005463607$~Ha.}
\label{tab:corresponding_mu_energies}
\begin{tabular}{cccc}
\toprule
$\mu$-iter & $E_{\mathrm{SQD}}$ (Ha) & $E_{\mathrm{QSCI-RBM}}$ (Ha) & $|\Delta E_{\mathrm{DMET-SQD}}|$ \quad  \quad $|\Delta E_{\mathrm{DMET-QSCI-RBM}}|$ \\
\midrule
1 & $-1573.7056145005$ & $-1573.7048228758$ & $0.50507$~Ha \quad  \quad $0.50428$~Ha \\
2 & $-1573.7309421834$ & $-1573.7306899848$ & $0.53040$~Ha \quad  \quad $0.53014$~Ha \\
3 & $-1573.1889628833$ & $-1573.2023632579$ & $0.01158$~Ha \quad  \quad $0.00182$~Ha \\
4 & $-1573.2010989415$ & $-1573.2004976784$ & $0.000553$~Ha \quad  \quad $0.000049$~Ha \\
\bottomrule
\end{tabular}
\end{table}

\begin{figure}[htbp]
    \centering
    \includegraphics[width=\linewidth]{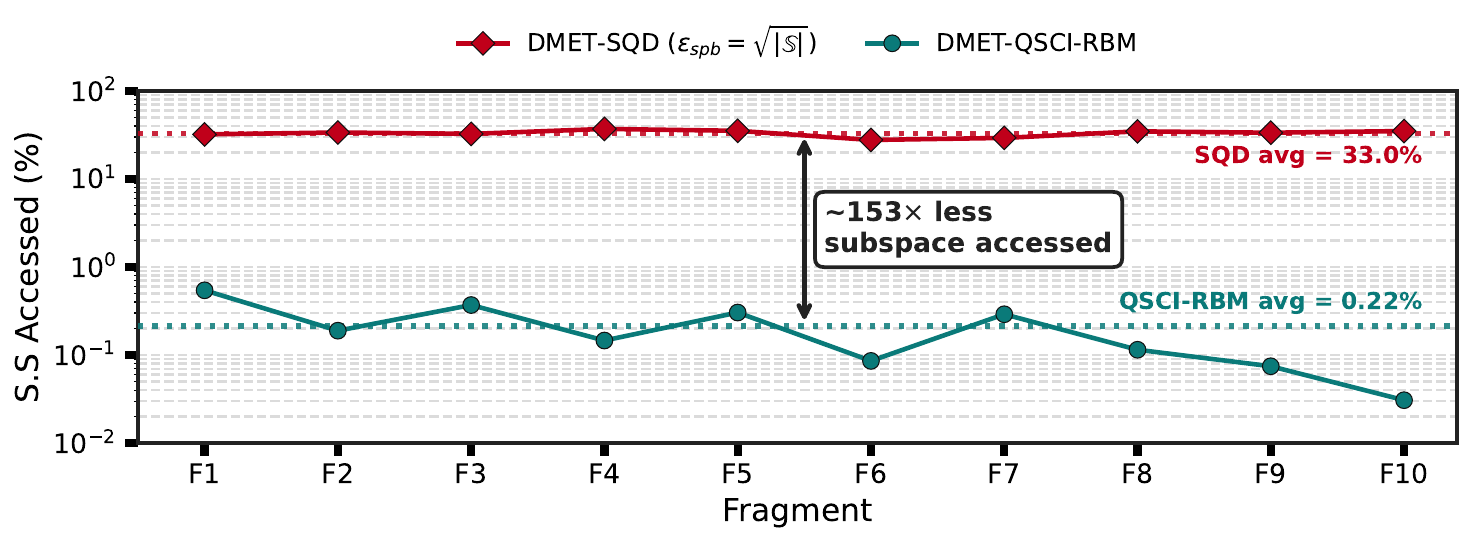}
    \caption{
        \textbf{Fragment-resolved symmetry-space ($\mathbb{S}$) accessed by DMET-SQD
        ($\varepsilon_{spb}=\sqrt{|\mathbb{S}|}$) and DMET-QSCI-RBM for the Carmofur/SARS-CoV-2
        M\textsuperscript{pro} complex (10 fragments, F1-F10).} Points show the mean symmetry-space percentage
        accessed across all four $\mu$-iterations for each fragment. Dotted horizontal lines mark each method's overall mean across all fragments and
        $\mu$-iterations (DMET-SQD: $33.0\%$; DMET-QSCI-RBM: $0.22\%$).
    }
    \label{fig:final_dmet_sqd_dmet_qsci}
\end{figure}

The practical significance of this accuracy advantage becomes clear when the associated subspace footprint is
examined. Figure~\ref{fig:final_dmet_sqd_dmet_qsci} shows the fragment-resolved symmetry space accessed by
each solver, averaged across all four $\mu$-iterations. DMET-SQD accesses $33.0\%$ of the symmetry-preserving
subspace on average across all ten fragments, essentially independent of fragment identity or $\mu$-iteration
(range: $26.4$--$38.4\%$). DMET-QSCI-RBM, by contrast, accesses only $0.22\%$ on average
(range: $0.03$--$0.60\%$) leading to a $\sim$153-fold reduction in subspace size relative to DMET-SQD, achieved while
simultaneously reaching tighter energy accuracy at the converged $\mu$.

On average, DMET-QSCI-RBM accesses $0.22\%$ of the $\mathbb{S}$, where this figure is evaluated \emph{post-proliferation}~\cite{Patra2026}, that is, on the dimension of the Cartesian product of unique $\alpha$ and $\beta$ spin strings that is actually diagonalized. The same post-proliferation definition is used for every subspace percentage reported for the standalone QSCI-RBM runs in Section~\ref{sec:small_molecules_result}, so that all quoted values refer to the dimension actually diagonalized.


Figures~\ref{fig:DMET_SQD_SS_all_iters} and \ref{fig:DMET_QSCI_RBM_SS_all_iters} in the Supplementary Information show the per-fragment
symmetry-space access broken out by individual $\mu$-iteration for DMET-SQD and DMET-QSCI-RBM,
respectively. For DMET-SQD, the four $\mu$-iteration curves lie almost on top of one another at every
fragment, with the fragment-to-fragment ordering essentially unchanged across iterations (26.4--38.4\%
range, overall mean 33.04\%). This iteration-independence indicates that DMET-SQD's subspace footprint is
governed primarily by the fixed $\varepsilon_{spb}$ truncation and the fragment's own sampled configuration
pool, rather than by the state of $\mu$-self-consistency -as the subspace accessed at the far-from-converged
$\mu_1$ is essentially indistinguishable from that at the converged $\mu_4$.

DMET-QSCI-RBM shows a qualitatively different pattern. While the four $\mu$-iteration curves again track
each other closely at most fragments, the absolute magnitude is nearly two orders of magnitude smaller
(0.03--0.60\% range, overall mean 0.22\%). F1 consistently accesses the largest subspace of all ten
fragments at every $\mu$-iteration, while F10 consistently accesses the smallest. Individual per-fragment,
per-$\mu$-iteration subspace-access plots for both DMET-SQD and DMET-QSCI-RBM are provided in
Section~\ref{app: ss_accessed} of the Supplementary Information for further investigation.

Bath-orbital fractionality, as a measure of orbital entanglement analogous to the single-orbital entropy
used to characterize multireference character within a fixed reference space, might be expected to serve as a predictor of the correlation strength captured by a given
active space. Even this related, upstream question, connecting an entanglement-based measure to active-space
quality, has been described as an open challenge in its own right~\cite{boguslawski2012}. Whether such a measure
can further predict the final subspace, or core space, required within an \emph{already-fixed} active
space, as considered in this work, is accordingly a still less established question. F1's elevated
subspace requirement and F10's minimal subspace requirement bracket the full range of RBM-guided subspace
access observed across the ten fragments, and each is explained by a different structural factor. F10's
minimal subspace access is consistent with bath-orbital fractionality: F10 exhibits the lowest bath-orbital
fractionality of all ten fragments (Figure~\ref{fig:carmofur_10f_eigenval_spectrum}), matching its rank as
the fragment with the smallest RBM-guided subspace requirement exactly. F1's elevated subspace requirement,
by contrast, is not explained by fractionality, since F3, not F1, exhibits the highest bath-orbital
fractionality among the ten fragments, yet it is F1 that consistently accesses the largest subspace under
RBM-guided expansion.

This decoupling is instead explained by F1's active-space-to-impurity ratio. This is because F1 is the
smallest fragment (Figure~\ref{fig:carmofur_pl}(c)), the fixed 16-orbital active space adopted uniformly
across all fragments covers a substantially larger fraction of F1's total impurity size than it does for
any other fragment, and this ratio, like F1's subspace access, is likewise the largest of the ten
fragments. Together, these two factors account for both extremes of the observed subspace-access range,
though the ranking of the remaining eight fragments is not fully explained by either factor alone, and
disentangling their relative and combined contributions across the full fragment set is left as an avenue
for future work. This persistent, iteration-independent fragment ranking, together with DMET-SQD's own
iteration-independent subspace footprint noted above, suggests that fragment-level subspace access under
both solvers is governed more by fragment-specific structural properties than by the progress of the outer
$\mu$-self-consistency loop.

Taken together, these results indicate that DMET-QSCI-RBM's compactness advantage does not come at the cost
of accuracy or convergence speed relative to DMET-SQD. This is because at closely corresponding chemical potentials, each
independently arrived at via the two solvers' own Newton-Secant self-consistency loops, and the same number
of $\mu$-iterations, DMET-QSCI-RBM reaches equal or tighter energy accuracy than DMET-SQD while accessing
roughly two orders of magnitude less of the symmetry-preserving subspace, fragment-for-fragment, across the
entire self-consistency trajectory.

\subsection{Quantum Resources Analysis for DMET-SQD and DMET-QSCI-RBM}
\label{sec: quantum_resources}

Figures~\ref{fig:circuit_depth_heatmap} and~\ref{fig:exec_duration_heatmap} in the Supplementary Information report the per-fragment
transpiled circuit depth and QPU execution time, respectively, across all four $\mu$-iterations of both
DMET-SQD ($\varepsilon_{spb}=\sqrt{|\mathbb{S}|}$) and DMET-QSCI-RBM on Carmofur-M\textsuperscript{pro}
(both executed on \texttt{ibm\_boston}). Circuit depths across both configurations and all fragments fall predominantly in
the range of 450 to 900, with occasional outliers such as DMET-SQD F7 at C4 (depth 883) and DMET-QSCI-RBM
F5 at C4 (depth 973), both attributable to per-circuit variation rather than any systematic trend across
solver configuration or $\mu$-iteration. Execution times are similarly consistent, clustering around
28 to 30~s per fragment per iteration, with isolated exceptions such as DMET-SQD F2 at C2 (34.43~s) and
DMET-QSCI-RBM F8 at C4 (35.69~s), likewise attributable to per-circuit variation rather than to any
dependence on the impurity's or solver's configuration.

These figures show that both solvers sample identical LUCJ circuits at the same shot budget and converge
within the same number of $\mu$-iterations (Section~\ref{sec:cross_method_comparison}), so the practical
benefit of DMET-QSCI-RBM lies entirely in the substantially more compact classical diagonalization subspace
constructed at each iteration, reducing per-fragment classical post-processing cost without any
corresponding change to the quantum sampling budget.


\section{Conclusion}
\label{sec:conclusion}

In this work, we introduced DMET-QSCI-RBM, a machine-learned compact subspace generation protocol that
integrates Restricted Boltzmann Machine-guided configuration recovery directly into the Quantum Selected
Configuration Interaction pipeline within a Density Matrix Embedding Theory framework. Rather than relying
on symmetry filtering alone or on perturbative anchoring schemes such as MP2-seeded configuration recovery,
DMET-QSCI-RBM learns the underlying distribution of dominant determinants directly from hardware-sampled
bitstrings, enabling the targeted generation of high-probability configurations without recourse to any
perturbative approximation. Validated first on three small-molecule benchmarks of increasing active-space
size, C\textsubscript{2}H\textsubscript{4} (sto-3g, 28 qubits), CH\textsubscript{5}NO (sto-3g, 40 qubits), and H\textsubscript{2}O (cc-pVDZ, 48
qubits), the framework was then applied to the simulation of the Carmofur/SARS-CoV-2
M\textsuperscript{pro} protein-ligand complex across 10 DMET fragments. There, DMET-QSCI-RBM and a
DMET-SQD baseline ($\varepsilon_{spb} = \sqrt{|\mathbb{S}|}$) both reached full chemical-potential
self-consistency within the same four $\mu$-iterations, with DMET-QSCI-RBM achieving tighter energy
accuracy relative to the DMET-FCI reference than DMET-SQD while accessing roughly two orders of magnitude
less of the symmetry-preserving configuration subspace at converged chemical potentials.

Beyond confirming the compactness advantage anticipated in the abstract, the results presented here surface
an additional insight that refines our understanding of when and why this compactness translates into
practical efficiency. Because both solvers converge within the same number of $\mu$-iterations and led to closely corresponding chemical potentials at every step, the accuracy and subspace-size
differences observed cannot be attributed to differing degrees of $\mu$-self-consistency or to any
difference in quantum hardware resource usage, since per-fragment circuit depth and execution time were
found to be essentially uniform across both solver configurations. The advantage of DMET-QSCI-RBM therefore
arises specifically from the compactness of the accumulated determinant memory constructed by RBM-guided
subspace expansion, isolating this as the source of its efficiency rather than any difference in outer-loop
convergence behaviour or quantum sampling cost. We also identified a fragment-specific anomaly, one
fragment (F1) consistently required a larger RBM-guided subspace than the other nine
despite not exhibiting the highest bath-orbital fractionality among them, a decoupling between
entanglement structure and subspace requirement warranting future investigation.

Taken together, these findings support compact, learned subspace generation as a practically meaningful
addition to fragmentation-based quantum embedding frameworks. As pharmaceutically and biologically relevant
targets grow beyond what any single quantum embedding fragment can directly address, techniques such as
DMET~\cite{sqd_first_paper} increasingly depend on keeping the classical diagonalization cost per fragment
low, since this cost is repeated across every fragment and every self-consistency iteration of the embedding
loop. The embedded wavefunction (EWF) framework~\cite{Shajan2026} similarly restricts quantum resource
requirements to fixed local fragment problems, and although it does not impose an analogous self-consistency
criterion between fragment and environment, its fragment problems still grow expensive to diagonalize as
active-space size increases. RBM-guided compact subspace expansion directly targets this shared bottleneck,
offering a natural addition to both DMET- and EWF-based fragmentation strategies rather than competing with
either scheme itself. Recent demonstrations have shown that EWF-based SQD workflows can already be scaled to
biological systems spanning several thousand
atoms~\cite{merz2026crossing12000atombarrierheterogeneous}, and RBM-guided compact subspace generation
offers a direct path toward extending such simulations further without sacrificing chemical accuracy,
helping pave the way toward classically tractable quantum simulations of protein-ligand complexes and other
biomolecular systems in their full pharmacological context, rather than in the truncated active-site models
to which such simulations are often restricted today.

Several directions remain open for future work. Chief among these is a comparative assessment of the
present framework alongside other compact subspace-generation strategies such as those that anchor
configuration recovery in perturbative estimates or exchange configuration data iteratively between quantum
and classical solvers~\cite{patra2025physicsinformedgenerativemachine, hi_vqe_2024}. In parallel, as noted in
Section~\ref{sec:small_molecules_result} and Section~\ref{sec:carmofur_results}, the RBM component of
QSCI-RBM, whose learning rate, hidden-layer size, Gibbs sampling chain length, and coefficient thresholds
were selected empirically rather than optimized systematically, is the source of the tunable
hyperparameters in this framework, along with the total number of QSCI-RBM iterations, which is currently a
fixed, user-supplied value rather than a threshold informed by the diagonalization resources available to a
given user. We also assessed the fragment-resolved subspace accessed against candidate structural
predictors, including the bath-orbital fractionality spectrum, the active-space-to-impurity ratio, and
transpiled circuit depth, with mixed results as the former two accounted for some, but not all, of the observed
fragment-to-fragment variation. A more systematic, quantitative treatment of these and other candidate
predictors, and their combined contribution to RBM-guided subspace access, is accordingly a further
direction for future work.

A dedicated hyperparameter search over the RBM's architecture and training settings, informed by
chemistry-based knowledge such as known features of the target system's correlation structure, together
with an adaptive or pre-assigned subspace-size threshold in the spirit of methods such as
HI-VQE~\cite{hi_vqe_2024}, represents a promising avenue for pushing subspace compactness and energy
accuracy further beyond the results reported here. A further direction lies in extending sample-based
algorithms of this kind, currently designed around NISQ-era sampling constraints, toward the fault-tolerant
quantum computing era~\cite{KSV2026Fujitsu}. Together, these directions point toward scalable, automated,
economically and commercially deployable molecular simulation pipelines, in which the fraction of the
subspace accessed or the classical diagonalization cost can be fixed a priori to suit the accuracy and
throughput demands of a specific industrial application.


\section*{Data Availability}
The data that support the findings of this study are available from the corresponding author upon reasonable request.

\section*{Acknowledgements}

We acknowledge the use of IBM Quantum Credits via the IBM Quantum Startups Program for this work. The views expressed are those of the authors and do not reflect the official policy or position of IBM or the IBM Quantum Platform team. 

The authors would also like to extend their appreciation to the advisors of Qclairvoyance Quantum Labs for their support and timely help.

\section*{Funding}
No funding was received for this research.

\section*{Competing interests}
R.M. is a paid consultant at Qclairvoyance Quantum Labs. The other authors declare no competing interests.

\bibliography{sample}

@misc{dmet_sqd,
      title={Towards quantum-centric simulations of extended molecules: sample-based quantum diagonalization enhanced with density matrix embedding theory}, 
      author={Akhil Shajan and Danil Kaliakin and Abhishek Mitra and Javier Robledo Moreno and Zhen Li and Mario Motta and Caleb Johnson and Abdullah Ash Saki and Susanta Das and Iskandar Sitdikov and Antonio Mezzacapo and Kenneth M. Merz Jr},
      year={2024},
      howpublished={2411.09861},
      archivePrefix={arXiv},
      primaryClass={quant-ph},
      url={https://arxiv.org/abs/2411.09861},
}

@article{Jin2020_carmofur,
  author  = {Jin, Zhenming and Zhao, Yao and Sun, Yuan and Zhang, Bing and
             Wang, Haofeng and Wu, Yan and Zhu, Yan and Zhu, Chen and
             Hu, Tianyu and Du, Xiaoyu and Duan, Yinkai and Yu, Jing and
             Yang, Xiaobao and Yang, Xiuna and Yang, Kailin and Liu, Xiang and
             Guddat, Luke W. and Xiao, Gengfu and Zhang, Leike and
             Yang, Haitao and Rao, Zihe},
  title   = {Structural basis for the inhibition of SARS-CoV-2 main protease by antineoplastic drug carmofur},
  journal = {Nature Structural \& Molecular Biology},
  year    = {2020},
  volume  = {27},
  number  = {6},
  pages   = {529--532},
  doi     = {10.1038/s41594-020-0440-6},
  issn    = {1545-9985},
  url     = {https://doi.org/10.1038/s41594-020-0440-6}
}

@article{Dunning1989ccpvdz,
    author = {Dunning, Thom H., Jr.},
    title = {Gaussian basis sets for use in correlated molecular calculations. I. The atoms boron through neon and hydrogen},
    journal = {The Journal of Chemical Physics},
    volume = {90},
    number = {2},
    pages = {1007-1023},
    year = {1989},
    month = {01},
    issn = {0021-9606},
    doi = {10.1063/1.456153},
    url = {https://doi.org/10.1063/1.456153},
    eprint = {https://pubs.aip.org/aip/jcp/article-pdf/90/2/1007/18974738/1007_1_online.pdf},
}

@ARTICLE{propane_fci_Gao2024,
  author = {Gao, Hong and Imamura, Satoshi and Kasagi, Akihiko and Yoshida, Eiji},
  title = {Distributed Implementation of Full Configuration Interaction for One Trillion Determinants},
  journal = {Journal of Chemical Theory and Computation},
  volume = {20},
  number = {3},
  pages = {1185--1192},
  year = {2024},
  month = {2},
  day = {13},
  doi = {10.1021/acs.jctc.3c01190},
  url = {https://doi.org/10.1021/acs.jctc.3c01190},
  publisher = {American Chemical Society},
  issn = {1549-9618}
}

@ARTICLE{kitaev1995abelian_qpe,
  author    = {A. Yu. Kitaev},
  title     = {Quantum Measurements and the Abelian Stabilizer Problem},
  journal   = {arXiv:quant-ph/9511026},
  year      = {1995},
  doi       = {10.48550/arXiv.quant-ph/9511026},
}

@ARTICLE{
sqd_first_paper,
author = {Javier Robledo-Moreno  and Mario Motta  and Holger Haas  and Ali Javadi-Abhari  and Petar Jurcevic  and William Kirby  and Simon Martiel  and Kunal Sharma  and Sandeep Sharma  and Tomonori Shirakawa  and Iskandar Sitdikov  and Rong-Yang Sun  and Kevin J. Sung  and Maika Takita  and Minh C. Tran  and Seiji Yunoki  and Antonio Mezzacapo },
title = {Chemistry beyond the scale of exact diagonalization on a quantum-centric supercomputer},
journal = {Science Advances},
volume = {11},
number = {25},
pages = {eadu9991},
year = {2025},
doi = {10.1126/sciadv.adu9991}}

@article{kanno2023qsci,
  title = {Quantum-selected configuration interaction: Classical diagonalization of Hamiltonians in subspaces selected by quantum computers},
  author = {Kanno, Keita and Kohda, Masaya and Imai, Ryosuke and Koh, Sho and Mitarai, Kosuke and Mizukami, Wataru and Nakagawa, Yuya O.},
  journal = {Phys. Rev. Res.},
  volume = {8},
  issue = {2},
  pages = {023268},
  numpages = {27},
  year = {2026},
  month = {Jun},
  publisher = {American Physical Society},
  doi = {10.1103/dmn4-snfx},
  url = {https://link.aps.org/doi/10.1103/dmn4-snfx}
}

@ARTICLE{Peruzzo2014_vqe,
  author = {Peruzzo, Alberto and McClean, Jarrod and Shadbolt, Peter and Yung, Man-Hong and Zhou, Xiao-Qi and Love, Peter J. and Aspuru-Guzik, Alán and O’Brien, Jeremy L.},
  title = {A variational eigenvalue solver on a photonic quantum processor},
  journal = {Nature Communications},
  year = {2014},
  volume = {5},
  number = {1},
  pages = {4213},
  issn = {2041-1723},
  doi = {10.1038/ncomms5213},
  url = {https://doi.org/10.1038/ncomms5213}
}

@ARTICLE{Belaloui2025_current_vqe,
  author = {Belaloui, Nacer Eddine and Tounsi, Abdellah and Khamadja, Abdelmouheymen Rabah and Louamri, Mohamed Messaoud and Benslama, Achour and Bernal Neira, David E. and Rouabah, Mohamed Taha},
  title = {Ground-State Energy Estimation on Current Quantum Hardware through the Variational Quantum Eigensolver: A Practical Study},
  journal = {Journal of Chemical Theory and Computation},
  year = {2025},
  volume = {21},
  number = {14},
  pages = {6777--6792},
  publisher = {American Chemical Society},
  issn = {1549-9618},
  doi = {10.1021/acs.jctc.4c01657},
  url = {https://doi.org/10.1021/acs.jctc.4c01657}
}

@ARTICLE{wiley_protein_ligand_interaction,
author = {Kirsopp, Josh J. M. and Di Paola, Cono and Manrique, David Zsolt and Krompiec, Michal and Greene-Diniz, Gabriel and Guba, Wolfgang and Meyder, Agnes and Wolf, Detlef and Strahm, Martin and Muñoz Ramo, David},
title = {Quantum computational quantification of protein–ligand interactions},
journal = {International Journal of Quantum Chemistry},
volume = {122},
number = {22},
pages = {e26975},
doi = {https://doi.org/10.1002/qua.26975},
url = {https://onlinelibrary.wiley.com/doi/abs/10.1002/qua.26975},
year = {2022}
}

@ARTICLE{critical_lims_sqd_2025,
   title={Critical Limitations in Quantum-Selected Configuration Interaction Methods},
   volume={21},
   ISSN={1549-9626},
   url={http://dx.doi.org/10.1021/acs.jctc.5c00375},
   DOI={10.1021/acs.jctc.5c00375},
   number={14},
   journal={Journal of Chemical Theory and Computation},
   publisher={American Chemical Society (ACS)},
   author={Reinholdt, Peter and Ziems, Karl Michael and Kjellgren, Erik Rosendahl and Coriani, Sonia and Sauer, Stephan P. A. and Kongsted, Jacob},
   year={2025},
   month=jun, pages={6811–6822} }

@misc{scipy_newton,
  title        = {scipy.optimize.newton --- SciPy v1.11.4 Manual},
  author       = {{SciPy Developers}},
  howpublished = {\url{https://docs.scipy.org/doc/scipy/reference/generated/scipy.optimize.newton.html}},
  year         = {2024},
  note         = {Accessed: 2025-10-28}
}

@misc{senicourt2022tangeloopensourcepythonpackage,
      title={Tangelo: An Open-source Python Package for End-to-end Chemistry Workflows on Quantum Computers}, 
      author={Valentin Senicourt and James Brown and Alexandre Fleury and Ryan Day and Erika Lloyd and Marc P. Coons and Krzysztof Bieniasz and Lee Huntington and Alejandro J. Garza and Shunji Matsuura and Rudi Plesch and Takeshi Yamazaki and Arman Zaribafiyan},
      year={2022},
      howpublished={2206.12424},
      archivePrefix={arXiv},
      primaryClass={quant-ph},
      url={https://arxiv.org/abs/2206.12424},
}

@manual{ffsim,
author = {{The ffsim developers}},
title = {{ffsim: Faster simulations of fermionic quantum circuits, v0.0.56}},
organization = {qiskit-community},
year = {2025},
url = {https://github.com/qiskit-community/ffsim},
note = {Accessed: 2025-11-11},
}

@misc{javadiabhari2024quantumcomputingqiskit,
      title={Quantum computing with Qiskit}, 
      author={Ali Javadi-Abhari and Matthew Treinish and Kevin Krsulich and Christopher J. Wood and Jake Lishman and Julien Gacon and Simon Martiel and Paul D. Nation and Lev S. Bishop and Andrew W. Cross and Blake R. Johnson and Jay M. Gambetta},
      year={2024},
      howpublished={2405.08810},
      archivePrefix={arXiv},
      primaryClass={quant-ph},
      url={https://arxiv.org/abs/2405.08810},
}

@article{wouters2016dmets_guide,
  author       = {Wouters, Sebastian and Jiménez-Hoyos, Carlos A. and Sun, Qiming and Chan, Garnet K.-L.},
  title        = {A Practical Guide to Density Matrix Embedding Theory in Quantum Chemistry},
  journal      = {Journal of Chemical Theory and Computation},
  volume       = {12},
  number       = {6},
  pages        = {2706--2719},
  year         = {2016},
  month        = jun,
  publisher    = {American Chemical Society},
  doi          = {10.1021/acs.jctc.6b00316},
  url          = {https://doi.org/10.1021/acs.jctc.6b00316}
}

@ARTICLE{bartlett_cc_theory_2007,
  author={Bartlett, Rodney J. and Musiał, Monika},
  journal={Reviews of Modern Physics}, 
  title={Coupled-Cluster Theory in Quantum Chemistry}, 
  year={2007},
  volume={79},
  number={1},
  pages={291--352},
  doi={10.1103/RevModPhys.79.291},
  url={https://link.aps.org/doi/10.1103/RevModPhys.79.291},
  publisher={American Physical Society}
}

@article{Kawashima2021_dmet_vqe_ion,
  author       = {Kawashima, Yukio and Lloyd, Erika and Coons, Marc P. and Nam, Yunseong and Matsuura, Shunji and Garza, Alejandro J. and Johri, Sonika and Huntington, Lee and Senicourt, Valentin and Maksymov, Andrii O. and Nguyen, Jason H. V. and Kim, Jungsang and Alidoust, Nima and Zaribafiyan, Arman and Yamazaki, Takeshi},
  title        = {Optimizing electronic structure simulations on a trapped-ion quantum computer using problem decomposition},
  journal      = {Communications Physics},
  year         = {2021},
  volume       = {4},
  number       = {1},
  pages        = {245},
  doi          = {10.1038/s42005-021-00751-9},
  url          = {https://doi.org/10.1038/s42005-021-00751-9}
}

@misc{sun2017pyscf,
      title={The Python-based Simulations of Chemistry Framework (PySCF)}, 
      author={Qiming Sun and Timothy C. Berkelbach and Nick S. Blunt and George H. Booth and Sheng Guo and Zhendong Li and Junzi Liu and James McClain and Elvira R. Sayfutyarova and Sandeep Sharma and Sebastian Wouters and Garnet Kin-Lic Chan},
      year={2017},
      howpublished={1701.08223},
      archivePrefix={arXiv},
      primaryClass={physics.chem-ph},
      url={https://arxiv.org/abs/1701.08223},
}

@book{szabo1996_modernqc,
  title        = {Modern Quantum Chemistry: Introduction to Advanced Electronic Structure Theory},
  author       = {Szabo, A. and Ostlund, N. S.},
  publisher    = {Dover Publications},
  year         = {1996},
  isbn         = {9780486691862},
  series       = {Dover Books on Chemistry},
  url          = {https://books.google.co.in/books?id=6mV9gYzEkgIC},
  note         = {See Section 2.4.1, pp.\ 89--95},
}

@article{papakonstantinou2013secant,
  title        = {Origin and Evolution of the Secant Method in One Dimension},
  author       = {Papakonstantinou, Joanna M. and Tapia, Richard A.},
  journal      = {The American Mathematical Monthly},
  volume       = {120},
  number       = {6},
  pages        = {500--518},
  year         = {2013},
  publisher    = {Mathematical Association of America},
  doi          = {10.4169/amer.math.monthly.120.06.500},
  url          = {http://www.jstor.org/stable/10.4169/amer.math.monthly.120.06.500}
}

@article{Preskill_2018,
   title={Quantum Computing in the NISQ era and beyond},
   volume={2},
   ISSN={2521-327X},
   url={http://dx.doi.org/10.22331/q-2018-08-06-79},
   DOI={10.22331/q-2018-08-06-79},
   journal={Quantum},
   publisher={Verein zur Forderung des Open Access Publizierens in den Quantenwissenschaften},
   author={Preskill, John},
   year={2018},
   month=aug, pages={79} }

@article{sci_evangelista_1,
    author = {Evangelista, Francesco A.},
    title = {Adaptive multiconfigurational wave functions},
    journal = {The Journal of Chemical Physics},
    volume = {140},
    number = {12},
    pages = {124114},
    year = {2014},
    month = {03},
    issn = {0021-9606},
    doi = {10.1063/1.4869192},
    url = {https://doi.org/10.1063/1.4869192},
}

@ARTICLE{2020SciPy-NMeth,
  author  = {Virtanen, Pauli and Gommers, Ralf and Oliphant, Travis E. and
            Haberland, Matt and Reddy, Tyler and Cournapeau, David and
            Burovski, Evgeni and Peterson, Pearu and Weckesser, Warren and
            Bright, Jonathan and {van der Walt}, St{\'e}fan J. and
            Brett, Matthew and Wilson, Joshua and Millman, K. Jarrod and
            Mayorov, Nikolay and Nelson, Andrew R. J. and Jones, Eric and
            Kern, Robert and Larson, Eric and Carey, C J and
            Polat, {\.I}lhan and Feng, Yu and Moore, Eric W. and
            {VanderPlas}, Jake and Laxalde, Denis and Perktold, Josef and
            Cimrman, Robert and Henriksen, Ian and Quintero, E. A. and
            Harris, Charles R. and Archibald, Anne M. and
            Ribeiro, Ant{\^o}nio H. and Pedregosa, Fabian and
            {van Mulbregt}, Paul and {SciPy 1.0 Contributors}},
  title   = {{{SciPy} 1.0: Fundamental Algorithms for Scientific
            Computing in Python}},
  journal = {Nature Methods},
  year    = {2020},
  volume  = {17},
  pages   = {261--272},
  adsurl  = {https://rdcu.be/b08Wh},
  doi     = {10.1038/s41592-019-0686-2},
}

@ARTICLE{Barison2025_QCExcitedStates_SQD,
  author    = {Barison, Stefano and Robledo Moreno, Javier and Motta, Mario},
  title     = {Quantum-Centric Computation of Molecular Excited States with Extended Sample-Based Quantum Diagonalization},
  journal   = {Quantum Science and Technology},
  volume    = {10},
  number    = {2},
  pages     = {025034},
  year      = {2025},
  month     = {February},
  doi       = {10.1088/2058-9565/adb781},
  url       = {https://doi.org/10.1088/2058-9565/adb781},
  publisher = {IOP Publishing},
  issn      = {2058-9565},
}

@misc{hi_vqe_2024,
      title={HIVQE: Handover Iterative Variational Quantum Eigensolver for Efficient Quantum Chemistry Calculations}, 
      author={Aidan Pellow-Jarman and Shane McFarthing and Doo Hyung Kang and Pilsun Yoo and Eyuel Eshetu Elala and Rowan Pellow-Jarman and P. Mai Nakliang and Jaewan Kim and June-Koo Kevin Rhee},
      year={2025},
      howpublished={2503.06292},
      archivePrefix={arXiv},
      primaryClass={quant-ph},
      url={https://arxiv.org/abs/2503.06292},
}

@misc{piccinelli2025_sqdrift,
      title={Quantum chemistry with provable convergence via randomized sample-based quantum diagonalization}, 
      author={Samuele Piccinelli and Alberto Baiardi and Max Rossmannek and Almudena Carrera Vazquez and Francesco Tacchino and Stefano Mensa and Edoardo Altamura and Ali Alavi and Mario Motta and Javier Robledo-Moreno and William Kirby and Kunal Sharma and Antonio Mezzacapo and Ivano Tavernelli},
      year={2025},
      howpublished={2508.02578},
      archivePrefix={arXiv},
      primaryClass={quant-ph},
      url={https://arxiv.org/abs/2508.02578}, 
}

@ARTICLE{Alexeev2024_QCSMaterials_sqd,
  author    = {Alexeev, Yuri and Amsler, Maximilian and Barroca, Marco Antonio and Bassini, Sanzio and Battelle, Torey and Camps, Daan and Casanova, David and Choi, Young Jay and Chong, Frederic T. and Chung, Charles and Codella, Christopher and C{\'o}rcoles, Antonio D. and Cruise, James and Di Meglio, Alberto and Duran, Ivan and Eckl, Thomas and Economou, Sophia and Eidenbenz, Stephan and Elmegreen, Bruce and Fare, Clyde and Faro, Ismael and Sanz Fern{\'a}ndez, Cristina and Neumann Barros Ferreira, Rodrigo and Fuji, Keisuke and Fuller, Bryce and Gagliardi, Laura and Galli, Giulia and Glick, Jennifer R. and Gobbi, Isacco and Gokhale, Pranav and de la Puente Gonzalez, Salvador and Greiner, Johannes and Gropp, Bill and Grossi, Michele and Gull, Emanuel and Healy, Burns and Hermes, Matthew R. and Huang, Benchen and Humble, Travis S. and Ito, Nobuyasu and Izmaylov, Artur F. and Javadi-Abhari, Ali and Jennewein, Douglas and Jha, Shantenu and Jiang, Liang and Jones, Barbara and de Jong, Wibe Albert and Jurcevic, Petar and Kirby, William and Kister, Stefan and Kitagawa, Masahiro and Klassen, Joel and Klymko, Katherine and Koh, Kwangwon and Kondo, Masaaki and K{\"u}rk{\c{c}}uog{\u{l}}u, Do{\u{g}}a Murat and Kurowski, Krzysztof and Laino, Teodoro and Landfield, Ryan and Leininger, Matt and Leyton-Ortega, Vicente and Li, Ang and Lin, Meifeng and Liu, Junyu and Lorente, Nicolas and Luckow, Andre and Martiel, Simon and Martin-Fernandez, Francisco and Martonosi, Margaret and Marvinney, Claire and Castaneda Medina, Arcesio and Merten, Dirk and Mezzacapo, Antonio and Michielsen, Kristel and Mitra, Abhishek and Mittal, Tushar and Moon, Kyungsun and Moore, Joel and Mostame, Sarah and Motta, Mario and Na, Young-Hye and Nam, Yunseong and Narang, Prineha and Ohnishi, Yu-ya and Ottaviani, Daniele and Otten, Matthew and Pakin, Scott and Pascuzzi, Vincent R. and Pednault, Edwin and Piontek, Tomasz and Pitera, Jed and Rall, Patrick and Ravi, Gokul Subramanian and Robertson, Niall and Rossi, Matteo A.C. and Rydlichowski, Piotr and Ryu, Hoon and Samsonidze, Georgy and Sato, Mitsuhisa and Saurabh, Nishant and Sharma, Vidushi and Sharma, Kunal and Shin, Soyoung and Slessman, George and Steiner, Mathias and Sitdikov, Iskandar and Suh, In-Saeng and Switzer, Eric D. and Tang, Wei and Thompson, Joel and Todo, Synge and Tran, Minh C. and Trenev, Dimitar and Trott, Christian and Tseng, Huan-Hsin and Tubman, Norm M. and Tureci, Esin and Garc{\'i}a Vali{\~n}as, David and Vallecorsa, Sofia and Wever, Christopher and Wojciechowski, Konrad and Wu, Xiaodi and Yoo, Shinjae and Yoshioka, Nobuyuki and Yu, Victor Wen-zhe and Yunoki, Seiji and Zhuk, Sergiy and Zubarev, Dmitry},
  title     = {Quantum-centric supercomputing for materials science: A perspective on challenges and future directions},
  journal   = {Future Generation Computer Systems},
  year      = {2024},
  volume    = {160},
  pages     = {666--710},
  month     = {04},
  doi       = {10.1016/j.future.2024.04.060},
  url       = {https://doi.org/10.1016/j.future.2024.04.060},
  issn      = {0167-739X},
  publisher = {Elsevier}
}

@ARTICLE{MacDonald1933_RayleighRitz,
  author    = {MacDonald, J. K. L.},
  title     = {Successive Approximations by the Rayleigh–Ritz Variation Method},
  journal   = {Phys. Rev.},
  year      = {1933},
  volume    = {43},
  number    = {10},
  pages     = {830--833},
  month     = {May},
  doi       = {10.1103/PhysRev.43.830},
  url       = {https://link.aps.org/doi/10.1103/PhysRev.43.830},
  publisher = {American Physical Society}
}

@ARTICLE{Knizia2012_DMET,
  author    = {Knizia, Gerald and Chan, Garnet Kin-Lic},
  title     = {Density Matrix Embedding: A Simple Alternative to Dynamical Mean-Field Theory},
  journal   = {Physical Review Letters},
  volume    = {109},
  number    = {18},
  pages     = {186404},
  year      = {2012},
  month     = {November},
  doi       = {10.1103/PhysRevLett.109.186404},
  url       = {https://doi.org/10.1103/PhysRevLett.109.186404},
  publisher = {American Physical Society},
  issn      = {0031-9007}
}

@ARTICLE{Georges1996_DMFT,
  author    = {Georges, Antoine and Kotliar, Gabriel and Krauth, Werner and Rozenberg, Marcelo J.},
  title     = {Dynamical Mean-Field Theory of Strongly Correlated Fermion Systems and the Limit of Infinite Dimensions},
  journal   = {Reviews of Modern Physics},
  volume    = {68},
  number    = {1},
  pages     = {13--125},
  year      = {1996},
  month     = {January},
  doi       = {10.1103/RevModPhys.68.13},
  url       = {https://doi.org/10.1103/RevModPhys.68.13},
  publisher = {American Physical Society},
  issn      = {0034-6861}
}

@article{mcclean2018barren,
  author       = {McClean, Jarrod R. and Boixo, Sergio and Smelyanskiy, Vadim N. and Babbush, Ryan and Neven, Hartmut},
  title        = {Barren plateaus in quantum neural network training landscapes},
  journal      = {Nature Communications},
  year         = {2018},
  volume       = {9},
  pages        = {1},
}

@ARTICLE{Davidson1975_iterative_eigensolver,
  author    = {Davidson, Ernest R.},
  title     = {The iterative calculation of a few of the lowest eigenvalues and corresponding eigenvectors of large real-symmetric matrices},
  journal   = {Journal of Computational Physics},
  year      = {1975},
  volume    = {17},
  number    = {1},
  pages     = {87--94},
  doi       = {10.1016/0021-9991(75)90065-0},
  url       = {https://doi.org/10.1016/0021-9991(75)90065-0},
  issn      = {0021-9991},
  publisher = {Elsevier}
}

@article{patra2025survey,
author = {Ashish Kumar Patra  and Anurag K S V  and Vikas Dattatraya Ghevade  and Sai Shankar P  and Ruchika Bhat  and Amit Saxena  and Raghavendra V  and Jaiganesh Gnana Sekaran },
title = {Survey of Quantum Algorithms: Foundations, Frameworks and Applications},
journal = {Authorea},
volume = {2025},
number = {1117},
pages = {},
year = {2025},
doi = {10.22541/au.176341037.72812424/v1}}

@article{Bowling2025_ProteinLigand_FragQC,
  author  = {Bowling, Paige E. and Broderick, Dustin R. and Herbert, John M.},
  title   = {Convergent Protocols for Computing Protein--Ligand Interaction Energies Using Fragment-Based Quantum Chemistry},
  journal = {Journal of Chemical Theory and Computation},
  year    = {2025},
  volume  = {21},
  number  = {2},
  pages   = {951--966},
  doi     = {10.1021/acs.jctc.4c01429},
  url     = {https://doi.org/10.1021/acs.jctc.4c01429},
  issn    = {1549-9618},
  publisher = {American Chemical Society}
}

@misc{patra2025physicsinformedgenerativemachine,
      title={Physics Informed Generative Machine Learning for Accelerated Quantum-centric Supercomputing}, 
      author={Chayan Patra and Dibyendu Mondal and Sonaldeep Halder and Dipanjali Halder and Mostafizur Rahaman Laskar and Richa Goel and Rahul Maitra},
      year={2025},
      eprint={2512.06858},
      archivePrefix={arXiv},
      primaryClass={quant-ph},
      url={https://arxiv.org/abs/2512.06858}, 
}

@article{Herzog_CIGen_2023,
  author       = {Herzog, Basile and Casier, Bastien and Leb{\`e}gue, S{\'e}bastien and Rocca, Dario},
  title        = {Solving the Schr{\"o}dinger Equation in the Configuration Space with Generative Machine Learning},
  journal      = {Journal of Chemical Theory and Computation},
  year         = {2023},
  month        = {Apr},
  volume       = {19},
  number       = {9},
  pages        = {2484--2490},
  publisher    = {American Chemical Society},
  doi          = {10.1021/acs.jctc.2c01216},
  url          = {https://doi.org/10.1021/acs.jctc.2c01216},
  issn         = {1549-9618}
}

@article{Coe_MLCI_2018,
  author       = {Coe, J. P.},
  title        = {Machine Learning Configuration Interaction},
  journal      = {Journal of Chemical Theory and Computation},
  year         = {2018},
  month        = {Oct},
  volume       = {14},
  number       = {11},
  pages        = {5739--5749},
  publisher    = {American Chemical Society},
  doi          = {10.1021/acs.jctc.8b00849},
  url          = {https://doi.org/10.1021/acs.jctc.8b00849},
  issn         = {1549-9618}
}

@article{Nomura_RBM_Learning_2017,
  title = {Restricted Boltzmann machine learning for solving strongly correlated quantum systems},
  author = {Nomura, Yusuke and Darmawan, Andrew S. and Yamaji, Youhei and Imada, Masatoshi},
  journal = {Phys. Rev. B},
  volume = {96},
  issue = {20},
  pages = {205152},
  numpages = {8},
  year = {2017},
  month = {Nov},
  publisher = {American Physical Society},
  doi = {10.1103/PhysRevB.96.205152},
  url = {https://link.aps.org/doi/10.1103/PhysRevB.96.205152}
}

@book{Helgaker2000,
author = {Helgaker, Trygve and Jørgensen, Poul and Olsen, Jeppe},
publisher = {John Wiley \& Sons, Ltd},
isbn = {9781119019572},
title = {Molecular Electronic‐Structure Theory},
edition = {1st},
doi = {10.1002/9781119019572.ch1},
url = {https://onlinelibrary.wiley.com/doi/abs/10.1002/9781119019572.ch1},
year = {2000},
}

@misc{merz2026crossing12000atombarrierheterogeneous,
      title={Crossing the 12,000-atom barrier with heterogeneous quantum-classical supercomputing: quantum chemistry of protein-ligand complexes}, 
      author={Kenneth M. Merz, Jr. and Akhil Shajan and Danil Kaliakin and Fangchun Liang and Yuichi Otsuka and Tomonori Shirakawa and Lukas Broers and Han Xu and Miwako Tsuji and Mitsuhisa Sato and Seiji Yunoki and Ryo Wakizaka and Yukio Kawashima and Jun Doi and Toshinari Itoko and Hiroshi Horii and Thaddeus Pellegrini and Javier Robledo Moreno and Kevin J. Sung and Ella Fejer and Robert Walkup and Seetharami Seelam and Mario Motta},
      year={2026},
      eprint={2605.01138},
      archivePrefix={arXiv},
      primaryClass={quant-ph},
      url={https://arxiv.org/abs/2605.01138}, 
}

@article{Vogiatzis2017,
    author = {Vogiatzis, Konstantinos D. and Ma, Dongxia and Olsen, Jeppe and Gagliardi, Laura and de Jong, Wibe A.},
    title = {Pushing configuration-interaction to the limit: Towards massively parallel MCSCF calculations},
    journal = {The Journal of Chemical Physics},
    volume = {147},
    number = {18},
    pages = {184111},
    year = {2017},
    month = {Nov},
    issn = {0021-9606},
    doi = {10.1063/1.4989858},
    url = {https://doi.org/10.1063/1.4989858},
}

@article{Shayit2025,
  author       = {Shayit, Agam and Liao, Can and Upadhyay, Shiv and Hu, Hang and Zhang, Tianyuan and DePrince III, A. Eugene and Yang, Chao and Li, Xiaosong},
  title        = {Numerically exact configuration interaction at quadrillion-determinant scale},
  journal      = {Nature Communications},
  month        = {Dec},
  year         = {2025},
  volume       = {16},
  number       = {1},
  pages        = {11016},
  doi          = {10.1038/s41467-025-65967-7},
  url          = {https://doi.org/10.1038/s41467-025-65967-7},
  issn         = {2041-1723}
}

@article{boguslawski2012,
    author = {Boguslawski, Katharina and Tecmer, Pawel̷ and Legeza, Örs and Reiher, Markus},
    title = {Entanglement Measures
for Single- and Multireference
Correlation Effects},
    journal = {The Journal of Physical Chemistry Letters},
    volume = {3},
    number = {21},
    pages = {3129-3135},
    year = {2012},
    month = {11},
    issn = {1948-7185},
    doi = {10.1021/jz301319v},
    url = {https://doi.org/10.1021/jz301319v},
    eprint = {https://pubs.acs.org/jpclcd/article-pdf/3/21/3129/19688466/jz301319v.pdf},
}

@article{KSV2025,
author = {Anurag, K. S. V. and Patra, Ashish Kumar and Ghevade, Vikas Dattatraya and Sai Shankar, P. and Bhat, Ruchika and Raghavendra, V. and Maitra, Rahul and Jaiganesh, G.},
title = {Resource Estimation for VQE on Small Molecules: Impact of Fermion Mappings and Hamiltonian Reductions},
journal = {Journal of Computational Chemistry},
volume = {47},
number = {11},
pages = {e70379},
doi = {https://doi.org/10.1002/jcc.70379},
url = {https://onlinelibrary.wiley.com/doi/abs/10.1002/jcc.70379},
year = {2026}
}

@misc{kirby2026observationimprovedaccuracyclassical,
      title={Observation of Improved Accuracy over Classical Sparse Ground-State Solvers using a Quantum Computer}, 
      author={William Kirby and Bibek Pokharel and Javier Robledo Moreno and Kevin C. Smith and Sergey Bravyi and Abhinav Deshpande and Constantinos Evangelinos and Bryce Fuller and James R. Garrison and Ben Jaderberg and Caleb Johnson and Petar Jurcevic and Su-un Lee and Simon Martiel and Mario Motta and Seetharami Seelam and Oles Shtanko and Kevin J. Sung and Minh Tran and Vinay Tripathi and Kazuhiro Seki and Kazuya Shinjo and Han Xu and Lukas Broers and Tomonori Shirakawa and Seiji Yunoki and Kunal Sharma and Antonio Mezzacapo},
      year={2026},
      eprint={2603.03496},
      archivePrefix={arXiv},
      primaryClass={quant-ph},
      url={https://arxiv.org/abs/2603.03496}, 
}

@article{Wang2021,
  author       = {Wang, Samson and Fontana, Enrico and Cerezo, M. and Sharma, Kunal and Sone, Akira and Cincio, Lukasz and Coles, Patrick J.},
  title        = {Noise-induced barren plateaus in variational quantum algorithms},
  journal      = {Nature Communications},
  year         = {2021},
  volume       = {12},
  number       = {1},
  pages        = {6961},
  month        = {Nov},
  doi          = {10.1038/s41467-021-27045-6},
  url          = {https://doi.org/10.1038/s41467-021-27045-6},
  issn         = {2041-1723},
}

@article{Nakagawa2024,
    author = {Nakagawa, Yuya O. and Kamoshita, Masahiko and Mizukami, Wataru and Sudo, Shotaro and Ohnishi, Yu-ya},
    title = {ADAPT-QSCI: Adaptive Construction of an Input State for Quantum-Selected Configuration Interaction},
    journal = {Journal of Chemical Theory and Computation},
    volume = {20},
    number = {24},
    pages = {10817-10825},
    month = {Dec},
    year = {2024},
    doi = {10.1021/acs.jctc.4c00846},
    URL = {https://doi.org/10.1021/acs.jctc.4c00846},
}

@article{Sugisaki2025,
    author ="Sugisaki, Kenji and Kanno, Shu and Itoko, Toshinari and Sakuma, Rei and Yamamoto, Naoki",
    title  ="Hamiltonian simulation-based quantum-selected configuration interaction for large-scale electronic structure calculations with a quantum computer",
    journal  ="Phys. Chem. Chem. Phys.",
    month ="Sep",
    year  ="2025",
    volume  ="27",
    issue  ="38",
    pages  ="20869-20884",
    publisher  ="The Royal Society of Chemistry",
    doi  ="10.1039/D5CP02202A",
    url  ="http://dx.doi.org/10.1039/D5CP02202A"
}

@article{Mikkelsen2025,
  title = {Quantum-selected configuration interaction with time-evolved state},
  author = {Mikkelsen, Mathias and Nakagawa, Yuya O.},
  journal = {Phys. Rev. Res.},
  volume = {7},
  issue = {4},
  pages = {043043},
  numpages = {17},
  year = {2025},
  month = {Oct},
  publisher = {American Physical Society},
  doi = {10.1103/75pv-hbrx},
  url = {https://link.aps.org/doi/10.1103/75pv-hbrx}
}

@misc{Shirakawa2025,
      title={Closed-loop calculations of electronic structure on a quantum processor and a classical supercomputer at full scale}, 
      author={Tomonori Shirakawa and Javier Robledo-Moreno and Toshinari Itoko and Vinay Tripathi and Kento Ueda and Yukio Kawashima and Lukas Broers and William Kirby and Himadri Pathak and Hanhee Paik and Miwako Tsuji and Yuetsu Kodama and Mitsuhisa Sato and Constantinos Evangelinos and Seetharami Seelam and Robert Walkup and Seiji Yunoki and Mario Motta and Petar Jurcevic and Hiroshi Horii and Antonio Mezzacapo},
      month={Oct},
      year={2025},
      eprint={2511.00224},
      archivePrefix={arXiv},
      primaryClass={quant-ph},
      url={https://arxiv.org/abs/2511.00224}, 
}

@article{Liepuoniute2025,
    author = {Liepuoniute, Ieva and Doney, Kirstin D. and Robledo Moreno, Javier and Job, Joshua A. and Friend, William S. and Jones, Gavin O.},
    title = {Quantum-Centric Computational Study of Methylene Singlet and Triplet States},
    journal = {Journal of Chemical Theory and Computation},
    volume = {21},
    number = {10},
    pages = {5062-5070},
    month = {May},
    year = {2025},
    doi = {10.1021/acs.jctc.5c00075},
    URL = {https://doi.org/10.1021/acs.jctc.5c00075}
}

@article{Kaliakin2025a,
    author = {Kaliakin, Danil and Shajan, Akhil and Liang, Fangchun and Merz, Kenneth M. Jr.},
    title = {Implicit Solvent Sample-Based Quantum Diagonalization},
    journal = {The Journal of Physical Chemistry B},
    volume = {129},
    number = {23},
    pages = {5788-5796},
    month = {June},
    year = {2025},
    doi = {10.1021/acs.jpcb.5c01030},
    URL = {https://doi.org/10.1021/acs.jpcb.5c01030}
}

@article{Kaliakin2025b,
  author       = {Kaliakin, Danil and Shajan, Akhil and Liang, Fangchun and Robledo Moreno, Javier and Li, Zhen and Mitra, Abhishek and Motta, Mario and Johnson, Caleb and Saki, Abdullah Ash and Das, Susanta and Sitdikov, Iskandar and Mezzacapo, Antonio and Merz Jr., Kenneth M.},
  title        = {Accurate quantum-centric simulations of intermolecular interactions},
  journal      = {Communications Physics},
  year         = {2025},
  volume       = {8},
  number       = {1},
  pages        = {396},
  month        = {Oct},
  doi          = {10.1038/s42005-025-02305-9},
  url          = {https://doi.org/10.1038/s42005-025-02305-9},
  issn         = {2399-3650}
}

@misc{Yu2025,
      title={Quantum-Centric Algorithm for Sample-Based Krylov Diagonalization}, 
      author={Jeffery Yu and Javier Robledo Moreno and Joseph T. Iosue and Luke Bertels and Daniel Claudino and Bryce Fuller and Peter Groszkowski and Travis S. Humble and Petar Jurcevic and William Kirby and Thomas A. Maier and Mario Motta and Bibek Pokharel and Alireza Seif and Amir Shehata and Kevin J. Sung and Minh C. Tran and Vinay Tripathi and Antonio Mezzacapo and Kunal Sharma},
      month={Jan},
      year={2025},
      eprint={2501.09702},
      archivePrefix={arXiv},
      primaryClass={quant-ph},
      url={https://arxiv.org/abs/2501.09702}, 
}

@article{Yoshioka2025,
  author       = {Yoshioka, Nobuyuki and Amico, Mirko and Kirby, William and Jurcevic, Petar and Dutt, Arkopal and Fuller, Bryce and Garion, Shelly and Haas, Holger and Hamamura, Ikko and Ivrii, Alexander and Majumdar, Ritajit and Minev, Zlatko and Motta, Mario and Pokharel, Bibek and Rivero, Pedro and Sharma, Kunal and Wood, Christopher J. and Javadi-Abhari, Ali and Mezzacapo, Antonio},
  title        = {Krylov diagonalization of large many-body Hamiltonians on a quantum processor},
  journal      = {Nature Communications},
  year         = {2025},
  volume       = {16},
  number       = {1},
  pages        = {5014},
  month        = {Jun},
  doi          = {10.1038/s41467-025-59716-z},
  url          = {https://doi.org/10.1038/s41467-025-59716-z},
  issn         = {2041-1723}
}

@misc{Asthana2025,
      title={Quantum Krylov algorithm using unitary decomposition for exact eigenstates of fermionic systems using quantum computers}, 
      author={Ayush Asthana},
      month={Dec},
      year={2025},
      eprint={2512.11788},
      archivePrefix={arXiv},
      primaryClass={quant-ph},
      url={https://arxiv.org/abs/2512.11788}, 
}

@article{Huggins2022,
  author  = {Huggins, William J. and O’Gorman, Bryan A. and Rubin, Nicholas C. and Reichman, David R. and Babbush, Ryan and Lee, Joonho},
  title   = {Unbiasing fermionic quantum Monte Carlo with a quantum computer},
  journal = {Nature},
  month   = {Mar},
  year    = {2022},
  volume  = {603},
  number  = {7901},
  pages   = {416--420},
  doi     = {10.1038/s41586-021-04351-z}
}

@misc{Yoshida2025,
      title={Auxiliary-field quantum Monte Carlo method with quantum selected configuration interaction}, 
      author={Yuichiro Yoshida and Luca Erhart and Takuma Murokoshi and Rika Nakagawa and Chihiro Mori and Takafumi Miyanaga and Toshio Mori and Wataru Mizukami},
      month={Feb},
      year={2025},
      eprint={2502.21081},
      archivePrefix={arXiv},
      primaryClass={quant-ph},
      url={https://arxiv.org/abs/2502.21081}, 
}

@misc{Danilov2025,
      title={Enhancing the accuracy and efficiency of sample-based quantum diagonalization with phaseless auxiliary-field quantum Monte Carlo}, 
      author={Don Danilov and Javier Robledo-Moreno and Kevin J. Sung and Mario Motta and James Shee},
      month={Mar},
      year={2025},
      eprint={2503.05967},
      archivePrefix={arXiv},
      primaryClass={quant-ph},
      url={https://arxiv.org/abs/2503.05967}, 
}

@misc{Walkup2026,
      title={Scaling Sample-Based Quantum Diagonalization on GPU-Accelerated Systems using OpenMP Offload}, 
      author={Robert Walkup and Juha Jäykkä and Igor Pasichnyk and Zachary Streeter and Kasia Świrydowicz and Mikko Tukiainen and Yasuko Eckert and Luke Bertels and Daniel Claudino and Peter Groszkowski and Travis S. Humble and Constantinos Evangelinos and Javier Robledo-Moreno and William Kirby and Antonio Mezzacapo and Antonio Córcoles and Seetharami Seelam},
      month={Jan},
      year={2026},
      eprint={2601.16169},
      archivePrefix={arXiv},
      primaryClass={cs.ET},
      url={https://arxiv.org/abs/2601.16169}, 
}

@misc{Doi2026,
      title={GPU-Accelerated Selected Basis Diagonalization with Thrust for SQD-based Algorithms}, 
      author={Jun Doi and Tomonori Shirakawa and Yukio Kawashima and Seiji Yunoki and Hiroshi Horii},
      month={Jan},
      year={2026},
      eprint={2601.16637},
      archivePrefix={arXiv},
      primaryClass={cs.DC},
      url={https://arxiv.org/abs/2601.16637}, 
}

@misc{Raisuddin2025,
      title={From Promise to Practice: Benchmarking Quantum Chemistry on Quantum Hardware}, 
      author={Osama M. Raisuddin and Haimeng Zhang and Mario Motta and Fabian M. Faulstich},
      month={Nov},
      year={2025},
      eprint={2512.01012},
      archivePrefix={arXiv},
      primaryClass={quant-ph},
      url={https://arxiv.org/abs/2512.01012}, 
}

@misc{Weaving2025,
      title={Towards Compact Wavefunctions from Quantum-Selected Configuration Interaction}, 
      author={Tim Weaving and Angus Mingare and Alexis Ralli and Peter V. Coveney},
      month={Sep},
      year={2025},
      eprint={2509.02525},
      archivePrefix={arXiv},
      primaryClass={quant-ph},
      url={https://arxiv.org/abs/2509.02525}, 
}

@misc{Yoo2026,
      title={Extending the Handover-Iterative VQE to Challenging Strongly Correlated Systems: $N_2$ and Fe-S Cluster}, 
      author={Pilsun Yoo and Kyungmin Kim and Eyuel E. Elala and Shane McFarthing and Aidan Pellow and Johanna I. Fuks and Doo Hyung Kang and Pratanphorn Nakliang and Jaewan Kim and Himadri Pathak and Tomonori Shirakawa and Seiji Yunoki and June-Koo Kevin Rhee},
      month={Jan},
      year={2026},
      archivePrefix={arXiv},
      primaryClass={quant-ph},
      url={https://arxiv.org/abs/2601.06935}, 
}

@misc{Ghasemi2026,
      title={Hybrid Quantum Algorithms for Computational Chemistry: Application to the Pyridine-Li ion Complex}, 
      author={Fatemeh Ghasemi and Yousung Kang and Yukio Kawashima and Kyungsun Moon},
      month={Jan},
      year={2026},
      archivePrefix={arXiv},
      primaryClass={physics.chem-ph},
      eprint={2601.10002},
      url={https://arxiv.org/abs/2601.10002}, 
}

@misc{Smith2025,
      title={Quantum-centric simulation of hydrogen abstraction by sample-based quantum diagonalization and entanglement forging}, 
      author={Tyler Smith and Tanvi P. Gujarati and Mario Motta and Ben Link and Ieva Liepuoniute and Triet Friedhoff and Hiromichi Nishimura and Nam Nguyen and Kristen S. Williams and Javier Robledo Moreno and Caleb Johnson and Kevin J. Sung and Abdullah Ash Saki and Marna Kagele},
      month={Aug},
      year={2025},
      eprint={2508.08229},
      archivePrefix={arXiv},
      primaryClass={quant-ph},
      url={https://arxiv.org/abs/2508.08229}, 
}

@misc{McFarthing2026,
      title={Noise-Resilient Quantum Chemistry with Half the Qubits}, 
      author={Shane McFarthing and Aidan Pellow-Jarman and Francesco Petruccione},
      month={Feb},
      year={2026},
      eprint={2602.01165},
      archivePrefix={arXiv},
      primaryClass={quant-ph},
      url={https://arxiv.org/abs/2602.01165}, 
}

@misc{Yoshida2026,
      title={Doubling the size of quantum selected configuration interaction based on seniority-zero space and its application to QC-QSCI-AFQMC}, 
      author={Yuichiro Yoshida and Takuma Murokoshi and Naoya Kuroda and Wataru Mizukami},
      month={Feb},
      year={2026},
      eprint={2602.07912},
      archivePrefix={arXiv},
      primaryClass={quant-ph},
      url={https://arxiv.org/abs/2602.07912}, 
}

@article{Xu2024,
    author = {Xu, Enhua and Shimomoto, Yuma and Ten-no, Seiichiro L. and Tsuchimochi, Takashi},
    title = {Many-Body-Expansion Based on Variational Quantum Eigensolver and Deflation for Dynamical Correlation},
    journal = {The Journal of Physical Chemistry A},
    volume = {128},
    number = {12},
    pages = {2507-2521},
    month = {Mar},
    year = {2024},
    doi = {10.1021/acs.jpca.4c00351},
    URL = {https://doi.org/10.1021/acs.jpca.4c00351},
}

@article{Hermes2020,
author = {Hermes, Matthew R. and Pandharkar, Riddhish and Gagliardi, Laura},
title = {Variational Localized Active Space Self-Consistent Field Method},
journal = {Journal of Chemical Theory and Computation},
volume = {16},
number = {8},
pages = {4923-4937},
month = {Aug},
year = {2020},
doi = {10.1021/acs.jctc.0c00222},
URL = {https://doi.org/10.1021/acs.jctc.0c00222},
}

@article{Otten2022,
    author = {Otten, Matthew and Hermes, Matthew R. and Pandharkar, Riddhish and Alexeev, Yuri and Gray, Stephen K. and Gagliardi, Laura},
    title = {Localized Quantum Chemistry on Quantum Computers},
    journal = {Journal of Chemical Theory and Computation},
    volume = {18},
    number = {12},
    pages = {7205-7217},
    month = {Dec},
    year = {2022},
    doi = {10.1021/acs.jctc.2c00388},
    URL = {https://doi.org/10.1021/acs.jctc.2c00388}
}

@article{Liu2023,
    author = {Liu, Yuan and Meitei, Oinam R. and Chin, Zachary E. and Dutt, Arkopal and Tao, Max and Chuang, Isaac L. and Van Voorhis, Troy},
    title = {Bootstrap Embedding on a Quantum Computer},
    journal = {Journal of Chemical Theory and Computation},
    volume = {19},
    number = {8},
    pages = {2230-2247},
    month = {Apr},
    year = {2023},
    doi = {10.1021/acs.jctc.3c00012},
    URL = {https://doi.org/10.1021/acs.jctc.3c00012},
}

@article{Cances2025,
    author = {Cancès, Eric and Faulstich, Fabian M. and Kirsch, Alfred and Letournel, Eloïse and Levitt, Antoine},
    title = {Analysis of density matrix embedding theory around the non-interacting limit},
    journal = {Communications on Pure and Applied Mathematics},
    volume = {78},
    number = {8},
    pages = {1359-1410},
    doi = {https://doi.org/10.1002/cpa.22244},
    url = {https://onlinelibrary.wiley.com/doi/abs/10.1002/cpa.22244},
    month = {Feb},
    year = {2025}
}

@misc{Negre2025,
      title={New perspectives on Density-Matrix Embedding Theory}, 
      author={Alicia Negre and Fabian Faulstich and Raehyun Kim and Thomas Ayral and Lin Lin and Eric Cancès},
      month={Mar},
      year={2025},
      eprint={2503.09881},
      archivePrefix={arXiv},
      primaryClass={cond-mat.str-el},
      url={https://arxiv.org/abs/2503.09881}, 
}

@article{Patra2026,
    author = {Patra, Ashish Kumar and K. S. V., Anurag and P., Sai Shankar and Bhat, Ruchika and V., Raghavendra and Maitra, Rahul and G., Jaiganesh},
    title = {Quantum simulation of ligand-like molecules through sample-based quantum diagonalization in density matrix embedding framework},
    journal = {The Journal of Chemical Physics},
    volume = {165},
    number = {15},
    pages = {154108},
    year = {2026},
    month = {10},
    issn = {0021-9606},
    doi = {10.1063/5.0338676},
    url = {https://doi.org/10.1063/5.0338676},
}

@misc{Wang2025,
      title={Sample-based quantum diagonalization as parallel fragment solver for the localized active space self-consistent field method}, 
      author={Qiaohong Wang and Mario Motta and Ruhee D'Cunha and Kevin J. Sung and Matthew R. Hermes and Tanvi Gujarati and Yukio Kawashima and Yu-ya Ohnishi and Gavin O. Jones and Laura Gagliardi},
      month={Dec},
      year={2025},
      eprint={2512.14936},
      archivePrefix={arXiv},
      primaryClass={physics.chem-ph},
      url={https://arxiv.org/abs/2512.14936}, 
}

@Article{Bierman2026,
    author ="Bierman, Joel and Liu, Yuan",
    title  ="Towards utility-scale electronic structure with sample-based quantum bootstrap embedding",
    journal  ="Digital Discovery",
    month ="Jan",
    year  ="2026",
    pages  ="-",
    publisher  ="RSC",
    doi  ="10.1039/D5DD00416K",
    url  ="http://dx.doi.org/10.1039/D5DD00416K",
}

@misc{Shajan2026,
      title={Molecular Quantum Computations on a Protein}, 
      author={Akhil Shajan and Danil Kaliakin and Fangchun Liang and Thaddeus Pellegrini and Hakan Doga and Subhamoy Bhowmik and Susanta Das and Antonio Mezzacapo and Mario Motta and Kenneth M. Merz Jr},
      month={Jan},
      year={2026},
      eprint={2512.17130},
      archivePrefix={arXiv},
      primaryClass={quant-ph},
      url={https://arxiv.org/abs/2512.17130}, 
}

@misc{ksv2026iqm,
      title={Towards Chemically Accurate and Scalable Quantum Simulations on IQM Quantum Hardware: A Quantum-HPC Hybrid Approach}, 
      author={Anurag, K. S. V. and Ashish Kumar, Patra and Manas, Mukherjee and Alok, Shukla and Sai, Shankar P. and Ruchika Bhat and Radhika T. S. L. and Jaiganesh G},
      year={2026},
      eprint={2604.01983},
      archivePrefix={arXiv},
      primaryClass={quant-ph},
      url={https://arxiv.org/abs/2604.01983}, 
}

@article{MFCC_Zhang_2003,
    author = {Zhang, Da W. and Zhang, J. Z. H.},
    title = {Molecular fractionation with conjugate caps for full quantum mechanical calculation of protein–molecule interaction energy},
    journal = {The Journal of Chemical Physics},
    volume = {119},
    number = {7},
    pages = {3599-3605},
    year = {2003},
    month = {Aug},
    issn = {0021-9606},
    doi = {10.1063/1.1591727},
    url = {https://doi.org/10.1063/1.1591727}
}

@article{Fock1930,
  author    = {Fock, V.},
  title     = {N{\"a}herungsmethode zur L{\"o}sung des quantenmechanischen Mehrk{\"o}rperproblems},
  journal   = {Zeitschrift f{\"u}r Physik},
  year      = {1930},
  volume    = {61},
  number    = {1},
  pages     = {126--148},
  doi       = {10.1007/BF01340294},
  url       = {https://doi.org/10.1007/BF01340294}
}

@article{dynamical_decoupling_Rahman,
  title = {Learning how to dynamically decouple by optimizing rotational gates},
  author = {Rahman, Arefur and Egger, Daniel J. and Arenz, Christian},
  journal = {Phys. Rev. Appl.},
  volume = {22},
  issue = {5},
  pages = {054074},
  numpages = {12},
  year = {2024},
  month = {Nov},
  publisher = {American Physical Society},
  doi = {10.1103/PhysRevApplied.22.054074},
  url = {https://link.aps.org/doi/10.1103/PhysRevApplied.22.054074}
}

@article{dibyendu2026dmetcompass,
    author = {Mondal, Dibyendu and Patra, Ashish Kumar and Maitra, Rahul},
    title = {Advancing practical quantum embedding simulations via operator commutativity-based state preparation for complex chemical systems},
    journal = {Chemical Science},
    year = {2026},
    month = {08},
    issn = {2041-6520},
    doi = {10.1039/d6sc03132c},
    url = {https://doi.org/10.1039/d6sc03132c},
    eprint = {https://pubs.rsc.org/sc/article-pdf/doi/10.1039/d6sc03132c/13662588/d6sc03132c.pdf},
}

@article{pauli_twirling_foundation_wallman_2016,
  title = {Noise tailoring for scalable quantum computation via randomized compiling},
  author = {Wallman, Joel J. and Emerson, Joseph},
  journal = {Phys. Rev. A},
  volume = {94},
  issue = {5},
  pages = {052325},
  numpages = {9},
  year = {2016},
  month = {Nov},
  publisher = {American Physical Society},
  doi = {10.1103/PhysRevA.94.052325},
  url = {https://link.aps.org/doi/10.1103/PhysRevA.94.052325}
}

@article{Moller1934,
  title = {Note on an Approximation Treatment for Many-Electron Systems},
  author = {M\o{}ller, Chr. and Plesset, M. S.},
  journal = {Phys. Rev.},
  volume = {46},
  issue = {7},
  pages = {618--622},
  numpages = {0},
  year = {1934},
  month = {Oct},
  publisher = {American Physical Society},
  doi = {10.1103/PhysRev.46.618},
  url = {https://link.aps.org/doi/10.1103/PhysRev.46.618}
}

@book{jensen2007introduction_fci_size,
  author    = {Frank Jensen},
  title     = {Introduction to Computational Chemistry},
  edition   = {2},
  year      = {2007},
  publisher = {Wiley},
  isbn      = {9780470058046},
  note      = {See Sec.~4.2.2, p.~141},
  url       = {https://books.google.co.in/books?id=RDIG48UcZfYC}
}

@article{Jordan1928,
  author  = {Jordan, P. and Wigner, E.},
  title   = {{\"U}ber das Paulische {\"A}quivalenzverbot},
  journal = {Zeitschrift f{\"u}r Physik},
  year    = {1928},
  volume  = {47},
  number  = {9},
  pages   = {631--651},
  doi     = {10.1007/BF01331938},
  url     = {https://doi.org/10.1007/BF01331938},
  issn    = {0044-3328}
}

@article{
carleo_2017,
author = {Giuseppe Carleo  and Matthias Troyer },
title = {Solving the quantum many-body problem with artificial neural networks},
journal = {Science},
volume = {355},
number = {6325},
pages = {602-606},
year = {2017},
doi = {10.1126/science.aag2302},
URL = {https://www.science.org/doi/abs/10.1126/science.aag2302},
eprint = {https://www.science.org/doi/pdf/10.1126/science.aag2302}}

@article{kowalski_cc_downfolding,
    author = {Bauman, Nicholas P. and Kowalski, Karol},
    title = {Coupled cluster downfolding methods: The effect of double commutator terms on the accuracy of ground-state energies},
    journal = {The Journal of Chemical Physics},
    volume = {156},
    number = {9},
    pages = {094106},
    year = {2022},
    month = {03},
    issn = {0021-9606},
    doi = {10.1063/5.0076260},
    url = {https://doi.org/10.1063/5.0076260},
}

@article{ewf_embedding_2022,
  title = {Systematic Improvability in Quantum Embedding for Real Materials},
  author = {Nusspickel, Max and Booth, George H.},
  journal = {Phys. Rev. X},
  volume = {12},
  issue = {1},
  pages = {011046},
  numpages = {15},
  year = {2022},
  month = {Mar},
  publisher = {American Physical Society},
  doi = {10.1103/PhysRevX.12.011046},
  url = {https://link.aps.org/doi/10.1103/PhysRevX.12.011046}
}

@misc{iijima2023accuratequantumchemicalcalculations,
      title={Towards Accurate Quantum Chemical Calculations on Noisy Quantum Computers}, 
      author={Naoki Iijima and Satoshi Imamura and Mikio Morita and Sho Takemori and Akihiko Kasagi and Yuhei Umeda and Eiji Yoshida},
      year={2023},
      eprint={2311.09634},
      archivePrefix={arXiv},
      primaryClass={quant-ph},
      url={https://arxiv.org/abs/2311.09634}, 
}

@misc{vaquerosabater2026noiseconfigurationrecoveryimpact,
      title={Noise and Configuration Recovery Impact on Quantum Selected Configuration Interaction}, 
      author={Nonia Vaquero-Sabater and Abel Carreras and Lukas Broers and Tomonori Shirakawa and Seiji Yunoki and David Casanova},
      year={2026},
      eprint={2605.23697},
      archivePrefix={arXiv},
      primaryClass={quant-ph},
      url={https://arxiv.org/abs/2605.23697}, 
}

@misc{elgammal2026additivebindingenergiesasphalt,
      title={Additive binding energies in asphalt on a quantum processor via quantum-selected configuration interaction (QSCI)}, 
      author={Karim Elgammal and Marc Maußner},
      year={2026},
      eprint={2605.27640},
      archivePrefix={arXiv},
      primaryClass={quant-ph},
      url={https://arxiv.org/abs/2605.27640}, 
}

@article{bauman2025coupledclusterdownfoldingtheory,
  author       = {Bauman, Nicholas P. and Zheng, Muqing and Liu, Chenxu and Myers, Nathan M. and Panyala, Ajay and Peng, Bo and Li, Ang and Kowalski, Karol},
  title        = {Coupled Cluster Downfolding Theory in Simulations of Chemical Systems on Quantum Hardware},
  journal      = {arXiv},
  year         = {2025},
  publisher    = {arXiv},
  doi          = {10.48550/arXiv.2507.01199},
  url          = {https://doi.org/10.48550/arXiv.2507.01199}
}

@article{kowalski2024resourceadaptivequantumflowalgorithms,
  author       = {Kowalski, Karol and Bauman, Nicholas P.},
  title        = {Resource-adaptive quantum flow algorithms for quantum simulations of many-body systems: sub-flow embedding procedures},
  journal      = {arXiv},
  year         = {2024},
  publisher    = {arXiv},
  doi          = {10.48550/arXiv.2410.11992},
  url          = {https://doi.org/10.48550/arXiv.2410.11992}
}

@article{FMO_VQE_Lim_2024,
  title   = {Fragment Molecular Orbital-Based Variational Quantum Eigensolver for Quantum Chemistry in the Age of Quantum Computing},
  author  = {Lim, Hocheol and Kang, Doo Hyung and Kim, Jeonghoon and Pellow-Jarman, Aidan and McFarthing, Shane and Pellow-Jarman, Rowan and Jeon, Hyeon-Nae and Oh, Byungdu and Rhee, June-Koo Kevin and No, Kyoung Tai},
  journal = {Scientific Reports},
  volume  = {14},
  number  = {1},
  pages   = {2422},
  year    = {2024},
  doi     = {10.1038/s41598-024-52926-3},
  issn    = {2045-2322},
}

@Article{jinlong_2023_mbe,
author ="Ma, Huan and Liu, Jie and Shang, Honghui and Fan, Yi and Li, Zhenyu and Yang, Jinlong",
title  ="Multiscale quantum algorithms for quantum chemistry",
journal  ="Chem. Sci.",
year  ="2023",
volume  ="14",
issue  ="12",
pages  ="3190-3205",
publisher  ="The Royal Society of Chemistry",
doi  ="10.1039/D2SC06875C",
url  ="http://dx.doi.org/10.1039/D2SC06875C"}

@article{blumenthal2025_dmft_intuition,
  author       = {Blumenthal, Emmy},
  title        = {Building Intuition for Dynamical Mean-Field Theory: A Simple Model and the Cavity Method},
  journal      = {arXiv},
  year         = {2025},
  publisher    = {arXiv},
  doi          = {10.48550/arXiv.2507.16654},
  url          = {https://doi.org/10.48550/arXiv.2507.16654}
}

@article{ye2020bootstrap,
  author = {Ye, Hong-Zhou and Tran, Henry K. and Van Voorhis, Troy},
  title = {Bootstrap Embedding For Large Molecular Systems},
  journal = {J. Chem. Theory Comput.},
  year = {2020},
  month = {June},
  volume = {16},
  number = {8},
  pages = {5035--5046},
  doi = {10.1021/acs.jctc.0c00438},
  url = {https://doi.org/10.1021/acs.jctc.0c00438},
  publisher = {American Chemical Society}
}

@inproceedings{Hardikar2024,
author = {Hardikar, Tarini and Heitritter, Kenneth and Brown, James and D'Cunha, Ruhee and Mitra, Abhishek and Weatherly, Shaun and Liu, Yuan and Otten, Matthew and Voorhis, Troy and Gagliardi, Laura and Setia, Kanav},
year = {2024},
month = {09},
booktitle = {2024 IEEE International Conference on Quantum Computing and Engineering (QCE)},
pages = {538-544 (2024)},
title = {Quanta-Bind: A Quantum Computing Pipeline for Modeling Strongly Correlated Metal-Protein Interactions},
doi = {10.1109/QCE60285.2024.00069}
}

@misc{KSV2026Fujitsu,
      title={Bridging the NISQ and Fault-Tolerant Regimes: Generative-ML-Assisted Quantum Selected CI for Molecular Simulations}, 
      author={Anurag, K. S. V. and Ashish Kumar, Patra and Manas, Mukherjee and Ruchika, Bhat and Sai Shankar, P. and Rahul, Maitra and Jaiganesh, G},
      year={2026},
      month = {June},
      eprint={2606.30551},
      archivePrefix={arXiv},
      primaryClass={quant-ph},
      url={https://arxiv.org/abs/2606.30551}, 
}

@article{cho2025quemb,
  author = {Cho, Minsik and Meitei, Oinam Romesh and Weisburn, Leah P. and Weser, Oskar and Weatherly, Shaun and Alexiu, Alexandra and Hanscam, Rebecca and Tran, Henry K. and Ye, Hong-Zhou and Welborn, Matthew and Ricke, Nathan and Tsuchimochi, Takashi and Trofimov, Aleksandr and Orkhon, Temujin and Whelpley, Noah and Luo, Carina and Van Voorhis, Troy},
  title = {QuEmb: A Toolbox for Bootstrap Embedding Calculations of Molecular and Periodic Systems},
  journal = {J. Phys. Chem. A},
  year = {2025},
  month = {July},
  volume = {129},
  number = {28},
  pages = {6538--6551},
  doi = {10.1021/acs.jpca.5c02983},
  url = {https://doi.org/10.1021/acs.jpca.5c02983},
  publisher = {American Chemical Society}
}

@article{fujii_2022_deep_vqe,
  title = {Deep Variational Quantum Eigensolver: A Divide-And-Conquer Method for Solving a Larger Problem with Smaller Size Quantum Computers},
  author = {Fujii, Keisuke and Mizuta, Kaoru and Ueda, Hiroshi and Mitarai, Kosuke and Mizukami, Wataru and Nakagawa, Yuya O.},
  journal = {PRX Quantum},
  volume = {3},
  issue = {1},
  pages = {010346},
  numpages = {12},
  year = {2022},
  month = {Mar},
  publisher = {American Physical Society},
  doi = {10.1103/PRXQuantum.3.010346},
  url = {https://link.aps.org/doi/10.1103/PRXQuantum.3.010346}
}

@misc{qiskit_ibm_runtime_2025,
  author       = {{IBM Quantum}},
  title        = {Qiskit IBM Runtime v0.36.1},
  year         = {2025},
  howpublished = {\url{https://github.com/Qiskit/qiskit-ibm-runtime}},
  note         = {Accessed: 2026-05-01}
}

@inproceedings{paszke2019pytorch,
  author    = {Paszke, Adam and Gross, Sam and Massa, Francisco and Lerer, Adam and
               Bradbury, James and Chanan, Gregory and Killeen, Trevor and Lin, Zeming and
               Gimelshein, Natalia and Antiga, Luca and Desmaison, Alban and K{\"o}pf, Andreas and
               Yang, Edward and DeVito, Zach and Raison, Martin and Tejani, Alykhan and
               Chilamkurthy, Sasank and Steiner, Benoit and Fang, Lu and Bai, Junjie and
               Chintala, Soumith},
  title     = {{PyTorch}: An Imperative Style, High-Performance Deep Learning Library},
  booktitle = {Advances in Neural Information Processing Systems 32},
  pages     = {8024--8035},
  year      = {2019},
  url       = {https://arxiv.org/abs/1912.01703}
}

\newpage

\appendix

\section*{SUPPLEMENTARY MATERIAL}


\section{Execution Dates and Hardware Calibration Data}
\label{app:execution_dates_calibration}

\setcounter{table}{0}
\renewcommand{\thetable}{A\arabic{table}}
\setcounter{figure}{0}
\renewcommand{\thefigure}{A\arabic{figure}}
\setcounter{equation}{0}
\renewcommand{\theequation}{A\arabic{equation}}

The DMET-SQD experiment, run with a single spin-bin threshold $\varepsilon_{spb} = \sqrt{\mathbb{S}}$, was
executed over four $\mu$-iterations spanning August~26--29, 2026, on the Heron R3 processor
(\texttt{ibm\_boston}). The DMET-QSCI-RBM experiments were run on \texttt{ibm\_boston} as well, on separate hardware sampling dates:
an initial pair of $\mu$-iterations on July~14, 2026, followed by the third and fourth, converged
$\mu$-iterations on August~31 and September~1, 2026 respectively. The H$_2$O (cc-pVDZ, 48
qubits) simulation was likewise run on \texttt{ibm\_boston}, on September~3, 2026. The hardware
calibration snapshots recorded immediately prior to each of these eight runs are provided in
Table~\ref{tab:calib_ibm_boston_combined} of the Supplementary Material. The methoxyamine (CH$_5$NO)
simulation was carried out on January~23, 2026, on the Heron R1 processor (\texttt{ibm\_torino}). As this
backend was retired on 1~April~2026, the pre-execution calibration snapshot for this specific run could not
be recovered from the IBM Quantum API post-retirement, and is therefore omitted from the Supplementary
Material. The C$_2$H$_4$ simulation was carried out on January~14, 2026, on the Heron processor
(\texttt{ibm\_fez}), with calibration data provided in Table~\ref{tab:calib_ibm_fez_20260114_122902} of the
Supplementary Material. As with the Boston runs discussed above, these calibration snapshots represent the
pre-execution device state and should not be interpreted as a continuous record of hardware performance
across the full duration of each experiment.

\begin{table*}[htbp]
\centering
\caption{Calibration metrics for the Heron processor (IBM Boston) recorded prior to the DMET-SQD
($\varepsilon_{spb} = \sqrt{\mathbb{S}}$, four $\mu$-iterations), DMET-QSCI-RBM, and H$_2$O experiments
discussed in Sec.~3.2 and Sec.~3.4. All runs used independent hardware sampling. Each column corresponds to
the device's pre-execution calibration snapshot for that date, including coherence times ($T_1$, $T_2$),
readout assignment errors, gate error rates, and gate durations.}
\label{tab:calib_ibm_boston_combined}
\resizebox{\textwidth}{!}{%
\begin{tabular}{lccccccccc}
\hline
& \multicolumn{4}{c}{DMET-SQD ($\varepsilon_{spb} = \sqrt{\mathbb{S}}$)} & \multicolumn{3}{c}{DMET-QSCI-RBM} & H$_2$O Molecule \\
\cmidrule(lr){2-5} \cmidrule(lr){6-8} \cmidrule(lr){9-9}
Parameter & Aug 26 & Aug 27 & Aug 28 & Aug 29 & Jul 14 & Aug 31 & Sep 1 & Sep 3 \\
\hline
$T_1$ ($\mu$s) & 276.982251 & 261.464460 & 276.275511 & 276.275511 & 266.562869 & 273.919810 & 249.275402 & 249.275402 \\
$T_2$ ($\mu$s) & 312.488372 & 309.832827 & 304.000000 & 304.000000 & 304.147150 & 312.213679 & 300.611997 & 300.611997 \\
Readout assg. error & 0.006348 & 0.005798 & 0.006226 & 0.006531 & 0.004150 & 0.006226 & 0.003418 & 0.003418 \\
$P(0\,|\,1)$ & 0.005859 & 0.005981 & 0.005371 & 0.005737 & 0.005859 & 0.005859 & 0.005371 & 0.005371 \\
$P(1\,|\,0)$ & 0.005615 & 0.004883 & 0.005371 & 0.005859 & 0.001953 & 0.006226 & 0.000977 & 0.000977 \\
Readout length (ns) & 2180.0 & 2180.0 & 2180.0 & 2180.0 & 2180.0 & 2180.0 & 2180.0 & 2180.0 \\
Identity gate error & 0.000158 & 0.000186 & 0.000171 & 0.000171 & 0.000152 & 0.000192 & 0.000198 & 0.000198 \\
$R_z$ gate error & 0.000000 & 0.000000 & 0.000000 & 0.000000 & 0.000000 & 0.000000 & 0.000000 & 0.000000 \\
$\sqrt{X}$ (Sx) gate error & 0.000158 & 0.000186 & 0.000171 & 0.000171 & 0.000152 & 0.000192 & 0.000198 & 0.000198 \\
Pauli-X gate error & 0.000158 & 0.000186 & 0.000171 & 0.000171 & 0.000152 & 0.000192 & 0.000198 & 0.000198 \\
CZ gate error & 0.001432 & 0.001528 & 0.001357 & 0.001357 & 0.001638 & 0.001482 & 0.001482 & 0.001510 \\
Gate time (CZ) (ns) & 68.0 & 68.0 & 68.0 & 68.0 & 68.0 & 68.0 & 68.0 & 68.0 \\
\hline
\end{tabular}%
}
\end{table*}

\begin{table}[h]
\centering
\caption{Calibration metrics for the Heron processor (IBM Fez) recorded on 14 January 2026. These values correspond to the device's pre-execution calibration snapshot, including coherence times ($T_1$, $T_2$), readout assignment errors, gate error rates, and gate durations.}
\label{tab:calib_ibm_fez_20260114_122902}
\begin{tabular}{lc}
\hline
Parameter & Value \\
\hline
$T_1$ ($\mu$s) & 145.142198  \\
$T_2$ ($\mu$s) & 104.489980 \\
Readout assignment error & 0.009460 \\
$P(0\,|\,1)$ & 0.012939 \\
$P(1\,|\,0)$ & 0.006592 \\
Readout length (ns) & 1560.0  \\
Identity gate error & 0.000283 \\
$R_z$ gate error & 0.000000 \\
$\sqrt{X}$ (Sx) gate error & 0.000283 \\
Pauli-X gate error & 0.000283 \\
CZ gate error & 0.002510 \\
Gate time (CZ) (ns) & 68.0  \\
\hline
\end{tabular}
\end{table}

\newpage

\section{Orbital Decomposition of Carmofur-M$^{pro}$ PL complex}\label{app: orb_decomp_pl}
\setcounter{table}{0}
\renewcommand{\thetable}{B\arabic{table}}
\setcounter{figure}{0}
\renewcommand{\thefigure}{B\arabic{figure}}
\setcounter{equation}{0}
\renewcommand{\theequation}{B\arabic{equation}}
\begin{table}[h]
\centering
\caption{Per-impurity orbital decomposition for fragments of the Carmofur–M\textsuperscript{pro} complex (cc-pVDZ basis, DMET-HF). For each fragment $y$, $A_y$ is the number of fragment (impurity) orbitals, $B_y$ is the number of bath orbitals (entangled environment orbitals with natural-orbital fractionality $\delta > \varepsilon_{\mathrm{occ}} = 10^{-13}$), $\mathrm{Cor}_y$ is the number of unentangled, doubly occupied environment orbitals, and $\mathrm{Vir}_y$ is the number of unentangled, unoccupied environment orbitals, such that $A_y + B_y + \mathrm{Cor}_y + \mathrm{Vir}_y$ equals the total number of spatial orbitals (494) for every fragment. For the current experiment, an active space containing 8 $A_y$ (HOMO) and 8 $B_y$ (LUMO) was selected, for each fragment.}
\label{tab:orbital_decomposition}
\begin{tabular}{lcccc}
\toprule
Fragment & $A_y$ & $B_y$ & $\mathrm{Cor}_y$ & $\mathrm{Vir}_y$ \\
\midrule
F1  & 19 & 19 & 82 & 374 \\
F2  & 43 & 43 & 58 & 350 \\
F3  & 61 & 61 & 40 & 332 \\
F4  & 47 & 47 & 54 & 346 \\
F5  & 66 & 66 & 35 & 327 \\
F6  & 52 & 52 & 49 & 341 \\
F7  & 57 & 57 & 44 & 336 \\
F8  & 48 & 48 & 53 & 345 \\
F9  & 48 & 48 & 53 & 345 \\
F10 & 53 & 53 & 51 & 340 \\
\bottomrule
\end{tabular}
\end{table}

\section{Derivation of the Worst-Case Samples-per-Batch Bound(s)}
\label{app:spb_derivation}

\setcounter{table}{0}
\renewcommand{\thetable}{C\arabic{table}}
\setcounter{figure}{0}
\renewcommand{\thefigure}{C\arabic{figure}}
\setcounter{equation}{0}
\renewcommand{\theequation}{C\arabic{equation}}

\paragraph{Setup.}
Consider a closed-shell fragment whose active space comprises $M$ spatial
orbitals and $N_e$ electrons per spin sector ($N_\alpha = N_\beta = N_e$).
The particle-number- and spin-$z$-conserving symmetry space onto which any
valid determinant must fall is
\begin{equation}
\mathbb{S} = \mathcal{A} \otimes \mathcal{B},
\qquad
|\mathbb{S}|
= \binom{M}{N_\alpha}\binom{M}{N_\beta}
= \binom{M}{N_e}^{2}
\equiv D^{2},
\label{eq:app_symspace}
\end{equation}
where $\mathcal{A}$ ($\mathcal{B}$) denotes the space of $\alpha$-
($\beta$-) spin strings of Hamming weight $N_e$, each of dimension
$D = \binom{M}{N_e}$. The closed-shell identity
\begin{equation}
D = \sqrt{|\mathbb{S}|}
\label{eq:app_sqrt_identity}
\end{equation}
is the structural fact from which the bound follows: the single-spin string
space has dimension exactly $\sqrt{|\mathbb{S}|}$.

\paragraph{Configuration retention.}
After configuration recovery, the post-selection and subsampling step
retains at most $N_s$ configurations per batch, where $N_s$ is the
samples-per-batch parameter $\varepsilon_{spb}$. Each retained configuration
is a spin-resolved pair $|\alpha_k\rangle \otimes |\beta_k\rangle$,
$k = 1, \dots, N_s$, from which the unique spin-string sets are extracted:
\begin{equation}
A_{\mathrm{uniq}} = \{ |\alpha_k\rangle \}_{k=1}^{N_s},
\qquad
B_{\mathrm{uniq}} = \{ |\beta_k\rangle \}_{k=1}^{N_s},
\qquad
|A_{\mathrm{uniq}}| \leq N_s,
\quad
|B_{\mathrm{uniq}}| \leq N_s .
\label{eq:app_uniq_sets}
\end{equation}

\paragraph{Proliferation.}
Following Eq.~(16), the diagonalization subspace is not spanned by the
sampled pairs themselves but by the tensor square of the pooled set
\begin{equation}
P_{\mathrm{uniq}} = A_{\mathrm{uniq}} \cup B_{\mathrm{uniq}},
\qquad
S_{\mathrm{proj}} = P_{\mathrm{uniq}} \otimes P_{\mathrm{uniq}},
\qquad
|S_{\mathrm{proj}}| = |P_{\mathrm{uniq}}|^{2}.
\label{eq:app_proliferation}
\end{equation}
By inclusion-exclusion,
\begin{equation}
|P_{\mathrm{uniq}}|
= |A_{\mathrm{uniq}}| + |B_{\mathrm{uniq}}|
- |A_{\mathrm{uniq}} \cap B_{\mathrm{uniq}}| .
\label{eq:app_inclusion_exclusion}
\end{equation}
For a closed-shell system one expects substantial overlap
$|A_{\mathrm{uniq}} \cap B_{\mathrm{uniq}}| > 0$, since a bitstring sampled
in the $\alpha$ sector is highly likely to also appear in the $\beta$
sector. The worst case for subspace inflation, however, is the fully
disjoint limit $A_{\mathrm{uniq}} \cap B_{\mathrm{uniq}} = \varnothing$
with no repeated strings within either sector, for which
Eqs.~\eqref{eq:app_uniq_sets} and \eqref{eq:app_inclusion_exclusion} give
\begin{equation}
|P_{\mathrm{uniq}}|^{\mathrm{wc}} = 2 N_s .
\label{eq:app_worst_case_pool}
\end{equation}
Independently of sampling, the pool can never exceed the single-spin string
space itself, so combining Eq.~\eqref{eq:app_worst_case_pool} with
Eq.~\eqref{eq:app_sqrt_identity},
\begin{equation}
|P_{\mathrm{uniq}}| \;\leq\; \min\!\left( 2 N_s,\; \sqrt{|\mathbb{S}|} \right),
\qquad
|S_{\mathrm{proj}}| \;\leq\;
\min\!\left( 2 N_s,\; \sqrt{|\mathbb{S}|} \right)^{2}.
\label{eq:app_hard_cap}
\end{equation}

\paragraph{Non-inflation condition.}
Requiring that the proliferated subspace never exceed the full symmetry
space even in the worst-case disjoint limit,
\begin{equation}
|S_{\mathrm{proj}}|^{\mathrm{wc}} = \left( 2 N_s \right)^{2}
= 4 N_s^{2} \;\leq\; |\mathbb{S}|,
\label{eq:app_noninflation}
\end{equation}
immediately yields the cap
\begin{equation}
\boxed{\;
N_s \;\leq\; \frac{\sqrt{|\mathbb{S}|}}{2}
\;=\; \varepsilon_{spb} \; } .
\label{eq:app_bound}
\end{equation}

\paragraph{Tightness and non-redundancy.}
The value $\varepsilon_{spb} = \sqrt{|\mathbb{S}|}/2$ is not merely a
sufficient cap but the unique threshold at which two regimes meet. For
$N_s \leq \sqrt{|\mathbb{S}|}/2$ the worst-case proliferated dimension
$4 N_s^{2} \leq |\mathbb{S}|$ is controlled by the retention budget, and
every retained configuration can, in the worst case, enlarge
$S_{\mathrm{proj}}$. Conversely, for $N_s > \sqrt{|\mathbb{S}|}/2$ the
minimum in Eq.~\eqref{eq:app_hard_cap} is attained by
$\sqrt{|\mathbb{S}|}$: the worst-case pool already saturates the full
single-spin string space, and any configuration retained beyond
$\sqrt{|\mathbb{S}|}/2$ is provably incapable of enlarging the worst-case
diagonalization subspace, constituting redundant over-allocation of the
diagonalization budget. At exactly $N_s = \sqrt{|\mathbb{S}|}/2$ the
worst-case pool satisfies $2 N_s = \sqrt{|\mathbb{S}|} = D$ and the
proliferated subspace saturates the symmetry space,
$|S_{\mathrm{proj}}|^{\mathrm{wc}} = |\mathbb{S}|$, with equality and no
excess. Hence $\varepsilon_{spb} = \sqrt{|\mathbb{S}|}/2$ is simultaneously
the largest retention guaranteeing
$|S_{\mathrm{proj}}| \leq |\mathbb{S}|$ under the disjoint worst case and
the smallest retention at which the worst-case pool exhausts the
single-spin space. We emphasize that this is a combinatorial engineering
bound on subspace allocation, independent of any energetic convergence
criterion; in practice, closed-shell $\alpha$-$\beta$ overlap in
Eq.~\eqref{eq:app_inclusion_exclusion} renders the bound conservative, so
that realized subspace dimensions typically satisfy
$|S_{\mathrm{proj}}| < |\mathbb{S}|$ strictly.

\paragraph{Opposite extremal limit and the relaxed bound
$\varepsilon_{spb} = \sqrt{|\mathbb{S}|}$.}
The bound $\sqrt{|\mathbb{S}|}/2$ derived above corresponds to the fully
disjoint extreme of Eq.~\eqref{eq:app_inclusion_exclusion}, in which the
sampled $\alpha$- and $\beta$-string sets share no elements. The opposite
extreme is the fully coincident limit, in which every recovered $\alpha$
string also appears as a $\beta$ string and vice versa, so that
\begin{equation}
A_{\mathrm{uniq}} = B_{\mathrm{uniq}},
\qquad
|A_{\mathrm{uniq}} \cap B_{\mathrm{uniq}}| = |A_{\mathrm{uniq}}| = N_s .
\label{eq:app_coincident}
\end{equation}
In this case the inclusion-exclusion relation
Eq.~\eqref{eq:app_inclusion_exclusion} collapses to
\begin{equation}
|P_{\mathrm{uniq}}|^{\mathrm{coin}}
= N_s + N_s - N_s = N_s ,
\label{eq:app_coincident_pool}
\end{equation}
and the non-inflation condition
$|S_{\mathrm{proj}}| = |P_{\mathrm{uniq}}|^{2} \leq |\mathbb{S}|$ becomes
\begin{equation}
N_s^{2} \;\leq\; |\mathbb{S}|
\qquad\Longrightarrow\qquad
N_s \;\leq\; \sqrt{|\mathbb{S}|} .
\label{eq:app_relaxed_bound}
\end{equation}
The pooled set thus grows only half as fast with $N_s$ in the coincident
limit as in the disjoint limit, and the retention budget guaranteeing
$|S_{\mathrm{proj}}| \leq |\mathbb{S}|$ is correspondingly relaxed by a
factor of two, from $\sqrt{|\mathbb{S}|}/2$ to $\sqrt{|\mathbb{S}|}$. For
closed-shell systems the spin-exchange near-degeneracy of the ground-state
wavefunction (if a configuration $|\alpha\rangle \otimes |\beta\rangle$
contributes significantly, so does
$|\beta\rangle \otimes |\alpha\rangle$) drives the recovered string sets
toward this coincident limit rather than the disjoint one, so that the
effective pool growth is governed by
Eq.~\eqref{eq:app_coincident_pool} in practice. Accordingly, for the
DMET-SQD experiments in this work we adopt the relaxed choice
\begin{equation}
\boxed{\;
\varepsilon_{spb} \;=\; \sqrt{|\mathbb{S}|} \; } ,
\label{eq:app_relaxed_choice}
\end{equation}
which corresponds to the coincident-limit bound
Eq.~\eqref{eq:app_relaxed_bound} and yields a larger, more thoroughly
sampled diagonalization subspace than the conservative disjoint-limit cap,
while remaining bounded by the symmetry space $|\mathbb{S}|$ under the
closed-shell coincidence that the recovered $\alpha$ and $\beta$ string
sets substantially coincide.

\section{Subspace accessed for the DMET-SQD and DMET-QSCI-RBM experiments}
\label{app: ss_accessed}

\setcounter{table}{0}
\renewcommand{\thetable}{D\arabic{table}}
\setcounter{figure}{0}
\renewcommand{\thefigure}{D\arabic{figure}}
\setcounter{equation}{0}
\renewcommand{\theequation}{D\arabic{equation}}

\begin{figure}
    \centering
    \includegraphics[width=\linewidth]{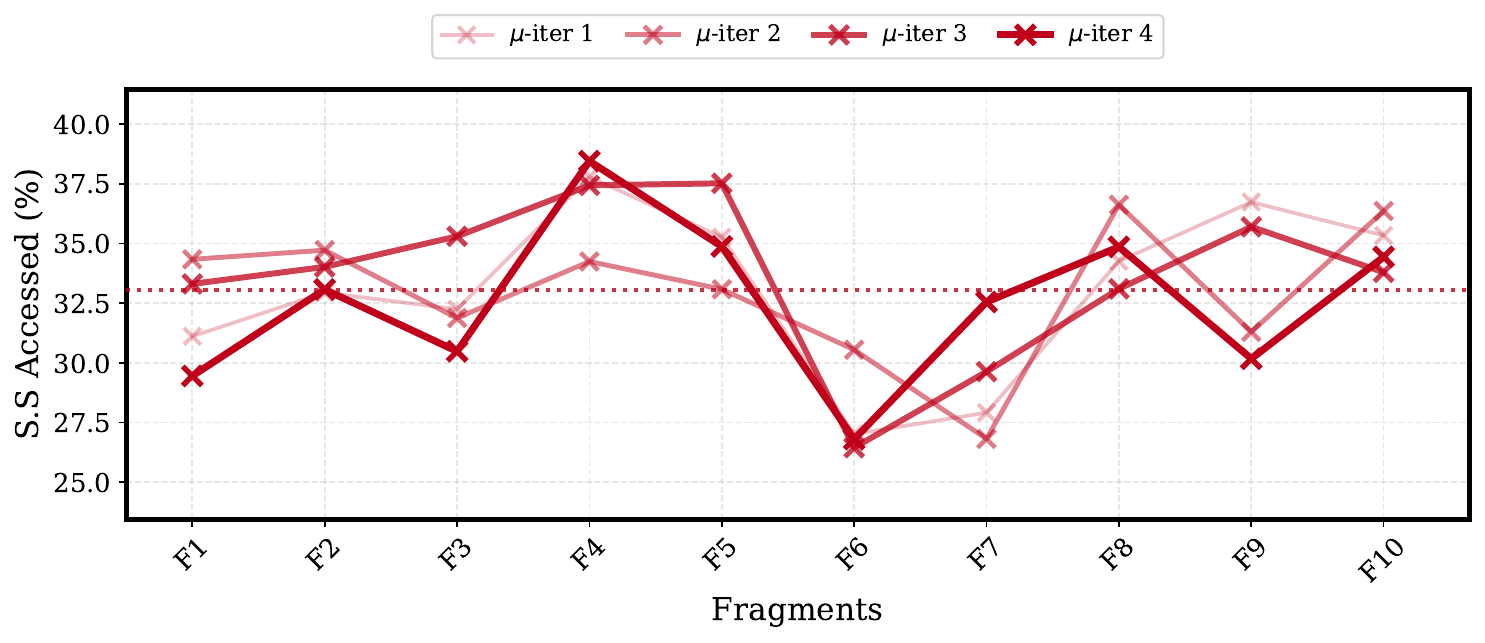}
    \caption{
    \textbf{Fragment-resolved symmetry space ($\mathbb{S}$) accessed by DMET-SQD
    ($\varepsilon_{spb}=\sqrt{|\mathbb{S}|}$, converged) across all four $\mu$-iterations for the
    Carmofur/SARS-CoV-2 M\textsuperscript{pro} complex (10 fragments, F1--F10).} Each curve
    corresponds to one $\mu$-iteration, with opacity and linewidth increasing progressively toward
    the final converged iteration. The dotted line marks the overall mean $\mathbb{S}$ accessed
    across all fragments and $\mu$-iterations ($33.04\%$).
    }
    \label{fig:DMET_SQD_SS_all_iters}
\end{figure}

\begin{figure}
    \centering
    \includegraphics[width=\linewidth]{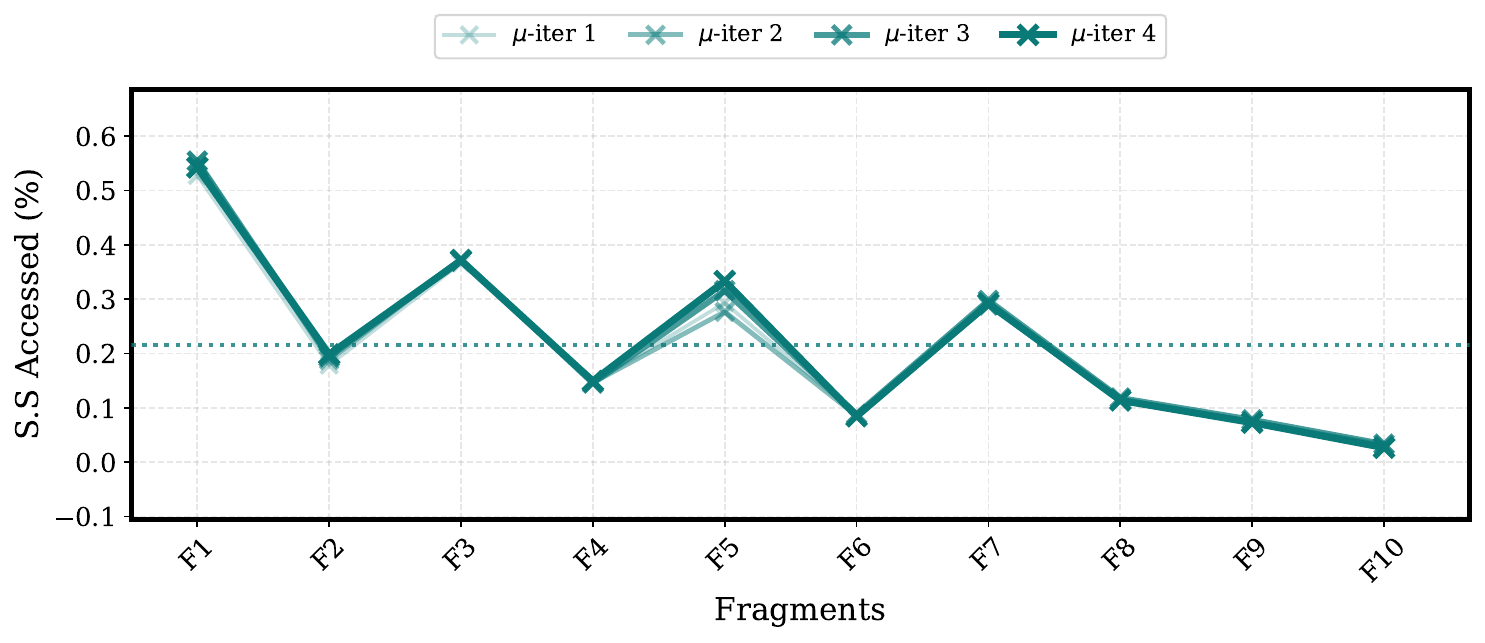}
    \caption{
    \textbf{Fragment-resolved symmetry space ($\mathbb{S}$) accessed by DMET-QSCI-RBM across all
    four $\mu$-iterations for the Carmofur/SARS-CoV-2 M\textsuperscript{pro} complex (10 fragments,
    F1--F10).} Each curve corresponds to one $\mu$-iteration, with opacity and linewidth increasing
    progressively toward the final converged iteration. The dotted line marks the overall mean
    $\mathbb{S}$ accessed across all fragments and $\mu$-iterations ($0.22\%$).
    }
    \label{fig:DMET_QSCI_RBM_SS_all_iters}
\end{figure}

\newpage

\section{Quantum Resource Utilization for DMET-SQD and DMET-QSCI-RBM experiments}

\setcounter{table}{0}
\renewcommand{\thetable}{F\arabic{table}}
\setcounter{figure}{0}
\renewcommand{\thefigure}{F\arabic{figure}}
\setcounter{equation}{0}
\renewcommand{\theequation}{F\arabic{equation}}

\begin{figure}[H]
    \centering
    \includegraphics[width=\linewidth]{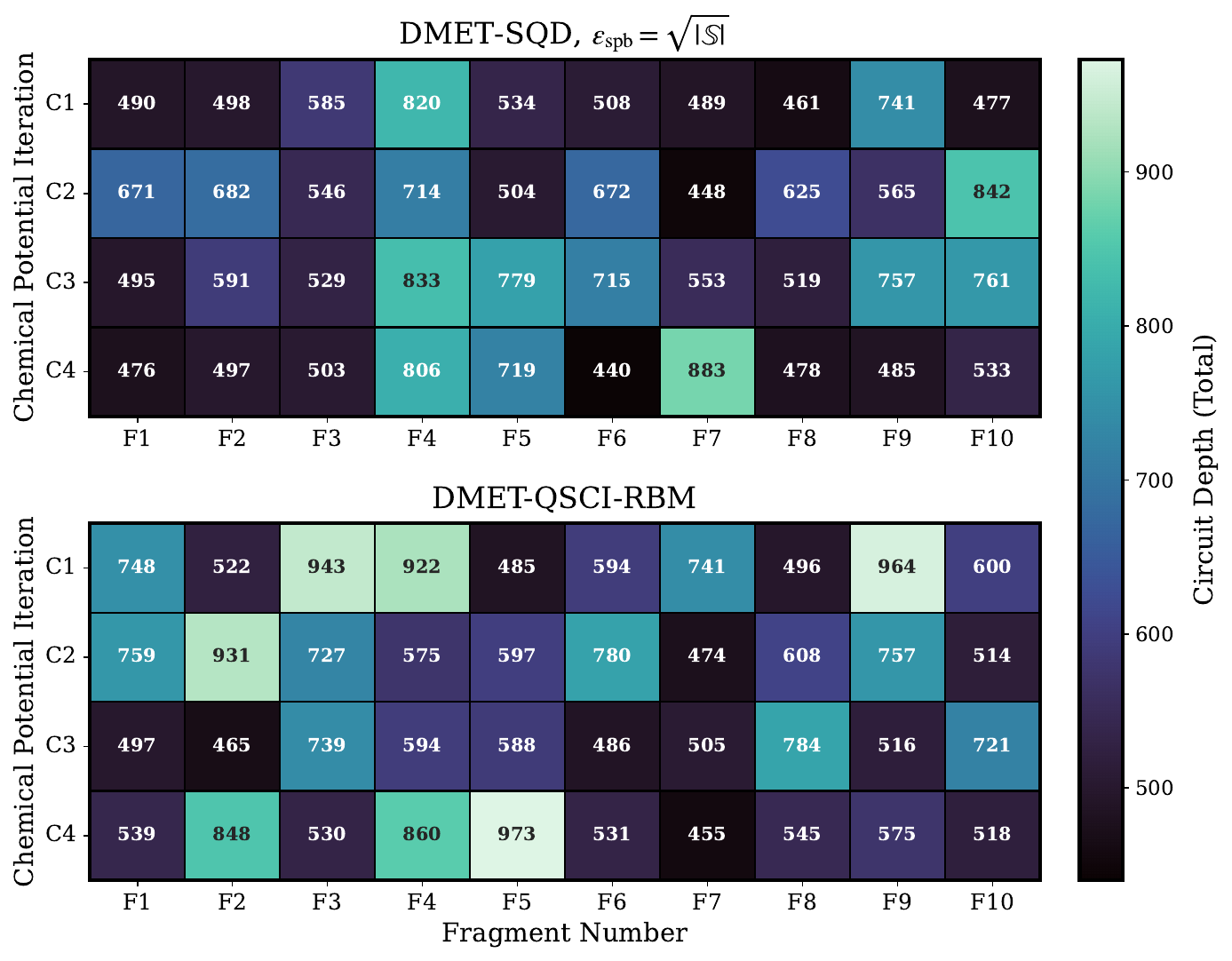}
\caption{
    \textbf{Transpiled circuit depth per fragment for DMET-SQD and DMET-QSCI-RBM on the
    Carmofur/SARS-CoV-2 M\textsuperscript{pro} complex (\texttt{ibm\_boston}).} \textit{Top}:
    DMET-SQD ($\varepsilon_{spb}=\sqrt{|\mathbb{S}|}$) across its four $\mu$-iterations (C1--C4).
    \textit{Bottom}: DMET-QSCI-RBM across its four $\mu$-iterations (C1--C4). Circuit depths fall
    predominantly in the 450--900 range for both solvers, with occasional fragment-specific outliers
    (e.g., DMET-SQD F7 at C4, depth 883; DMET-QSCI-RBM F5 at C4, depth 973) reflecting
    fragment-specific LUCJ layer requirements rather than any systematic trend across solver
    configuration or $\mu$-iteration.
}
\label{fig:circuit_depth_heatmap}
\end{figure}

\begin{figure}[H]
    \centering
    \includegraphics[width=\linewidth]{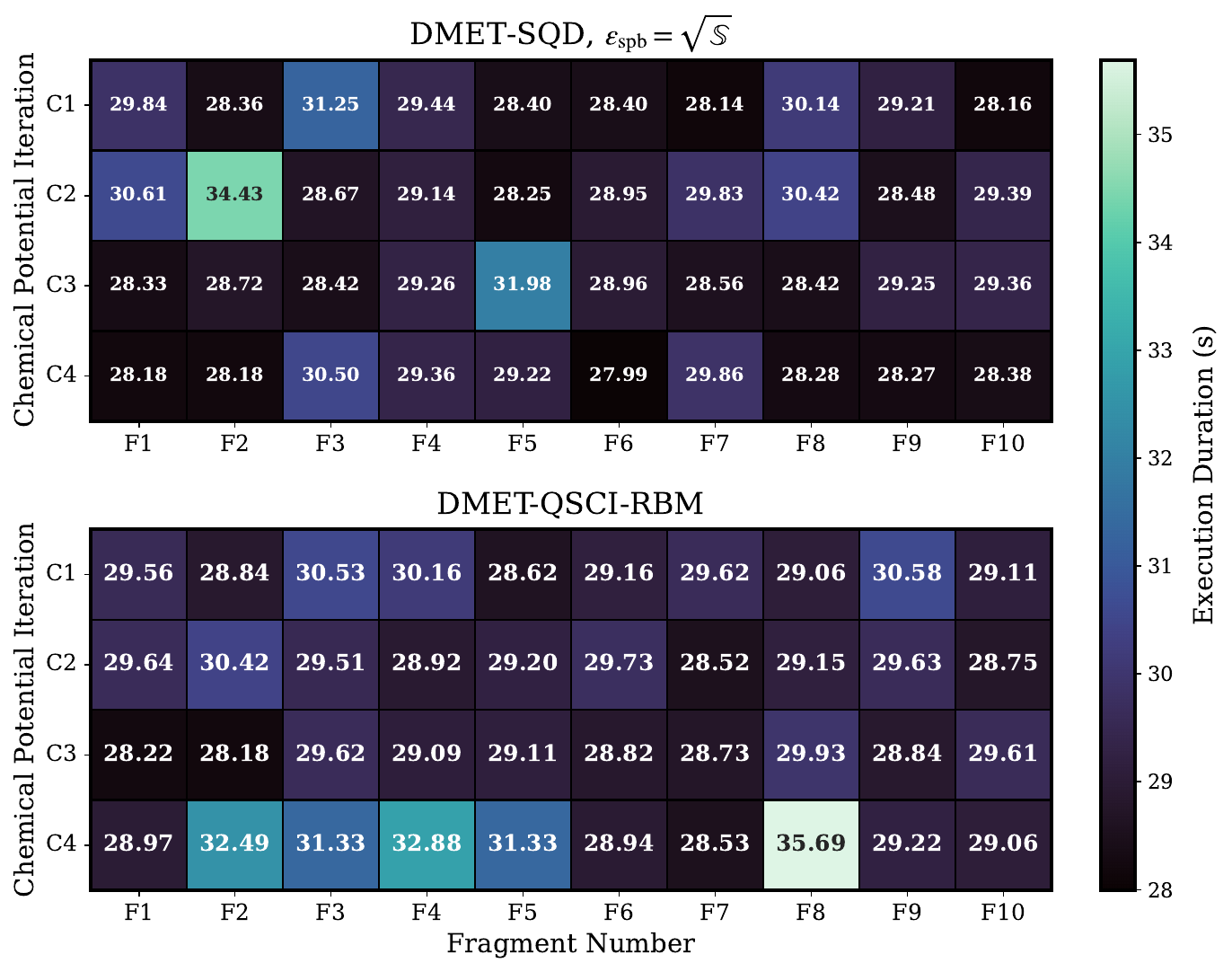}
\caption{
    \textbf{QPU execution duration per fragment for DMET-SQD and DMET-QSCI-RBM on the
    Carmofur/SARS-CoV-2 M\textsuperscript{pro} complex (\texttt{ibm\_boston}).} \textit{Top}:
    DMET-SQD ($\varepsilon_{spb}=\sqrt{\mathbb{S}}$) across its four $\mu$-iterations (C1--C4).
    \textit{Bottom}: DMET-QSCI-RBM across its four $\mu$-iterations (C1--C4). Execution durations
    are consistent across most fragment/iteration combinations ($\sim$28-30\,s), with isolated
    outliers (e.g., DMET-SQD F2 at C2, 34.43\,s; DMET-QSCI-RBM F8 at C4, 35.69\,s) attributable to
    queue-time variability on shared hardware rather than any systematic dependence on solver
    configuration or $\mu$-iteration.
}
\label{fig:exec_duration_heatmap}
\end{figure}


\section{Theoretical Framework}

\setcounter{table}{0}
\renewcommand{\thetable}{G\arabic{table}}
\setcounter{figure}{0}
\renewcommand{\thefigure}{G\arabic{figure}}
\setcounter{equation}{0}
\renewcommand{\theequation}{G\arabic{equation}}

\subsection{Electronic Structure Hamiltonian and Selected Configuration Interaction}

The electronic Hamiltonian in second quantization~\cite{Jordan1928, Fock1930, KSV2025} is written as
\begin{equation}
\hat{H}
=
\sum_{pq} h_{pq} a_p^\dagger a_q + \frac{1}{2} \sum_{pqrs} h_{pqrs} a_p^\dagger a_q^\dagger a_r a_s,
\end{equation}
where $h_{pq}$ and $h_{pqrs}$ are the one- and two-electron integrals in the
molecular orbital basis, and $a_p^\dagger, a_p$ are fermionic creation and
annihilation operators~\cite{Jordan1928, szabo1996_modernqc}. The exact
ground-state wavefunction expands over Slater determinants as
$\ket{\Psi} = \sum_i c_i \ket{D_i}$~\cite{szabo1996_modernqc, Helgaker2000};
Full Configuration Interaction (FCI) diagonalizes $\hat{H}$ exactly in this
basis but scales exponentially with system size, while Selected CI (SCI)
methods instead construct a compact determinant subspace capturing the
dominant correlation~\cite{sci_evangelista_1}.

Quantum Selected Configuration Interaction (QSCI)~\cite{kanno2023qsci,
sqd_first_paper} constructs this subspace by sampling bitstrings from a
quantum device preparing an approximate trial state $\ket{\tilde\Psi}$,
\begin{equation}
\mathbb{S} = \mathrm{span}\{\ket{D_i}\},
\end{equation}
projecting the Hamiltonian onto $\mathbb{S}$, and diagonalizing classically
via the Davidson algorithm~\cite{Davidson1975_iterative_eigensolver} to obtain
an upper bound on the ground-state energy by the
MacDonald-Rayleigh-Ritz theorem~\cite{MacDonald1933_RayleighRitz}. Because
the quantum device is used purely for sampling rather than variational
optimization, QSCI's accuracy is governed entirely by the quality of the
sampled subspace $\mathbb{S}$, and critical limitations of this dependence
- sensitivity to subspace dimension and shot budget - have recently been
characterised in detail~\cite{critical_lims_sqd_2025}.

\subsection{Configuration Recovery}
Hardware noise causes a fraction of sampled bitstrings to violate the
particle-number and spin-projection symmetries expected of physically valid
configurations. Configuration recovery techniques~\cite{sqd_first_paper}
restore physical validity by projecting or filtering sampled configurations
into the correct symmetry sector, but symmetry filtering alone does not
guarantee that the retained configurations are the energetically dominant
ones. Recovered subspaces may be unnecessarily inflated, and hardware noise
can systematically bias sampling away from high-excitation-order
configurations that require deeper, more entangling circuits to
prepare~\cite{critical_lims_sqd_2025}.

In practice, the number of configurations admitted from the recovered set
$\mathbb{S}_{CR}$ into the subsequent proliferation and diagonalization steps
is controlled by a user-defined sampling threshold $\varepsilon_{spb}$. Given
the recovered configurations ranked in descending order of their observed
shot probability (or shot count) $p_i$,
\begin{equation}
p_1 \geq p_2 \geq \cdots \geq p_{|\mathbb{S}_{CR}|},
\end{equation}
only the top-$\varepsilon_{spb}$ configurations are retained for the
proliferation step,
\begin{equation}
\mathbb{S}_{CR}^{\,\varepsilon_{spb}}
=
\left\{
\ket{\phi_i} \in \mathbb{S}_{CR} \;\middle|\; i \leq \varepsilon_{spb}
\right\},
\qquad
\varepsilon_{spb} \leq |\mathbb{S}_{CR}|.
\end{equation}
Because $\varepsilon_{spb}$ is fixed independently of the actual size of
$\mathbb{S}_{CR}$, it functions as a hard, tunable cap on the number of
dominant configurations carried forward. A small $\varepsilon_{spb}$
aggressively restricts recovery to only the highest-probability
configurations, while a large $\varepsilon_{spb}$ (in the limit
$\varepsilon_{spb} \to |\mathbb{S}_{CR}|$) imposes no effective truncation
and admits the full recovered set.

The retained configurations are not diagonalized directly. Each
configuration is a product $\ket{\phi_i} = \ket{\alpha_i}\otimes\ket{\beta_i}$
of an $\alpha$-spin and a $\beta$-spin occupation string, and following the
SQD construction~\cite{sqd_first_paper} the unique $\alpha$ strings
$\mathcal{A}_{\mathrm{uniq}}$ and unique $\beta$ strings
$\mathcal{B}_{\mathrm{uniq}}$ of $\mathbb{S}_{CR}^{\,\varepsilon_{spb}}$ are
pooled, and the diagonalization subspace is formed as the full Cartesian
product of the pool with itself (spin-string proliferation),
\begin{equation}
\mathcal{P}_{\mathrm{uniq}} = \mathcal{A}_{\mathrm{uniq}} \cup \mathcal{B}_{\mathrm{uniq}},
\qquad
\mathcal{S}_{\mathrm{proj}} = \mathcal{P}_{\mathrm{uniq}} \otimes \mathcal{P}_{\mathrm{uniq}},
\qquad
\dim (\mathcal{S}_{\mathrm{proj}}) = |\mathcal{P}_{\mathrm{uniq}}|^{2}.
\label{eq:proj}
\end{equation}
Proliferation surfaces cross-paired $\alpha$ and $\beta$ combinations that
were never sampled, but it also makes the diagonalized dimension
$|\mathcal{P}_{\mathrm{uniq}}|^{2}$ generally much larger than the number of
retained configurations. Subspace sizes in this work are therefore reported
post-proliferation, as the fraction
$100\,|\mathcal{P}_{\mathrm{uniq}}|^{2}/\binom{n_{\mathrm{orb}}}{N_\alpha}^{2}$
of the symmetry space.

Neither the truncation nor the proliferation guarantees that the
retained configurations are the energetically dominant ones. This motivates
augmenting or replacing configuration-recovery heuristics of this kind with a
generative model that learns and concentrates on the dominant configuration
distribution directly from hardware
samples~\cite{Herzog_CIGen_2023, Coe_MLCI_2018,
patra2025physicsinformedgenerativemachine}, as detailed in
Section~\ref{sec:rbm_theory} below.

\subsection{Density Matrix Embedding Theory}\label{app: dmet_theory}

Density Matrix Embedding Theory (DMET)~\cite{Knizia2012_DMET,
wouters2016dmets_guide} partitions the full electronic Hilbert space into an
impurity and an environment, $\mathbb{H} = \mathbb{H}_{\mathrm{imp}} \oplus
\mathbb{H}_{\mathrm{env}}$, where the impurity captures strong local
correlation and the environment is represented by a finite set of bath
orbitals obtained from the Schmidt decomposition of the mean-field reference
state $|\Phi_0\rangle$~\cite{Knizia2012_DMET}. Given a user-defined set of
fragment orbitals $A_y$, the remaining environment orbitals $\mathrm{Env}_y$
decompose as
\begin{equation}
\mathrm{Env}_y = B_y \cup \mathrm{Cor}_y \cup \mathrm{Vir}_y,
\end{equation}
where $B_y$ are the entangled bath orbitals reproducing fragment-environment
entanglement at the mean-field level, and $\mathrm{Cor}_y$, $\mathrm{Vir}_y$
are unentangled occupied and virtual orbitals, respectively, whose
contributions enter only as a mean-field correction and are excluded from the
embedded problem. The embedded impurity Hamiltonian is constructed by
projecting the full electronic Hamiltonian onto the fragment$+$bath subspace
$\hat{P}$,
\begin{equation}
\hat{H}_{\mathrm{emb}}
=
\hat{P}
\left(
\sum_{pq} h_{pq} \, a_p^\dagger a_q
+
\frac{1}{2}
\sum_{pqrs} v_{pqrs} \, a_p^\dagger a_q^\dagger a_s a_r
\right)
\hat{P}.
\end{equation}
In standard DMET, $\hat{H}_{\mathrm{emb}}$ is solved via FCI or high-level
classical solvers such as CCSD~\cite{bartlett_cc_theory_2007}; in this work it
is instead solved with QSCI-RBM, enabling larger embedded active spaces while
preserving tractability through compact subspace
selection~\cite{dmet_sqd, kanno2023qsci}.

We adopt the \emph{one-shot} DMET embedding
strategy~\cite{Knizia2012_DMET, wouters2016dmets_guide}, in which bath
orbitals are constructed once from the initial mean-field reference and are
not iteratively updated. Self-consistency is instead enforced through a
single scalar chemical potential $\mu$ introduced into the embedded
Hamiltonian, $\hat{H}_{\mathrm{emb}}(\mu_{chem}) = \hat{H}_{\mathrm{emb}} - \mu_{chem}
\sum_{p \in \mathrm{frag}} a_p^\dagger a_p$, adjusted via a secant-method
root-find~\cite{papakonstantinou2013secant} so that the trace of the embedded
1-RDM over the fragment orbitals,
\begin{equation}
\sum_{p \in \mathrm{frag}} \gamma_{pp}^{\mathrm{emb}}(\mu_{chem})
=
\sum_{p \in \mathrm{frag}} \langle \Psi_{\mathrm{emb}}(\mu_{chem}) | a_p^\dagger a_p | \Psi_{\mathrm{emb}}(\mu_{chem}) \rangle,
\end{equation}
matches the target active-electron count for that fragment, where
$|\Psi_{\mathrm{emb}}(\mu_{chem})\rangle$ is the correlated impurity ground state
obtained from QSCI-RBM diagonalization of $\hat{H}_{\mathrm{emb}}(\mu_{chem})$. Summing
this fragment-resolved trace over all fragments $y$ gives the total embedded
electron count $N_{\mathrm{tot}}(\mu_{chem})$, and self-consistency across the full
DMET loop is reached once this quantity matches the true total electron count
$N_{\mathrm{true}}$ of the system to within a tolerance $\varepsilon_{\mathrm{con}}$,
\begin{equation}
N_{\mathrm{tot}}(\mu_{chem}) - N_{\mathrm{true}}
=
\sum_{y} \sum_{p \in \mathrm{frag}_y} \gamma_{pp}^{\mathrm{emb},\,y}(\mu_{chem})
\; - \;
N_{\mathrm{true}}
\; < \; \varepsilon_{\mathrm{con}}.
\end{equation}
Recent theoretical analyses have further characterised the convergence behaviour of this self-consistency
loop in the non-interacting limit~\cite{Cances2025, Negre2025}. In practice, the self-consistent chemical
potential $\mu_{chem}^*$ satisfying this condition can be arrived at using methods such as the Newton-Secant
method, which iteratively proposes successive $\mu_{chem}$ values by treating the residual
$N_{\mathrm{tot}}(\mu_{chem}) - N_{\mathrm{true}}$ as the quantity to be zeroed, converging toward
$\mu_{chem}^*$.

\subsection{Restricted Boltzmann Machines for Configuration Generation}
\label{sec:rbm_theory}

Restricted Boltzmann Machines (RBMs) are energy-based generative models
comprising a layer of visible units $\vec{v} \in \{0,1\}^{n_v}$ and a layer of
hidden units $\vec{h} \in \{0,1\}^{n_h}$, with no intra-layer
connections~\cite{carleo_2017, Nomura_RBM_Learning_2017}. The joint energy
function is,
\begin{equation}
E(\vec{v}, \vec{h})
=
-\vec{a} \cdot \vec{v}
-\vec{b} \cdot \vec{h}
-\vec{v}^{\,T} W \vec{h},
\end{equation}
where $\vec{a}$ and $\vec{b}$ are visible and hidden bias vectors respectively,
and $W \in \mathbb{R}^{n_v \times n_h}$ is the weight matrix. The marginal
distribution over visible units,
\begin{equation}
P(\vec{v})
=
\frac{1}{Z}
\sum_{\vec{h}}
e^{-E(\vec{v},\vec{h})},
\end{equation}
is trained to approximate the ground-state amplitude distribution of the quantum
state~\cite{carleo_2017},
\begin{equation}
P(\vec{v})
\approx
|c_{\vec{v}}|^2.
\end{equation}
The RBM is trained on the symmetry-valid hardware samples via contrastive
divergence~\cite{Nomura_RBM_Learning_2017}, and once trained, new configurations
are generated by Gibbs sampling from the learned distribution. Because the RBM
concentrates probability mass on the dominant correlation subspace identified by
the hardware, it enables compact subspace construction that goes beyond the raw
symmetry-filtered shot list. In particular, this leads to the capturing of high-excitation-order
determinants that carry significant weight in multi-reference systems but may
appear infrequently in finite hardware shot budgets~\cite{patra2025physicsinformedgenerativemachine}.

\setcounter{table}{0}
\renewcommand{\thetable}{G\arabic{table}}
\setcounter{figure}{0}
\renewcommand{\thefigure}{G\arabic{figure}}
\setcounter{equation}{0}
\renewcommand{\theequation}{G\arabic{equation}}

\section{Choice of Hyper-parameters}\label{app : hyperparams}

The hyper-parameters in Table~\ref{tab:hyperparams} were obtained by manual optimization, with the compactness of the subspace and the accuracy of the final energy used as the primary indicators of a good setting. Their appropriate values depend on several factors, including the size of the symmetry space, the size of the core space within it, and the geometry of the molecule, which sets the correlation between the orbitals, together with other chemical characteristics. We therefore found it necessary to repeat the manual optimization whenever we moved to a different system size. By contrast, the single set of hyperparameters fixed on fragments F3 and F10 gave compact subspaces across all ten fragments of the DMET calculation, which share the same active-space size. This suggests that the system size may be a stronger determinant of the relevant hyperparameters than the chemical character of the individual fragment, although the present data are limited and we regard this as a hypothesis to be tested in future work. The same question has been raised in parallel by other groups working on RBM-based configuration recovery. In PIGen-SQD~\cite{patra2025physicsinformedgenerativemachine}, the RBM learning rate, the layer sizes and, by default, the fraction of the symmetry space generated per cycle were fixed, while the optimal numbers of training and Gibbs-sampling steps were acknowledged to be potentially system dependent and left to explicit hyperparameter optimization in future work.

\begin{table*}[htbp]
\centering
\caption{Sampling and RBM subspace-expansion hyperparameters used in the standalone QSCI-RBM and DMET-QSCI-RBM experiments.}
\label{tab:hyperparams}
\resizebox{\textwidth}{!}{%
\begin{tabular}{lcccc}
\toprule
Parameter & C$_2$H$_4$ & Carmofur (DMET) & CH$_5$NO & H$_2$O\\
\midrule
Qubits used for simulation & $28$ & $32$ & $40$ & $48$ \\
Symmetry-space dimension $\binom{n_{\mathrm{orb}}}{N_\alpha}^{2}$ & $9.02\times10^{6}$ & $1.66\times10^{8}$ & $6.01\times10^{9}$ & $1.81\times10^{9}$ \\
Hardware shots & $1\times10^{5}$ & $1\times10^{5}$ & $1\times10^{5}$ & $2.5\times10^{5}$ \\
Training samples drawn from memory (\texttt{inp\_size}) & $1\times10^{3}$ & $5\times10^{3}$ & $1\times10^{5}$ & $5\times10^{4}$ \\
Random Gibbs seeds drawn (\texttt{ran\_size}) & $1\times10^{3}$ & $1\times10^{4}$ & $1\times10^{6}$ & $8\times10^{4}$ \\
RBM learning rate & $1\times10^{-2}$ & $1\times10^{-2}$ & $5\times10^{-3}$ & $1\times10^{-2}$ \\
Coefficient threshold $\tau_{\mathrm{sv}}$ & $1\times10^{-7}$ & $1\times10^{-8}$ & $1\times10^{-9}$ & $1\times10^{-8}$ \\
Max.\ new determinants per iteration $K$ & $1\times10^{3}$ & $5\times10^{3}$ & $2\times10^{4}$ & $4\times10^{3}$ \\
RBM iterations $N_{\mathrm{iter}}$ & $50$ & $50$ & $40$ & $80$ \\
\bottomrule
\end{tabular}}
\end{table*}

\setcounter{table}{0}
\renewcommand{\thetable}{H\arabic{table}}
\setcounter{figure}{0}
\renewcommand{\thefigure}{H\arabic{figure}}
\setcounter{equation}{0}
\renewcommand{\theequation}{H\arabic{equation}}

\section{Classical cost breakdown for the 48-qubit water molecule}\label{app: water_classical_memory}

We performed an analysis of the post-processing of the CAS(10,24) water molecule (cc-pVDZ basis), run for 80 iterations on an AMD Ryzen Threadripper PRO 9995WX (96 cores, 256~GB RAM) with 16 PySCF threads and the RBM on the CPU with 2 torch threads, recording the wall-clock time and the memory used by each stage above its baseline, $\Delta M$, separately for each stage (Table~\ref{tab:h2o_cost}).
The memory was measured with a profiler, that reads the process-status interface of the Linux operating system. The memory held by the program was recorded as the baseline immediately before each stage was called, after releasing memory no longer in use, and the highest memory reached by the program during the stage was recorded. $\Delta M$ is the difference between the two, and the value reported for each stage is the largest $\Delta M$ over all iterations. Memory here refers to the RAM held by the running program.

\begin{figure}[htbp]
    \centering
    \includegraphics[width=\columnwidth]{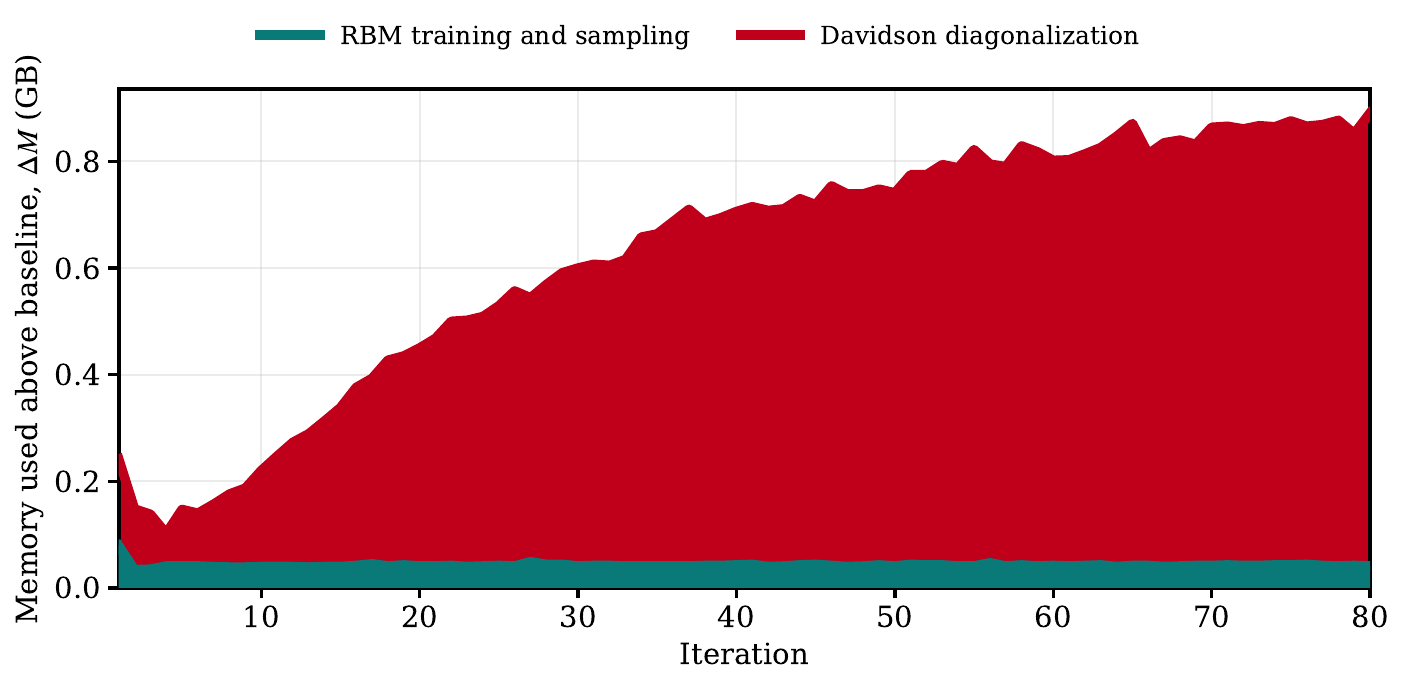}
    \caption{Memory used above baseline, $\Delta M$, per iteration for RBM training and sampling (teal) and the Davidson diagonalization (red) in the 48-qubit H$_2$O post-processing. $\Delta M$ is the additional RAM required by a stage beyond the memory already held when the stage began. The RBM requirement is nearly constant, whereas that of the diagonalization grows with the subspace dimension.}
    \label{fig:memory_per_iteration}
\end{figure}

The profiled stages total $2069$~s.
The Davidson diagonalization consumed $1908$~s ($92.2\%$) with $\Delta M = 0.89$~GB at the final subspace of $2.42\times10^{6}$ determinants, whereas RBM training, sampling and filtering together consumed $141$~s ($6.8\%$) with $\Delta M = 0.079$~GB. The memory used per iteration for RBM training and sampling, Davidson diagonalization are given in Figure~\ref{fig:memory_per_iteration}.
The reduced density matrices, which are needed only after the final iteration and were evaluated once, added $12$~s ($0.6\%$) with $\Delta M = 0.33$~GB.
The diagonalization therefore dominates both time and memory, at more than thirteen times the RBM time and about eleven times the RBM $\Delta M$.
The two costs also scale differently, since the RBM cost per iteration stays near $1.7$~s regardless of the subspace size, whereas the diagonalization cost per iteration grows with the subspace dimension, from $6.0$~s in the first iteration to $34$~s at the final iteration.
The absolute RBM time depends on the implementation, here a compiled CPU mini-batch contrastive divergence trainer with $16$ hidden units, while the diagonalization cost is set by the dimension of the subspace.
 
\begin{table}[t]
\centering
\caption{Measured classical cost of the 48-qubit H$_2$O post-processing over 80 iterations. Shares are fractions of the profiled time. The density matrices were evaluated once, after the final iteration. $\Delta M$ is the memory used by the stage above baseline, that is, the largest amount of additional RAM that the stage required beyond the memory already held by the program when the stage began (libraries, integrals and stored determinants), taken over the iterations.}
\label{tab:h2o_cost}
\begin{tabular}{lccc}
\toprule
Stage & Time (s) & Share (\%) & $\Delta M$ (GB) \\
\midrule
RBM training and sampling & 141.2 & 6.8 & 0.079 \\
Davidson diagonalization & 1907.6 & 92.2 & 0.891 \\
RDM evaluation (once) & 12.1 & 0.6 & 0.329 \\
Other (thresholding, bookkeeping) & 8.0 & 0.4 & 0.028 \\
\midrule
Total (profiled) & 2069.0 & 100 & n/a \\
\bottomrule
\end{tabular}
\end{table}

\end{document}